\documentclass[aps,amssymb,amsmath,prd,twocolumn,
showpacs,preprintnumbers,superscriptaddress,nofootinbib,floatfix]{revtex4-1}
\usepackage{graphicx}
\usepackage{orcidlink} 
\usepackage{color}
\usepackage{times}
\usepackage{amsmath}
\usepackage[normalem]{ulem}
\usepackage{inputenc}
\usepackage{bm}
\usepackage{multirow}
\usepackage{float}
\usepackage{url}
\usepackage{natbib}
\usepackage{wrapfig}

\hypersetup{colorlinks=true,linkcolor=red,urlcolor=blue,citecolor=blue}

\newcommand{\apjl}{Astrophys. J. Lett.}
\newcommand{\apjs}{Astrophys. J., Supp.}
\newcommand{\aap}{Astron. Astrophys.}

\newcommand{\mnras}{Mon. Not. R. Astron. Soc.}

\newcommand{\nphysa}{Nuclear Physics A}
\newcommand{\jcap}{ J. Cosmol. Astropart. Phys.}
\newcommand{\physrep}{Physics Reports}

\newcommand{\Msun}{\,{M}_{\odot}}
\def\fm3{\;\text{fm}^{-3}}

\newcommand{\eg}{e.g.,~}

\begin{document}

\title{Limitations in constraining neutron star radii and nuclear properties from inspiral gravitational wave detections}

\author{Zhenyu Zhu \orcidlink{0000-0001-9189-860X}}
\author{Richard O'Shaughnessy \orcidlink{0000-0001-5832-8517}}
\affiliation{Center for Computational Relativity and Gravitation, Rochester Institute of Technology, Rochester, NY 14623, USA;\\}

\begin{abstract}
    We investigate the constraints on the neutron star equation of state (EoS) and nuclear properties achievable with third-generation gravitational wave detectors using the Fisher information matrix approach within the relativistic mean field (RMF) theory. Assuming an optimistic binary neutron star (BNS) merger rate, we generate simulated inspiral gravitational wave (GW) signals corresponding to one year of observation. From these simulated data, we compute the covariance matrix and posterior distributions for nuclear properties and EoS. Our results show that the EoS can be tightly constrained, particularly in the density range between one and four times nuclear saturation density. However, due to the scarcity of low-mass neutron stars in the GW sample, the EoS at sub-saturation densities remains poorly constrained. Thus, in turn, leads to weaker constraints on neutron star radii, as the radii are sensitive to the low-density EoS. 
    Additionally, we present the expected correlations among nuclear parameters in general and plots of the inferred symmetry energy in particular, which represent degeneracies in their influence on the EoS and make them difficult to be constrained through GW observations alone.
    These highlights inherent limitations of inspiral GW signals in probing dense matter properties. Therefore, precise radius measurements, post-merger GW observations, and supplementary constraints from terrestrial nuclear experiments remain essential for a comprehensive understanding of dense matter. 
\end{abstract}

\maketitle

\section{Introduction}
Gravitational wave (GW) signals from binary neutron star (BNS) mergers have been considered as a promising tool for constraining the neutron star (NS) equation of state (EoS) and probing the properties of dense matter. Indeed, the first detection of the BNS merger event GW170817~\citep{2017PhRvL.119p1101A, 2019PhRvX...9a1001A} provided an independent constraint on the tidal deformability (TD) and EoS of NS, significantly advancing our understanding of dense matter~\citep{2018PhRvL.121p1101A, 2018PhRvL.120q2703A}. The tidal deformability, which is uniquely determined by the EoS, is encoded in the inspiral phase of the GW signal through its impact on the orbital dynamics~\citep{2019PhRvD.100d4003D}. The upcoming third-generation GW detectors, such as the Einstein Telescope (ET)~\citep{2010CQGra..27h4007P, 2012CQGra..29l4013S, 2020JCAP...03..050M} and Cosmic Explorer (CE)~\citep{2019BAAS...51g..35R}, are expected to detect a large number of BNS mergers with sufficiently high signal-to-noise ratio (SNR) to resolve the effects of tidal deformability. By accumulating the information from numerous BNS merger GWs, it may become possible to place tight constraints on the mass-tidal deformability relation, thereby enabling accurate inference of the EoS and nuclear properties of dense matter~\citep{2022arXiv220501182G, 2022PhRvD.106l3529G, 2024PhRvD.110d3013W}. On the other hand, progress in the measurements of neutron star mass, radius, and moment of inertia~\citep{2019ApJ...887L..24M, 2019ApJ...887L..21R, 2021ApJ...918L..28M, 2021ApJ...918L..27R, 2020MNRAS.497.3118H, 2021ApJ...915L..12F, 2021PhRvX..11d1050K} along with recent advancements in nuclear experiments~\citep{2021PhRvL.126q2502A, 2022PhRvL.129d2501A}, has also provided valuable insights into the properties of dense matter. The combination of more complementary observations and experiments is expected to yield a more comprehensive understanding of dense matter.

Relativistic mean-field (RMF) theory is a widely used framework for modeling the nuclear interactions and EoS of dense matter. It describes interactions among nucleons through the exchange of mesons, which couple to nucleons and mediate the nuclear force \citep{1974AnPhy..83..491W, 1998NuPhA.637..435S, 2008PhR...464..113L, 2022PhRvC.106e5804A}. RMF models have been successfully applied to describe the EoS of dense matter across the range of density and temperature relevant for NSs, BNS mergers, and core-collapse supernovae~\citep{2011ApJS..194...39F, 2013ApJ...774...17S, 2021PhRvD.104h3004Z, 2021PhRvD.104f3016R, 2022MNRAS.516.4760C, 2022MNRAS.513.1317Z, 2025arXiv250617569Z}. Moreover, RMF models offer flexible EoS parameterizations and direct connections to nuclear properties~\citep{2018PhRvC..97c5805Z, 2018ApJ...862...98Z, 2023PhRvC.108b5809Z}, enabling their inference from the observational and experimental data~\citep{2020ApJ...897..165T,2023ApJ...943..163Z, 2025arXiv250507677L}.

The symmetry energy is a key quantity in both nuclear physics and astrophysics, which characterizing the energy cost associated with varying neutron-proton asymmetry in nuclear matter. It influences a wide range of phenomena, including the neutron-skin thickness of nuclei~\citep{2021PhRvL.126q2502A, 2022PhRvL.129d2501A}, the radii of neutron stars, and the bulk viscosity during BNS mergers~\citep{2025arXiv250407805Y, 2025PhRvD.111d4074C, 2025PhRvL.134g1402C}. It also plays crucial role in the out-of-equilibrium behavior of EoS in BNS mergers~\citep{2019ApJ...875...12R}, thereby affecting neutrino processes and matter ejection. Given its significance, GW observations of BNS mergers have been regarded as a promising way to constrain the symmetry energy and other nuclear matter properties~\citep{2018ApJ...859...90Z, 2021PhRvL.127s2701E}.

Given a set of observational data, one can estimate the parameters of events with the Bayesian inference using for example the Markov Chain
Monte Carlo (MCMC) method~\citep{2015PhRvD..91d2003V,2020MNRAS.499.3295R, 2023PhRvD.107b4040W} or equivalent quadrature \cite{2018arXiv180510457L}. This approach is most reliable and accurate to obtain the posterior distributions of parameters from the real data. However, it is computationally expensive and impractical for the forecast of parameter estimation from simulated data of future detectors and observations. An alternative and efficient approach is the Fisher information matrix (FIM) method, which provides the covariance matrix of the Bayesian posterior probability distribution of parameters~\citep{1994PhRvD..49.2658C,1995PhRvD..52..848P,2013PhRvD..87b4004C, 2014PhRvD..89f4048O, 2020ApJS..250....6W, 2022ApJS..263....2I, 2022ApJ...941..208I, 2024PhRvD.109j3035H}. Consider the sensitivity of the 3rd generation GW detectors, we may expect, in an optimistic scenario, hundreds of BNS merger events to be detected in one day~\citep{2023PhRvX..13a1048A}. FIM method have been an efficient and practical way to investigate the achievable constraints on EoS and nuclear properties by ET and CE detectors~\cite{2019PhRvD.100h3010F,2023PhRvD.108l2006I,2024PhRvD.109j3035H}.

In this letter, we forecast the constraints on the EoS and nuclear properties using FIM method, based on the inspiral GW signals expected to be detected by the third-generation GW detectors over one year of observation. We generate simulated GW signals with an injected RMF EoS, a realistic NS mass distribution~\citep{2018MNRAS.478.1377A}, and an uniform BNS merger rate with an optimistic value of $p=1000\ {\rm Gpc}^{-3}\ {\rm yr}^{-1}$. Current estimates from existing data suggest a BNS merger rate in the range of $\sim 10$–$1700\ {\rm Gpc}^{-3}\ {\rm yr}^{-1}$~\citep{2023PhRvX..13a1048A}. The adopted value $p=1000\ {\rm Gpc}^{-3}\ {\rm yr}^{-1}$ represents an optimistic scenario, corresponding to a best-case outlook for constraining the EoS and nuclear properties by third-generation detectors. The FIM is computed from the simulated data with respect to nuclear parameters, rather than tidal deformabilities. The posterior distributions are obtained from the covariance matrix given by the inverse of the FIM.

\section{Methodology}
We compute the EoS of NS with the RMF model, and adopt the Lagrangian of the form as~\citep{2018PhRvC..97c5805Z, 2018ApJ...862...98Z, 2023PhRvC.108b5809Z, 2023ApJ...943..163Z}
\begin{eqnarray}
  \label{eq:Lagrangian}
    \mathcal{L}& = & \overline{\psi}\left(i\gamma_\mu \partial^\mu - M_N + g_\sigma \sigma - g_{\omega}\omega\gamma^0 - g_{\rho}\rho\tau_{3}\gamma^0\right)\psi  \nonumber \\
           && -\frac{1}{2}(\nabla\sigma)^2 - \frac{1}{2}m_\sigma^2 \sigma^2 - \frac{1}{3} g_2\sigma^3 - \frac{1}{4}g_3\sigma^4 \nonumber \\
           && + \frac{1}{2}(\nabla\omega)^2 + \frac{1}{2}m_\omega^2\omega^2 + \frac{1}{4}c_3 \omega^4\nonumber \\
           & & + \frac{1}{2}(\nabla\rho)^2 + \frac{1}{2}m_\rho^2\rho^2\ + \frac{1}{2}g_{\omega}^2\omega^2 \Lambda_v g_{\rho}^2\rho^2.
\end{eqnarray}
Where $\psi$, $\sigma$, $\omega$, and $\rho$ denote the field of nucleon, $\sigma$, $\omega$, and $\rho$ meson, respectively. The model contains 7 coupling constants, $g_\sigma$, $g_\omega$, $g_\rho$, $g_2$, $g_3$, $c_3$, and $\Lambda_v$, which are determined by nuclear properties at saturation density. Specifically, we use 7 quantities at saturation density to determine them: the saturation density $n_0$, the binding energy $E_0$, incompressibility $K_0$, skewness $J_0$, symmetry energy $E_{\rm sym}$, symmetry energy slope $L_{\rm sym}$, and the Dirac effective mass $M^*_N$. Among these, $n_0=0.16\,{\rm fm}^{-3}$ and $E_0=-16\,{\rm MeV}$ are fixed, as they are well constrained by experiments. The remaining 5 quantities are treated as free parameters and will be constrained by GW data. The injected EoS parameters are listed in Table~\ref{tab:parameters}.

\begin{table}
\centering
   \vskip-3mm
   \caption{The parameters of the EoS injection and the prior covariance of each parameters.}
    \begin{tabular}{ccccc}
        \hline\hline
        Parameters & $n_0$[${\rm fm}^{-3}$] & $E_0$[${\rm MeV}$] & $E_{\rm sym}$[${\rm MeV}$] & $L_{\rm sym}$[${\rm MeV}$] \\
        \hline
        Injection & $0.16$ & $-16$  & $31$ & $40$ \\
        prior deviation & --- & --- & $3$ & $30$ \\
        \hline
        \hline
        Parameters & $K_0$[${\rm MeV}$] & $J_0$[${\rm MeV}$] & $M^*_N/M_N$ \\
         \hline
        Injection & $260$ & $-350$ & $0.7$ \\
        prior deviation & $40$ & $300$ & $0.2$ \\
        \hline\hline
    \end{tabular}
    
  \label{tab:parameters}
\end{table}

The mass, radius and tidal deformability of a NS can be obtained by solving the Tolman-Oppenheimer-Volkoff (TOV) and tidal equations~\citep{2010PhRvD..81l3016H, 2020PhRvD.102h4058Z} with a given EoS. In the context FIM analyses, it is necessary to compute the derivatives of GW signals with respect to both the mass and EoS-dependent tidal deformability. To directly infer the posteriors of the EoS and nuclear properties, we compute the derivatives with respect to the EoS parameters listed in Table~\ref{tab:parameters}, rather than the tidal deformability.  To connect derivatives with respect to EOS parameters to (functional) derivatives of observables with respect to the EOS,  we implemented recently-developed analytical formulations for the variational TOV and tidal equations~\citep{2025PhRvD.111g4026L}, which provide more robust and efficient derivatives than the numerical methods. However, the original equations expressed in terms of radial coordinate $r$ have an initially unknown integration domain, which may lead to problematic exterior boundary condition for some EoS, and also makes the evaluation of variations with respect to EoS infeasible. To address these issues, we reformulate the TOV and tidal equations in terms of log-enthalpy $\log h=\log((e+p)/\rho)$~\citep{1992ApJ...398..569L, 1998PhRvD..58b4008L, 2008ApJ...677.1216H, 2010PhDT........18P, steil2017structure}, which provides a well-defined stellar surface and exterior boundary
\begin{eqnarray}
  \label{eq:tt_h}
  \frac{dr^2}{d\log h} & = & \mathcal{K}_{r^2} = -\frac{2r^2(1-2z)}{4\pi r^2 p + z}, \\
  \frac{dz}{d\log h} & = & \mathcal{K}_{z} = \left(2\pi e - \frac{z}{2r^2}\right)\frac{dr^2}{d\log h}, \\
  \frac{d^2 H}{d\log h^2} & = & A_H \frac{dH}{d\log h} - B_H H.
\end{eqnarray}
Where $R=r(0)$ and $M=R z(0)$ represent the radius and mass of the NS, respectively. The energy density, pressure and rest mass density are denoted by $e$, $p$, and $\rho$. The function $H$ is the perturbed metric function induced by the tidal field, and its relation to the tidal deformability is detailed in Ref.~\citep{2020PhRvD.102h4058Z}. The coefficients $A_H$ and $B_H$ are given by
\begin{eqnarray}
\label{eq:tidal_coef}
    A_H & = & \frac{4\pi r^2(1-2z)(e+3p)}{(4\pi r^2 p + z)^2}, \\
    B_H & = & \frac{A_H}{e+3p}\left(4e+8p+(p+e)(1+\kappa) - \frac{3}{2\pi r^2} \right)- 4, \nonumber \\
\end{eqnarray}
where $\kappa=1/c_s^2$. The corresponding variational equations can then be written as
\begin{eqnarray}
  \label{eq:tov_tidal_var}
  \frac{d \Delta_{r^2}}{d\log h} & = & \mathcal{K}_{r^2, r^2} \Delta_{r^2} + \mathcal{K}_{r^2, z} \Delta_{z} + \mathcal{K}_{r^2, p} \Delta_{p} \\
  \frac{d \Delta_{z}}{d\log h} & = & \mathcal{K}_{z, r^2} \Delta_{r^2} + \mathcal{K}_{z, z} \Delta_{z} + \mathcal{K}_{z, p} \Delta_{p} + \mathcal{K}_{z, e} \Delta_{e}, \\
  \frac{d \Delta_{H}}{d\log h} & = & \Delta_{H'}, \\
  \frac{d \Delta_{H'}}{d\log h} & = & A_H \Delta_{H'} - B_H \Delta_{H} + \sum_{i} (A_{H, i} H' - B_{H, i} H) \Delta_{i}, \nonumber \\
\end{eqnarray}
where $\mathcal{K}_{r^2,i}$, $\mathcal{K}_{z,i}$, $A_{H, i}$, and $B_{H, i}$ denote the corresponding derivatives with respect to $i=r^2,z,e,p,\kappa$. The quantities $\Delta_e$, $\Delta_p$, and $\Delta_\kappa$ are the functions of $\log h$ and represent the variations of EoS, which will be computed with numerical differentiation.

\begin{figure*}[t!]
\vspace{-0.3cm}
{\centering
\includegraphics[width=0.33\textwidth]{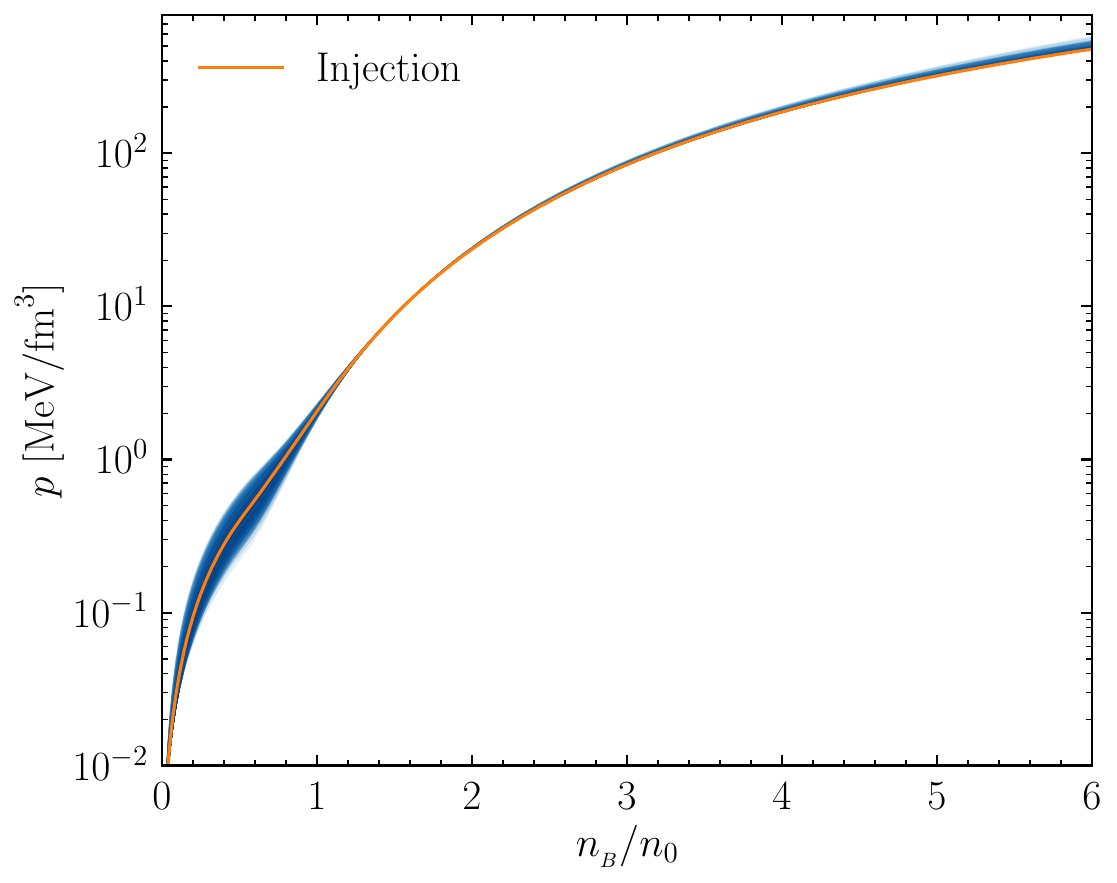}
\includegraphics[width=0.33\textwidth]{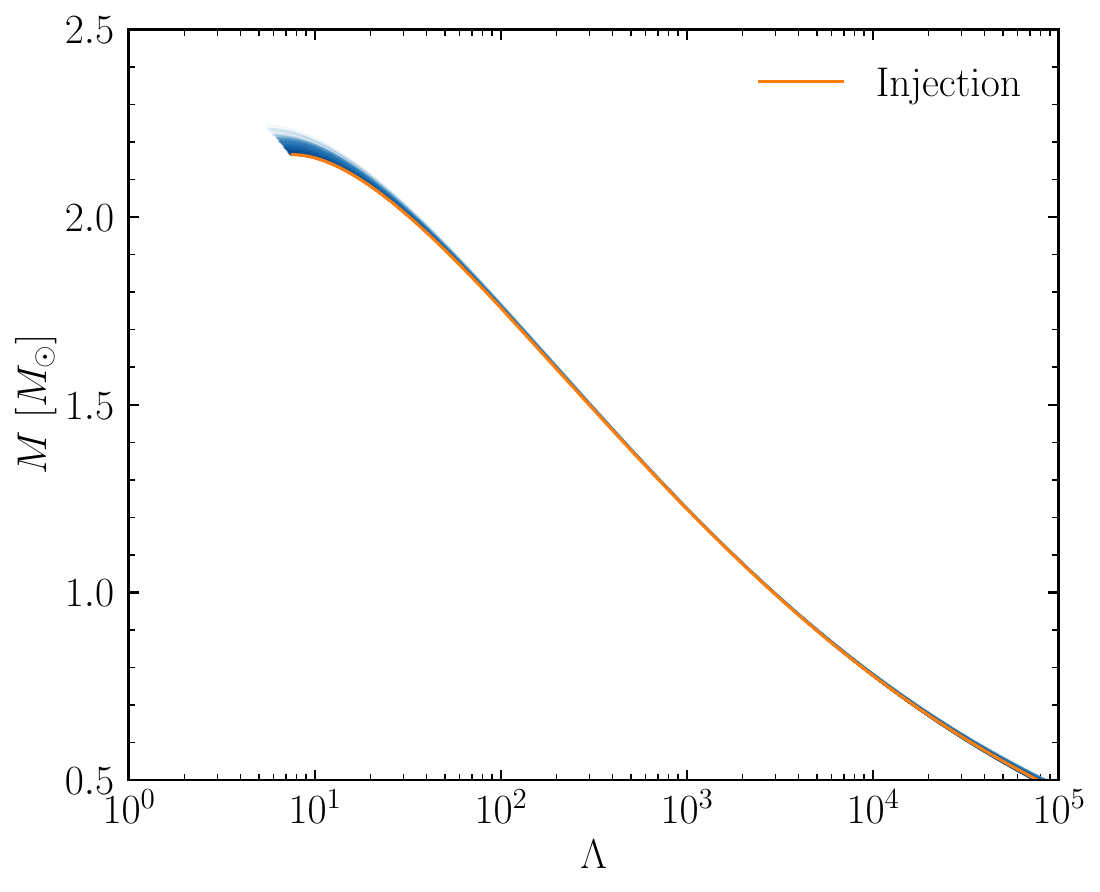}
\raisebox{-0.07cm}{\includegraphics[width=0.33\textwidth]{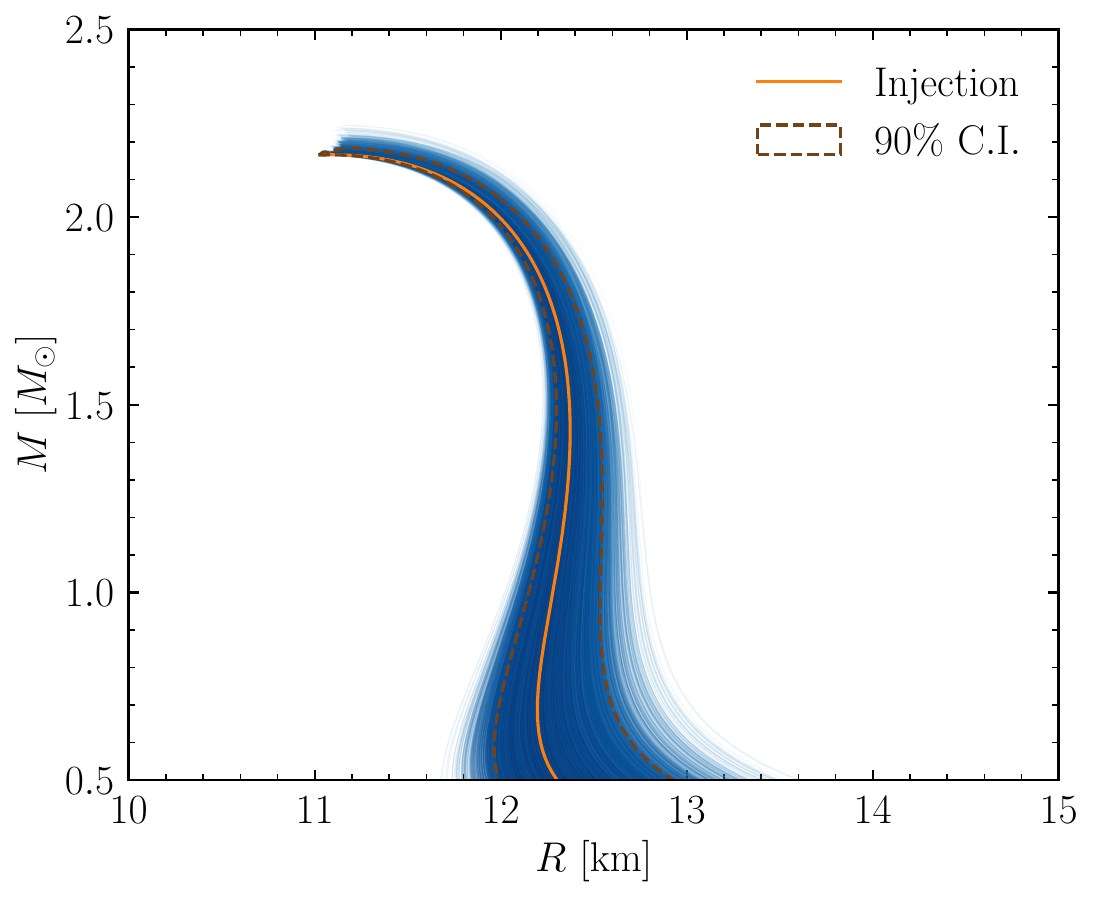}}}
\caption{The posteriors of the neutron star EoS (left panel), mass-$\Lambda$ relation (middle panel) and mass-radius relation (right panel). The blue lines represent our posterior samples, while the orange line indicates the injected results. We observe tight constraints on tidal deformabilities and EoS, particularly in the density range between $1$ and $4$ times nuclear saturation density. However, the radii of NS remain weakly constrained by the GW observations.}
\label{fig:posteriors}
\vspace{-0.1cm}
\end{figure*}

In contrast to the original formulation, which solely relies on the relation of $e(p)$, the log-enthalpy form of the TOV and tidal equations implicitly applies the first law of thermodynamics, and thus requires thermodynamic consistency of EoS. Deviations from these thermodynamic relations in EoS can lead to inaccurate results. Such inconsistencies often arise when matching core and crust EoS derived from different models. To avoid this issue, we parameterize the inner-crust EoS using a polynomial function that smoothly and consistently connects both the outer-crust BPS EoS~\citep{1971ApJ...170..299B} and our RMF-based core EoS~\citep{zhu2025inprep}.
\begin{eqnarray}
  \label{eq:ic_param}
  \log h(t) & = & [\log h(1) - \log h(0)]t + \log h(0) \nonumber \\
            &   & + t(1-t)(A_0 + A_1 t + A_2 t^2 ),
\end{eqnarray}
where $t=(\log n - \log n_{\rm io})/(\log n_{\rm cc} - \log n_{\rm io})$ ranges from 0 to 1, and $n_{\rm io}$ and $n_{\rm cc}$ represent the number densities at inner-outer crust and core-crust interfaces, respectively. The parameters $A_0$, $A_1$, and $A_2$ are determined by requiring continuity of sound speed and pressure at the interfaces. We set the core-crust interface at $n_{\rm cc}=0.08\,{\rm fm}^{-3}$.
It is remarkable that we impose the monotonicity of sound speed and thermodynamic consistency in the inner-crust EoS through this parameterization. 
These conditions cannot be satisfied by arbitrary combinations of nuclear parameters and core EoS, \eg a core EoS with small $E_{\rm sym}$ and large $L_{\rm sym}$ cannot be smoothly stitched to the outer-crust BPS EoS, regardless of how the inner-crust EoS is adjusted. Therefore, these conditions effectively act as a prior on the nuclear parameters.

We randomly generate simulated BNS merger signals over one year with vanishing spin using the \texttt{IMRPhenomD\_NRTidalv2} waveform template~\citep{2016PhRvD..93d4007K, 2019PhRvD.100d4003D}, considering events up to redshift $z=5$. 
Note that \texttt{NRtidalv2} model only accounts for the equal-mass tidal effects and neglects the asymmetric
tidal parameter. However, given the relatively small mass ratio in our BNS population, approximately $95\%$ of BNSs
have mass ratios $q < 1.5$, the impact of unequal-mass tidal deformabilities on the waveform is
minimal~\citep{2019PhRvD.100d4003D}. While future, more sophisticated waveform models may provide more information (e.g., via more
tidal information and particularly via postmerger and viscous physics), this modeling approach allows us to construct
a conservative estimate of the EOS information accessible via a third-generation BNS census, and likely has a
negligible influence on our qualitative conclusions.

On the other hand, there exist systematic uncertainties between the true astrophysical waveforms and the waveform model being used. These uncertainties can shift the inferred EoS away from the truth and introduce biases and errors. However, they can be mitigated by comparing results obtained with multiple waveform models, at the cost of some accuracy. Such losses in accuracy can easily be compensated by accumulating a larger number of events over longer observing periods.

A uniform spatial distribution of BNS merger events is used under the Planck 2018 model, along with an optimistic merger rate of $p=1000\ {\rm Gpc}^{-3}\ {\rm yr}^{-1}$, which is consistent with the current estimation between $10$ and $1700$ ${\rm Gpc}^{-3}\ {\rm yr}^{-1}$~\citep{2023PhRvX..13a1048A}. The component masses of BNS systems are assumed to be independent and drawn from a two-component Gaussian mixture distribution with a sharp cut-off at maximum mass of NS~\citep{2018MNRAS.478.1377A}
\begin{eqnarray}
  \label{eq:mass_dist}
  P(m) & = & \sum_{i=1,2} r_i \mathcal{N}(m; \mu_i, \sigma_i) \Theta(m-m_{\rm max})/\Phi_i,
\end{eqnarray}
where $\mathcal{N}(m; \mu_i, \sigma_i)$ denotes a Gaussian distribution with mean $\mu_i$ and standard deviation $\sigma_i$, and $\Theta$ is the step function introducing a cut-off at $m_{\rm max}=2.17\,\Msun$, corresponding to the NS maximum mass for our injected EoS. The weights $r_1=0.65$, $r_2=1-r_1$, and normalization factors $\Phi_i$ ensure that $P(m)$ can be normalized to 1. We adopt the best-fit parameters $\mu_1=1.34$, $\sigma_1=0.07$, and $\mu_2=1.80$, $\sigma_2=0.21$ from Ref.~\citep{2018MNRAS.478.1377A}. 

In summary, we generate simulated GW signals from BNS systems with the injected EoS, the mass distribution described above, and a uniform spatial distribution with an optimistic merger rate. In our following FIM analyses, we treat the likelihood of each individual events independently and accumulate the information on the EoS parameters from all observable events. However, the measurements also carry information about the underlying mass distribution, and can therefore contribute to constraints of NS maximum mass, yielding tighter EoS uncertainties. In this work, we focus on constraints from tidal deformabilities and leave constraints from the mass distribution for future work.

The likelihood of EoS parameters after accumulating all events can be expressed as
\begin{eqnarray}
  \label{eq:likelihood_eos}
  \mathcal{L}_{\rm net}(\theta_{\rm EOS}) & = & \prod_k \int d\theta' \mathcal{L}_k(\theta', \theta_{\rm EOS}),
\end{eqnarray}
where $\theta_{\rm EOS}$ and $\theta'$ denote the EoS parameters and the rest parameters, respectively. $\mathcal{L}_k$ denote the likelihood of each individual event. The integral over $\theta'$ marginalizes the likelihood over all other parameters except EoS parameters. The likelihoods of each individual event are approximated by a multivariate Gaussian distribution with the covariance matrix given by the inverse of the FIM. The FIM is defined as
\begin{eqnarray}
  \label{eq:fisher}
  F_{ij} & = & \sum_{\rm d} F_{ij}^{\rm d} = \sum_{\rm d} \left< \frac{\partial h^{\rm d}(f, \vec{\theta})}{\partial \theta_i} \Big| \frac{\partial h^{\rm d}(f, \vec{\theta})}{\partial \theta_j} \right>_{\rm d},
\end{eqnarray}
where $h^{\rm d}(f, \theta)$ denotes the GW strain in the frequency domain for detector $d$, and $\theta$ represents the set of parameters describing the GW signal. The inner product $\left<a|b\right>$ is defined as
\begin{eqnarray}
  \label{eq:inner_product}
  \left<a|b\right>_{\rm d} & = & 2\int_{-\infty}^{\infty} \frac{a^\ast(f)b(f) + a(f)b^\ast(f)}{S_n^{\rm d}(|f|)} df,
\end{eqnarray}
where $S_n^{\rm d}(f)$ denotes the one-sided noise power spectral density of detector $d$. Finally, we put Eq.~\ref{eq:fisher} into Eq.~\ref{eq:likelihood_eos} we obtain the FIM for the EoS parameters after accumulating all events.
\begin{eqnarray}
  \label{eq:fisher_eos_tot}
  F^{\rm net}_{ij} & = & \sum_{\rm k} F_{ij}^{\rm k}(\theta_{\rm EOS}; \theta'),\ \ \ (\theta_i, \theta_j \in \theta_{\rm EOS})
\end{eqnarray}

Using the simulated GW signals, we compute the signal-noise ratio (SNR), derivatives of GW strains, as well as the Fisher matrix with respect to the detectors ET in Sardinia and CE1 in Idaho by using the \texttt{GWFast} package~\citep{2022ApJ...941..208I, 2022ApJS..263....2I}. In our analyses, the EoS-dependent tidal deformability in the waveform model is replaced by the EoS parameters. We then marginalize over the extrinsic parameters and retain only the Fisher matrix components corresponding to the EoS parameters. The total Fisher matrix is accumulated over all events with SNR larger than 8, resulting in $\sim$ 250,000 events. 
Given the large number of events and the high SNR threshold, the linearized-signal approximation with respect to the EoS parameters is sufficiently accurate for applying the FIM method~\citep{2008PhRvD..77d2001V}.
To incorporate prior knowledge on the saturation properties, we introduce a prior Fisher matrix, defined as the inverse of the diagonal covariance matrix whose standard deviations are given in Table~\ref{tab:parameters}. 
Note that we fixed the saturation density $n_0$ and binding energy $E_0$ in our analyses, as they are well constrained by nuclear experiments comparing with other parameters, and have negligible impact on the EoS under their current constraints. The prior of other parameters are chosen to be consistent with current experimental and observational constraints~\citep{2013PhLB..727..276L, 2023PhRvC.108b5809Z, 2023ApJ...943..163Z}.
The posterior Fisher matrix is obtained by summing the prior and event Fisher matrices. Finally, the posterior distributions of EoS parameters are derived with the injection values as the mean and the inverse of the Fisher matrix as the covariance.

\begin{figure*}
\vspace{-0.3cm}
{\centering
\includegraphics[width=0.7\textwidth]{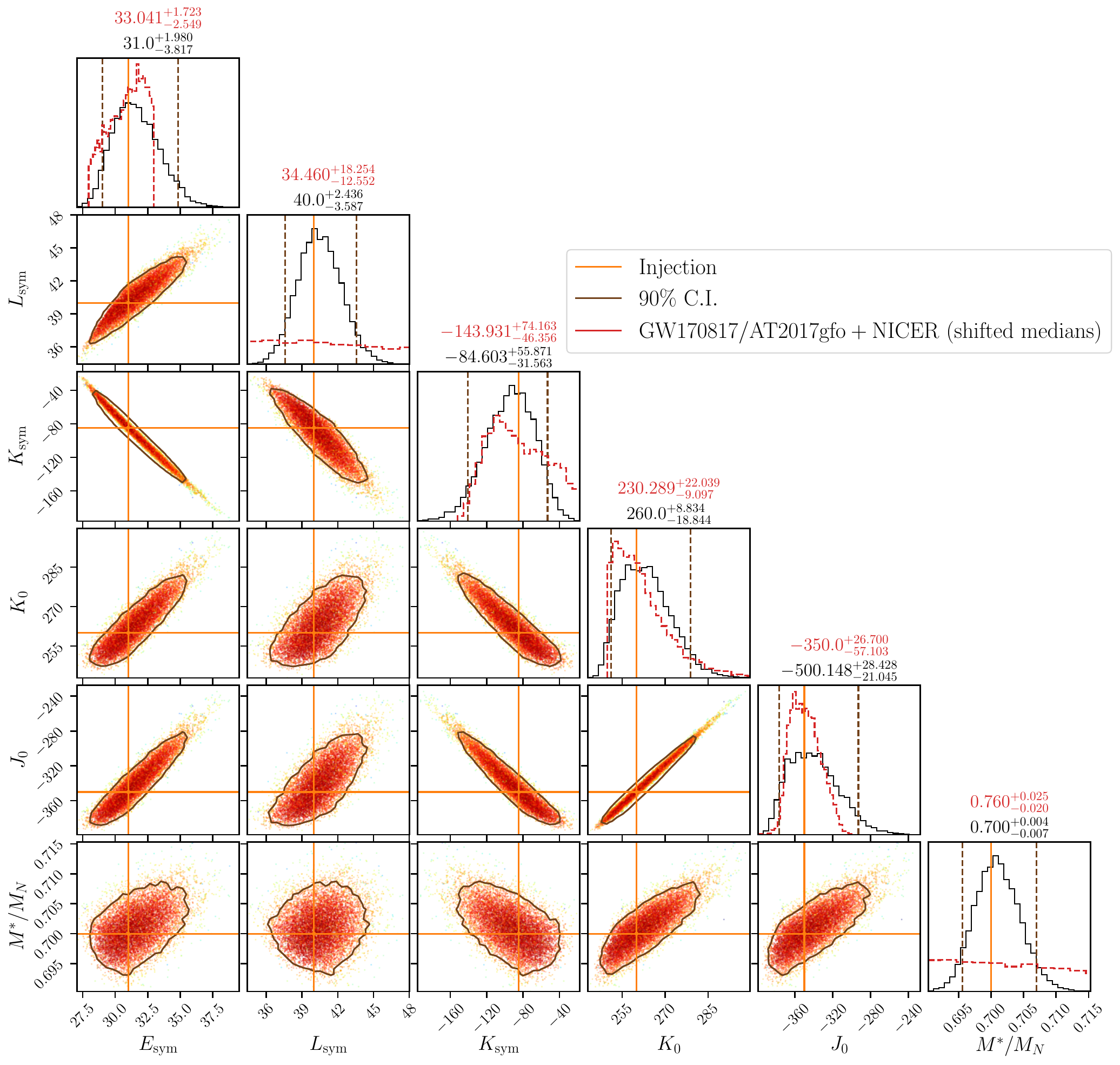}}
\caption{The corner plot for the nuclear properties. The injection and $90\%$ confidence intervals are denoted by the orange and brown lines, respectively. The posteriors from the Bayesian inference of GW170817/AT2017gfo and NICER data~\citep{2023ApJ...943..163Z} are also included as red lines in the diagonal 1D marginalized plots for comparison. Their median values are shifted to align with our injected values in order to facilitate the uncertainties comparison. The median values and $90\%$ confidence intervals are annotated at the top of each 1D marginalized plot for both posteriors of fisher matrix and GW170817/AT2017gfo + NICER. We observe little improvements in the constraints and strong correlations among most of nuclear properties, indicating their degeneracies in the neutron star EoS and tidal deformabilities.}
\label{fig:corner}
\vspace{-0.1cm}
\end{figure*}

\section{Results}

We observe tight constraints on the tidal deformabilities of NS with masses between $1.0$ and $2.0\,\Msun$ in Fig.~\ref{fig:posteriors}. Although there are few observations of NSs with mass below $1.2\,\Msun$ in our mass distribution~\citep{2018MNRAS.478.1377A}, their TDs---possessing larger values, exerting stronger impact on the waveform, and being relatively insensitive to the low-density EoS---lead to negligible uncertainties in this mass range. In contrast, the TDs of massive NSs (\( \gtrsim 2.0\,\Msun \)) are significantly smaller and merely have weak influences on the waveform, resulting in larger uncertainties. These uncertainties may be mitigated by precise mass measurement of massive neutron stars, which can provide constraints on the maximum mass~\citep{2021ApJ...915L..12F}. These tight constraints on TDs between $1.0$ and $2.0\,\Msun$ also translate into small uncertainties in EoS at the density between $2n_0$ and $4n_0$, which correspond to the central densities of NSs in this mass range. Furthermore, the EoS between $n_0$ and $2n_0$ is also well constrained, as the tidal deformabilities in this mass range are relatively sensitive to it.

In contrast to the supra-saturation EoS, we observe large uncertainties in the EoS at sub-saturation density and inner crust region. Due to the insensitivity of TDs of medium-mass ($1.2$-$2.0\,\Msun$) NSs to the low-density EoS, the effects of sub-saturation and crust EoS become degenerate in the GW observations. This degeneracy, combined with the scarcity of low-mass (below  $1.2\,\Msun$) NSs in the observed population, results in poorly constrained sub-saturation EoS. However, the radii of neutron stars in the $1.2$-$2.0\,\Msun$ range are sensitive to the low-density EoS~\citep{2016PhRvC..94c5804F, 2020PhRvC.101a5806P, 2021PhRvC.104a5801S}. Consequently, neutron star radii remain weakly constrained by inspiral GW observations, as illustrated in the right panel of Fig.~\ref{fig:posteriors}. As an example, the radius of $1.4\,\Msun$ NS, $R_{1.4}$, shows a spread of up to $500$ meters between the maximum and minimum posterior samples, while its tidal deformability exhibits inconspicuous uncertainty. 

Notably, the sub-saturation density and crust EoS are already well constrained by nuclear physics~\citep{2016PhRvC..93e4314D, 2024A&A...687A..44D, 2024PhRvD.109j3015S, 2025arXiv250516929K}. Incorporating this knowledge can place additional constraints on nuclear properties, narrow the EoS uncertainties in these regions, and thereby reduce the uncertainties in neutron star radii. 
However, based on our results, the GW observations alone are unable to provide strong constraints in this
  density regime, emphasizing both their limitations and the importance of continued nuclear experiments and precise
  radius measurements; see, e.g. \cite{2025JPhG...52e0501B}.

\section{Degeneracy on nuclear properties}

The tidal deformability measurements of NS from the inspiral GW observations exhibits its limitations in constraining the radii. Due to the degeneracy of sub-saturation EoS effects on TDs, these constraints do not effectively extend to the low-density region of EoS. Moreover, degeneracies also present among the nuclear properties: different combinations of nuclear properties can produce very similar EoS in the density range between $n_0$ and $4n_0$, along with nearly identical tidal deformabilities for NSs with masses between $1.2\,\Msun$ and $2.0\,\Msun$, despite exhibiting diverse behaviors at sub-saturation densities.
This degeneracy arises from the fact that the nuclear properties influence both the symmetric and asymmetric
  nuclear matter EoSs, which are two-dimensional functions depending on both density and proton fraction. However, the
  NS EoS is barotropic due to the $\beta$-equilibrium condition; see more discussions in Appendix  \ref{sec:appendix}.

We illustrate these degeneracies among nuclear properties in Fig.~\ref{fig:corner}, which displays the corner plot of the posterior distributions for nuclear parameters. The injection are denoted by the orange lines, while the $90\%$ confidence intervals are represented by the brown solid lines for the 2D joint posteriors and dashed lines for the 1D marginalized posteriors along the diagonal. In the framework of Fisher information matrix, the posteriors are expected to follow a multivariate Gaussian distribution. However, When generating posterior samples from this Gaussian distribution, the RMF model itself imposes physical constraints by excluding certain combinations of parameters that yield unphysical EoS for NS or nuclear matter. 
Some combinations of parameters can lead to non-convergence when solving RMF equation of motion, and to a non-positive sound speed in the EoS, such cases are unphysical and are therefore automatically excluded from our analyses.

As a result, the posteriors in Fig.~\ref{fig:corner} slightly deviate from a pure Gaussian shape.
Additionally, we also include the curvature of the symmetry energy, $K_{\rm sym}$, in the corner plot. Although this parameter is not a free input in our framework, it is derived from the nuclear matter EoS. We include it to facilitate comparisons with other studies that often report constraints on $K_{\rm sym}$~\citep{2018ApJ...859...90Z}. On the other hand, the effective mass $M^\ast$ is displayed due to its importance in RMF theory and its impact on the EoS stiffness and neutron star maximum mass~\citep{2018PhRvC..98f5804H}.

We also present the posterior distributions of each nuclear properties from the Bayesian inference of GW170817/AT2017gfo and NICER data in the 1D marginalized plots as red lines~\citep{2023ApJ...943..163Z}. Since their median values differ from our injection, we shift these distributions to align with our injected values to facilitate a direct comparison of uncertainties. The $90\%$ confidence intervals and median values are annotated at the top of each 1D marginalized plot, together with the uncertainties from our Fisher matrix posteriors.

The posteriors from GW170817/AT2017gfo and NICER data represent the current level of constraints on nuclear properties with existing observations, which include one BNS merger event~\citep{2017PhRvL.119p1101A, 2017ApJ...851L..21V} and two NS radius measurements~\citep{2019ApJ...887L..24M, 2019ApJ...887L..21R, 2021ApJ...918L..28M, 2021ApJ...918L..27R}. However, we note that even with $\sim 250,000$ BNS merger events detected by third-generation GW observatories, the constraints on most nuclear properties, except for the symmetry energy slope and Dirac effective mass, show little improvement. Although most nuclear properties vary over a wide range, their combinations yield very similar EoS in the density range between $n_0$ and $4n_0$, implying their degeneracies in the neutron star EoS and highlighting the limitations of inspiral GW signals in constraining nuclear properties. These degeneracies were also reported and discussed in Ref.~\citep{2022PhRvD.105h3016M, 2023PhRvD.108l2006I, 2024PhRvC.109b5804I, 2025arXiv250415893W}, where it was found that these properties cannot be recovered with a precision comparable to that of the EoS itself.
On the other hand, the uncertainties in the symmetry energy slope $L_{\rm sym}$ and Dirac effective mass $M^\ast/M_N$ are significantly reduced. The red lines in the 1D plots for $L_{\rm sym}$ and $M^\ast/M_N$ shown flat across the axis range, indicating much broader distributions compared to the Fisher matrix posteriors. 
Since $L_{\rm sym}$ represents the slope of symmetry energy and is more sensitive to the high-density, asymmetric NS EoS, it is expected to be better constrained by GW signals than $E_{\rm sym}$. Additionally, the $E_{\rm sym}$ uncertainty is already small given our current nuclear-physics knowledge, and thus the improvement from GW signals is limited.

The degeneracies can also be observed in these 2D joint posteriors. The 2D marginalized posteriors show strong correlations among the nuclear properties, particularly between $E_{\rm sym}$ and $K_{\rm sym}$, as well as between $K_0$ and $J_0$. These correlations indicate that their impacts on neutron star EoS and TDs are similar, allowing certain combinations to produce nearly identical results. In contrast, the correlations involving $L_{\rm sym}$ and $M^\ast/M_N$ are relatively weak, which is consistent with their weak degeneracies and tighter constraints from GW observations.

The true EoS of NS may not be fully captured by the RMF Lagrangian in Eq.~\ref{eq:Lagrangian}, and adopting different nuclear models or Lagrangians may lead to different posteriors of nuclear properties and their correlations. However, the degeneracies and correlations among them are expected to persist, as they arise from the intrinsic degeneracies of nuclear matter. This can be understood by noting that the cold, $\beta$-equilibrated NS EoS is effectively a one-dimensional function of density, whereas the nuclear-matter EoS is a two-dimensional function of both density and proton fraction. Moreover, a more complicated nuclear model could even strengthen these degeneracies, since it introduces additional parameters.

\begin{figure}
\vspace{-0.3cm}
{\centering
\includegraphics[width=0.49\textwidth]{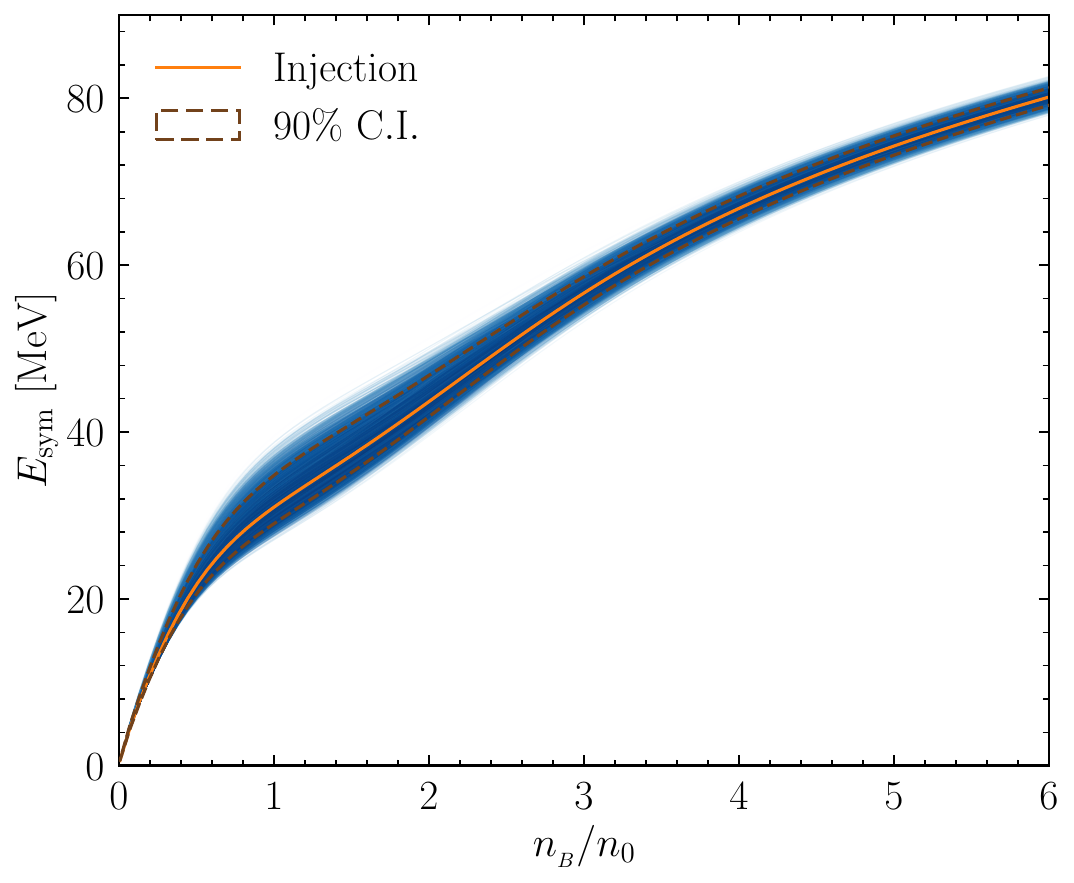}}
\caption{Posterior symmetry energy as a function of number density, with notations as in Fig.~\ref{fig:posteriors}.}
\label{fig:Esym_post}
\vspace{-0.1cm}
\end{figure}

Finally, we show the symmetry energy $E_{\rm sym}(n_{_B})$ of nuclear matter as a function of density in Fig.~\ref{fig:Esym_post}. The tight constraint on the slope $L_{\rm sym}$ leads to a well-determined trend in the symmetry energy posteriors at supra-saturation densities, reflecting improved constraints at high densities. Nevertheless, compared to the EoS, whose relative uncertainties remain below $3\%$ between $n_0$ and $4n_0$, the symmetry energy exhibits significantly larger uncertainties over the same density range. These uncertainties typically range from $5\%$ to $10\%$ and can reach up to $20\%$ near $n_0$, indicating that the symmetry energy remains less well constrained than the EoS. This highlights the limitations of using inspiral GW signals alone to precisely determine the density dependence of the symmetry energy.

\section{Conclusion}
We investigated the achievable constraints on the neutron star EoS, radii and nuclear properties using the inspiral GW signals from BNS mergers detected by third-generation GW detectors over one year of observation. Adopting an optimistic BNS merger rate of $p=1000\ {\rm Gpc}^{-3}\ {\rm yr}^{-1}$, we generated simulated GW signals observable by ET and CE detectors. Using the Fisher information matrix approach, we compute the posterior distributions of EoS, mass-radius relation and nuclear properties. Due to the optimistic merger rate, our results represent a best-case scenario for constraining these quantities with inspiral GW signals by third-generation detectors. We found that the EoS can be tightly constrained, particularly in the density range between one and four times nuclear saturation density. However, the scarcity of low-mass NSs in the observed population leads to poor constraints on the EoS at sub-saturation densities, which in turn limits the precision of NS radius.
Additionally, we present the correlations among most of the nuclear properties in the corner posterior distributions, which represent their intrinsic degeneracies on determining NS EoS and result in weak constraints and little precision improvement over current observational bounds.
As a consequence, the symmetry energy remains poorly constrained, even with data from third-generation detectors. 
Moreover, these degeneracies would persist, and could even be strengthened, if different nuclear models were adopted for the analyses, since they remain even when the EoS is tightly constrained by 3G GW observations.

Previous studies have also implied these degeneracies~\citep{2021PhRvC.104f5804E, 2022PhRvD.105h3016M, 2023PhRvD.108l2006I, 2024PhRvC.109b5804I, 2025arXiv250415893W}. \eg Iacovelli et al.\citep{2023PhRvD.108l2006I} reported multi-peaked posterior structures for the nuclear parameters, but they did not present the correlations or quantify the precision of the inferred EoS; in particular, the weak constraints at sub-saturation and inner-crust densities through GW observations found in this work were not explicitly discussed. Meanwhile, Wouters et al~\citep{2025arXiv250415893W} displayed similar degeneracies by recovering the nuclear parameters from EoS samples, but they focused more on optimization issues than on the physical implications; Imam et al~\citep{2024PhRvC.109b5804I} discussed different correlations that between nuclear parameters and tidal deformability, but did not address the correlations among nuclear parameters that lead to degeneracies.

Overall, our findings highlight the inherent limitations of inspiral GW signals in probing dense matter properties, emphasizing the importance of continued precise radius measurements, post-merger GW observations, and supplementary constraints from terrestrial nuclear experiments.

\begin{acknowledgements}
  ROS and ZZ acknowledge support from NSF AAG 2206321. ROS also acknowledges support from NSF PHY 2309172 and 2207920. This material is based upon work supported by NSF's LIGO Laboratory which is a major facility fully funded by the National Science Foundation.
  The authors are grateful for computational resources provided by the LIGO Laboratory and supported by National Science Foundation Grants PHY-0757058 and PHY-0823459.
\end{acknowledgements}

\bibliographystyle{apsrev4-1}
\bibliography{fisher}

\begin{thebibliography}{87}%
\makeatletter
\providecommand \@ifxundefined [1]{%
 \@ifx{#1\undefined}
}%
\providecommand \@ifnum [1]{%
 \ifnum #1\expandafter \@firstoftwo
 \else \expandafter \@secondoftwo
 \fi
}%
\providecommand \@ifx [1]{%
 \ifx #1\expandafter \@firstoftwo
 \else \expandafter \@secondoftwo
 \fi
}%
\providecommand \natexlab [1]{#1}%
\providecommand \enquote  [1]{``#1''}%
\providecommand \bibnamefont  [1]{#1}%
\providecommand \bibfnamefont [1]{#1}%
\providecommand \citenamefont [1]{#1}%
\providecommand \href@noop [0]{\@secondoftwo}%
\providecommand \href [0]{\begingroup \@sanitize@url \@href}%
\providecommand \@href[1]{\@@startlink{#1}\@@href}%
\providecommand \@@href[1]{\endgroup#1\@@endlink}%
\providecommand \@sanitize@url [0]{\catcode `\\12\catcode `\$12\catcode
  `\&12\catcode `\#12\catcode `\^12\catcode `\_12\catcode `\%12\relax}%
\providecommand \@@startlink[1]{}%
\providecommand \@@endlink[0]{}%
\providecommand \url  [0]{\begingroup\@sanitize@url \@url }%
\providecommand \@url [1]{\endgroup\@href {#1}{\urlprefix }}%
\providecommand \urlprefix  [0]{URL }%
\providecommand \Eprint [0]{\href }%
\providecommand \doibase [0]{http://dx.doi.org/}%
\providecommand \selectlanguage [0]{\@gobble}%
\providecommand \bibinfo  [0]{\@secondoftwo}%
\providecommand \bibfield  [0]{\@secondoftwo}%
\providecommand \translation [1]{[#1]}%
\providecommand \BibitemOpen [0]{}%
\providecommand \bibitemStop [0]{}%
\providecommand \bibitemNoStop [0]{.\EOS\space}%
\providecommand \EOS [0]{\spacefactor3000\relax}%
\providecommand \BibitemShut  [1]{\csname bibitem#1\endcsname}%
\let\auto@bib@innerbib\@empty
\bibitem [{\citenamefont {{Abbott}}\ \emph {et~al.}(2017)\citenamefont
  {{Abbott}}, \citenamefont {{Abbott}}, \citenamefont {{Abbott}}, \citenamefont
  {{Acernese}}, \citenamefont {{Ackley}}, \citenamefont {{Adams}},
  \citenamefont {{Adams}}, \citenamefont {{Addesso}}, \citenamefont
  {{Adhikari}}, \citenamefont {{Adya}},\ and\ \citenamefont
  {et~al.}}]{2017PhRvL.119p1101A}%
  \BibitemOpen
  \bibfield  {author} {\bibinfo {author} {\bibfnamefont {B.~P.}\ \bibnamefont
  {{Abbott}}}, \bibinfo {author} {\bibfnamefont {R.}~\bibnamefont {{Abbott}}},
  \bibinfo {author} {\bibfnamefont {T.~D.}\ \bibnamefont {{Abbott}}}, \bibinfo
  {author} {\bibfnamefont {F.}~\bibnamefont {{Acernese}}}, \bibinfo {author}
  {\bibfnamefont {K.}~\bibnamefont {{Ackley}}}, \bibinfo {author}
  {\bibfnamefont {C.}~\bibnamefont {{Adams}}}, \bibinfo {author} {\bibfnamefont
  {T.}~\bibnamefont {{Adams}}}, \bibinfo {author} {\bibfnamefont
  {P.}~\bibnamefont {{Addesso}}}, \bibinfo {author} {\bibfnamefont {R.~X.}\
  \bibnamefont {{Adhikari}}}, \bibinfo {author} {\bibfnamefont {V.~B.}\
  \bibnamefont {{Adya}}}, \ and\ \bibinfo {author} {\bibnamefont {et~al.}}
  (\bibinfo {collaboration} {LIGO Scientific Collaboration and Virgo
  Collaboration}),\ }\href {\doibase 10.1103/PhysRevLett.119.161101} {\bibfield
   {journal} {\bibinfo  {journal} {Phys. Rev. Lett.}\ }\textbf {\bibinfo
  {volume} {119}},\ \bibinfo {eid} {161101} (\bibinfo {year} {2017})},\ \Eprint
  {http://arxiv.org/abs/1710.05832} {arXiv:1710.05832 [gr-qc]} \BibitemShut
  {NoStop}%
\bibitem [{\citenamefont {{Abbott}}\ \emph {et~al.}(2019)\citenamefont
  {{Abbott}}, \citenamefont {{Abbott}}, \citenamefont {{Abbott}}, \citenamefont
  {{Acernese}}, \citenamefont {{Ackley}}, \citenamefont {{Adams}},
  \citenamefont {{Adams}}, \citenamefont {{Addesso}}, \citenamefont
  {{Adhikari}}, \citenamefont {{Adya}},\ and\ \citenamefont
  {et~al.}}]{2019PhRvX...9a1001A}%
  \BibitemOpen
  \bibfield  {author} {\bibinfo {author} {\bibfnamefont {B.~P.}\ \bibnamefont
  {{Abbott}}}, \bibinfo {author} {\bibfnamefont {R.}~\bibnamefont {{Abbott}}},
  \bibinfo {author} {\bibfnamefont {T.~D.}\ \bibnamefont {{Abbott}}}, \bibinfo
  {author} {\bibfnamefont {F.}~\bibnamefont {{Acernese}}}, \bibinfo {author}
  {\bibfnamefont {K.}~\bibnamefont {{Ackley}}}, \bibinfo {author}
  {\bibfnamefont {C.}~\bibnamefont {{Adams}}}, \bibinfo {author} {\bibfnamefont
  {T.}~\bibnamefont {{Adams}}}, \bibinfo {author} {\bibfnamefont
  {P.}~\bibnamefont {{Addesso}}}, \bibinfo {author} {\bibfnamefont {R.~X.}\
  \bibnamefont {{Adhikari}}}, \bibinfo {author} {\bibfnamefont {V.~B.}\
  \bibnamefont {{Adya}}}, \ and\ \bibinfo {author} {\bibnamefont {et~al.}},\
  }\href {\doibase 10.1103/PhysRevX.9.011001} {\bibfield  {journal} {\bibinfo
  {journal} {Physical Review X}\ }\textbf {\bibinfo {volume} {9}},\ \bibinfo
  {eid} {011001} (\bibinfo {year} {2019})},\ \Eprint
  {http://arxiv.org/abs/1805.11579} {arXiv:1805.11579 [gr-qc]} \BibitemShut
  {NoStop}%
\bibitem [{\citenamefont {{Abbott}}\ \emph {et~al.}(2018)\citenamefont
  {{Abbott}}, \citenamefont {{Abbott}}, \citenamefont {{Abbott}}, \citenamefont
  {{Acernese}}, \citenamefont {{Ackley}}, \citenamefont {{Adams}},
  \citenamefont {{Adams}}, \citenamefont {{Addesso}}, \citenamefont
  {{Adhikari}}, \citenamefont {{Adya}},\ and\ \citenamefont
  {et~al.}}]{2018PhRvL.121p1101A}%
  \BibitemOpen
  \bibfield  {author} {\bibinfo {author} {\bibfnamefont {B.~P.}\ \bibnamefont
  {{Abbott}}}, \bibinfo {author} {\bibfnamefont {R.}~\bibnamefont {{Abbott}}},
  \bibinfo {author} {\bibfnamefont {T.~D.}\ \bibnamefont {{Abbott}}}, \bibinfo
  {author} {\bibfnamefont {F.}~\bibnamefont {{Acernese}}}, \bibinfo {author}
  {\bibfnamefont {K.}~\bibnamefont {{Ackley}}}, \bibinfo {author}
  {\bibfnamefont {C.}~\bibnamefont {{Adams}}}, \bibinfo {author} {\bibfnamefont
  {T.}~\bibnamefont {{Adams}}}, \bibinfo {author} {\bibfnamefont
  {P.}~\bibnamefont {{Addesso}}}, \bibinfo {author} {\bibfnamefont {R.~X.}\
  \bibnamefont {{Adhikari}}}, \bibinfo {author} {\bibfnamefont {V.~B.}\
  \bibnamefont {{Adya}}}, \ and\ \bibinfo {author} {\bibnamefont {et~al.}}
  (\bibinfo {collaboration} {LIGO Scientific Collaboration and Virgo
  Collaboration}),\ }\href {\doibase 10.1103/PhysRevLett.121.161101} {\bibfield
   {journal} {\bibinfo  {journal} {Physical Review Letters}\ }\textbf {\bibinfo
  {volume} {121}},\ \bibinfo {eid} {161101} (\bibinfo {year} {2018})},\ \Eprint
  {http://arxiv.org/abs/1805.11581} {arXiv:1805.11581 [gr-qc]} \BibitemShut
  {NoStop}%
\bibitem [{\citenamefont {{Annala}}\ \emph {et~al.}(2018)\citenamefont
  {{Annala}}, \citenamefont {{Gorda}}, \citenamefont {{Kurkela}},\ and\
  \citenamefont {{Vuorinen}}}]{2018PhRvL.120q2703A}%
  \BibitemOpen
  \bibfield  {author} {\bibinfo {author} {\bibfnamefont {E.}~\bibnamefont
  {{Annala}}}, \bibinfo {author} {\bibfnamefont {T.}~\bibnamefont {{Gorda}}},
  \bibinfo {author} {\bibfnamefont {A.}~\bibnamefont {{Kurkela}}}, \ and\
  \bibinfo {author} {\bibfnamefont {A.}~\bibnamefont {{Vuorinen}}},\ }\href
  {\doibase 10.1103/PhysRevLett.120.172703} {\bibfield  {journal} {\bibinfo
  {journal} {\prl}\ }\textbf {\bibinfo {volume} {120}},\ \bibinfo {eid}
  {172703} (\bibinfo {year} {2018})},\ \Eprint
  {http://arxiv.org/abs/1711.02644} {arXiv:1711.02644 [astro-ph.HE]}
  \BibitemShut {NoStop}%
\bibitem [{\citenamefont {{Dietrich}}\ \emph {et~al.}(2019)\citenamefont
  {{Dietrich}}, \citenamefont {{Samajdar}}, \citenamefont {{Khan}},
  \citenamefont {{Johnson-McDaniel}}, \citenamefont {{Dudi}},\ and\
  \citenamefont {{Tichy}}}]{2019PhRvD.100d4003D}%
  \BibitemOpen
  \bibfield  {author} {\bibinfo {author} {\bibfnamefont {T.}~\bibnamefont
  {{Dietrich}}}, \bibinfo {author} {\bibfnamefont {A.}~\bibnamefont
  {{Samajdar}}}, \bibinfo {author} {\bibfnamefont {S.}~\bibnamefont {{Khan}}},
  \bibinfo {author} {\bibfnamefont {N.~K.}\ \bibnamefont {{Johnson-McDaniel}}},
  \bibinfo {author} {\bibfnamefont {R.}~\bibnamefont {{Dudi}}}, \ and\ \bibinfo
  {author} {\bibfnamefont {W.}~\bibnamefont {{Tichy}}},\ }\href {\doibase
  10.1103/PhysRevD.100.044003} {\bibfield  {journal} {\bibinfo  {journal}
  {\prd}\ }\textbf {\bibinfo {volume} {100}},\ \bibinfo {eid} {044003}
  (\bibinfo {year} {2019})},\ \Eprint {http://arxiv.org/abs/1905.06011}
  {arXiv:1905.06011 [gr-qc]} \BibitemShut {NoStop}%
\bibitem [{\citenamefont {{Punturo}}\ \emph {et~al.}(2010)\citenamefont
  {{Punturo}}, \citenamefont {{Abernathy}}, \citenamefont {{Acernese}},
  \citenamefont {{Allen}}, \citenamefont {{Andersson}}, \citenamefont {{Arun}},
  \citenamefont {{Barone}}, \citenamefont {{Barr}}, \citenamefont
  {{Barsuglia}}, \citenamefont {{Beker}},\ and\ \citenamefont
  {et~al.}}]{2010CQGra..27h4007P}%
  \BibitemOpen
  \bibfield  {author} {\bibinfo {author} {\bibfnamefont {M.}~\bibnamefont
  {{Punturo}}}, \bibinfo {author} {\bibfnamefont {M.}~\bibnamefont
  {{Abernathy}}}, \bibinfo {author} {\bibfnamefont {F.}~\bibnamefont
  {{Acernese}}}, \bibinfo {author} {\bibfnamefont {B.}~\bibnamefont {{Allen}}},
  \bibinfo {author} {\bibfnamefont {N.}~\bibnamefont {{Andersson}}}, \bibinfo
  {author} {\bibfnamefont {K.}~\bibnamefont {{Arun}}}, \bibinfo {author}
  {\bibfnamefont {F.}~\bibnamefont {{Barone}}}, \bibinfo {author}
  {\bibfnamefont {B.}~\bibnamefont {{Barr}}}, \bibinfo {author} {\bibfnamefont
  {M.}~\bibnamefont {{Barsuglia}}}, \bibinfo {author} {\bibfnamefont
  {M.}~\bibnamefont {{Beker}}}, \ and\ \bibinfo {author} {\bibnamefont
  {et~al.}},\ }\href {\doibase 10.1088/0264-9381/27/8/084007} {\bibfield
  {journal} {\bibinfo  {journal} {Classical and Quantum Gravity}\ }\textbf
  {\bibinfo {volume} {27}},\ \bibinfo {eid} {084007} (\bibinfo {year}
  {2010})}\BibitemShut {NoStop}%
\bibitem [{\citenamefont {{Sathyaprakash}}\ \emph {et~al.}(2012)\citenamefont
  {{Sathyaprakash}}, \citenamefont {{Abernathy}}, \citenamefont {{Acernese}},
  \citenamefont {{Ajith}}, \citenamefont {{Allen}}, \citenamefont
  {{Amaro-Seoane}}, \citenamefont {{Andersson}}, \citenamefont {{Aoudia}},
  \citenamefont {{Arun}}, \citenamefont {{Astone}},\ and\ \citenamefont
  {et~al.}}]{2012CQGra..29l4013S}%
  \BibitemOpen
  \bibfield  {author} {\bibinfo {author} {\bibfnamefont {B.}~\bibnamefont
  {{Sathyaprakash}}}, \bibinfo {author} {\bibfnamefont {M.}~\bibnamefont
  {{Abernathy}}}, \bibinfo {author} {\bibfnamefont {F.}~\bibnamefont
  {{Acernese}}}, \bibinfo {author} {\bibfnamefont {P.}~\bibnamefont {{Ajith}}},
  \bibinfo {author} {\bibfnamefont {B.}~\bibnamefont {{Allen}}}, \bibinfo
  {author} {\bibfnamefont {P.}~\bibnamefont {{Amaro-Seoane}}}, \bibinfo
  {author} {\bibfnamefont {N.}~\bibnamefont {{Andersson}}}, \bibinfo {author}
  {\bibfnamefont {S.}~\bibnamefont {{Aoudia}}}, \bibinfo {author}
  {\bibfnamefont {K.}~\bibnamefont {{Arun}}}, \bibinfo {author} {\bibfnamefont
  {P.}~\bibnamefont {{Astone}}}, \ and\ \bibinfo {author} {\bibnamefont
  {et~al.}},\ }\href {\doibase 10.1088/0264-9381/29/12/124013} {\bibfield
  {journal} {\bibinfo  {journal} {Classical and Quantum Gravity}\ }\textbf
  {\bibinfo {volume} {29}},\ \bibinfo {eid} {124013} (\bibinfo {year}
  {2012})},\ \Eprint {http://arxiv.org/abs/1206.0331} {arXiv:1206.0331 [gr-qc]}
  \BibitemShut {NoStop}%
\bibitem [{\citenamefont {{Maggiore}}\ \emph {et~al.}(2020)\citenamefont
  {{Maggiore}}, \citenamefont {{Van Den Broeck}}, \citenamefont {{Bartolo}},
  \citenamefont {{Belgacem}}, \citenamefont {{Bertacca}}, \citenamefont
  {{Bizouard}}, \citenamefont {{Branchesi}}, \citenamefont {{Clesse}},
  \citenamefont {{Foffa}}, \citenamefont {{Garc{\'\i}a-Bellido}}, ,\ and\
  \citenamefont {et~al.}}]{2020JCAP...03..050M}%
  \BibitemOpen
  \bibfield  {author} {\bibinfo {author} {\bibfnamefont {M.}~\bibnamefont
  {{Maggiore}}}, \bibinfo {author} {\bibfnamefont {C.}~\bibnamefont {{Van Den
  Broeck}}}, \bibinfo {author} {\bibfnamefont {N.}~\bibnamefont {{Bartolo}}},
  \bibinfo {author} {\bibfnamefont {E.}~\bibnamefont {{Belgacem}}}, \bibinfo
  {author} {\bibfnamefont {D.}~\bibnamefont {{Bertacca}}}, \bibinfo {author}
  {\bibfnamefont {M.~A.}\ \bibnamefont {{Bizouard}}}, \bibinfo {author}
  {\bibfnamefont {M.}~\bibnamefont {{Branchesi}}}, \bibinfo {author}
  {\bibfnamefont {S.}~\bibnamefont {{Clesse}}}, \bibinfo {author}
  {\bibfnamefont {S.}~\bibnamefont {{Foffa}}}, \bibinfo {author} {\bibfnamefont
  {J.}~\bibnamefont {{Garc{\'\i}a-Bellido}}}, , \ and\ \bibinfo {author}
  {\bibnamefont {et~al.}},\ }\href {\doibase 10.1088/1475-7516/2020/03/050}
  {\bibfield  {journal} {\bibinfo  {journal} {\jcap}\ }\textbf {\bibinfo
  {volume} {2020}},\ \bibinfo {eid} {050} (\bibinfo {year} {2020})},\ \Eprint
  {http://arxiv.org/abs/1912.02622} {arXiv:1912.02622 [astro-ph.CO]}
  \BibitemShut {NoStop}%
\bibitem [{\citenamefont {{Reitze}}\ \emph {et~al.}(2019)\citenamefont
  {{Reitze}}, \citenamefont {{Adhikari}}, \citenamefont {{Ballmer}},
  \citenamefont {{Barish}}, \citenamefont {{Barsotti}}, \citenamefont
  {{Billingsley}}, \citenamefont {{Brown}}, \citenamefont {{Chen}},
  \citenamefont {{Coyne}}, \citenamefont {{Eisenstein}}, \citenamefont
  {{Evans}}, \citenamefont {{Fritschel}}, \citenamefont {{Hall}}, \citenamefont
  {{Lazzarini}}, \citenamefont {{Lovelace}}, \citenamefont {{Read}},
  \citenamefont {{Sathyaprakash}}, \citenamefont {{Shoemaker}}, \citenamefont
  {{Smith}}, \citenamefont {{Torrie}}, \citenamefont {{Vitale}}, \citenamefont
  {{Weiss}}, \citenamefont {{Wipf}},\ and\ \citenamefont
  {{Zucker}}}]{2019BAAS...51g..35R}%
  \BibitemOpen
  \bibfield  {author} {\bibinfo {author} {\bibfnamefont {D.}~\bibnamefont
  {{Reitze}}}, \bibinfo {author} {\bibfnamefont {R.~X.}\ \bibnamefont
  {{Adhikari}}}, \bibinfo {author} {\bibfnamefont {S.}~\bibnamefont
  {{Ballmer}}}, \bibinfo {author} {\bibfnamefont {B.}~\bibnamefont {{Barish}}},
  \bibinfo {author} {\bibfnamefont {L.}~\bibnamefont {{Barsotti}}}, \bibinfo
  {author} {\bibfnamefont {G.}~\bibnamefont {{Billingsley}}}, \bibinfo {author}
  {\bibfnamefont {D.~A.}\ \bibnamefont {{Brown}}}, \bibinfo {author}
  {\bibfnamefont {Y.}~\bibnamefont {{Chen}}}, \bibinfo {author} {\bibfnamefont
  {D.}~\bibnamefont {{Coyne}}}, \bibinfo {author} {\bibfnamefont
  {R.}~\bibnamefont {{Eisenstein}}}, \bibinfo {author} {\bibfnamefont
  {M.}~\bibnamefont {{Evans}}}, \bibinfo {author} {\bibfnamefont
  {P.}~\bibnamefont {{Fritschel}}}, \bibinfo {author} {\bibfnamefont {E.~D.}\
  \bibnamefont {{Hall}}}, \bibinfo {author} {\bibfnamefont {A.}~\bibnamefont
  {{Lazzarini}}}, \bibinfo {author} {\bibfnamefont {G.}~\bibnamefont
  {{Lovelace}}}, \bibinfo {author} {\bibfnamefont {J.}~\bibnamefont {{Read}}},
  \bibinfo {author} {\bibfnamefont {B.~S.}\ \bibnamefont {{Sathyaprakash}}},
  \bibinfo {author} {\bibfnamefont {D.}~\bibnamefont {{Shoemaker}}}, \bibinfo
  {author} {\bibfnamefont {J.}~\bibnamefont {{Smith}}}, \bibinfo {author}
  {\bibfnamefont {C.}~\bibnamefont {{Torrie}}}, \bibinfo {author}
  {\bibfnamefont {S.}~\bibnamefont {{Vitale}}}, \bibinfo {author}
  {\bibfnamefont {R.}~\bibnamefont {{Weiss}}}, \bibinfo {author} {\bibfnamefont
  {C.}~\bibnamefont {{Wipf}}}, \ and\ \bibinfo {author} {\bibfnamefont
  {M.}~\bibnamefont {{Zucker}}},\ }in\ \href {\doibase
  10.48550/arXiv.1907.04833} {\emph {\bibinfo {booktitle} {Bulletin of the
  American Astronomical Society}}},\ Vol.~\bibinfo {volume} {51}\ (\bibinfo
  {year} {2019})\ p.~\bibinfo {pages} {35},\ \Eprint
  {http://arxiv.org/abs/1907.04833} {arXiv:1907.04833 [astro-ph.IM]}
  \BibitemShut {NoStop}%
\bibitem [{\citenamefont {{Gupta}}\ \emph {et~al.}(2022)\citenamefont
  {{Gupta}}, \citenamefont {{Puecher}}, \citenamefont {{Pang}}, \citenamefont
  {{Janquart}}, \citenamefont {{Koekoek}},\ and\ \citenamefont {{Van Den
  Broeck}}}]{2022arXiv220501182G}%
  \BibitemOpen
  \bibfield  {author} {\bibinfo {author} {\bibfnamefont {P.~K.}\ \bibnamefont
  {{Gupta}}}, \bibinfo {author} {\bibfnamefont {A.}~\bibnamefont {{Puecher}}},
  \bibinfo {author} {\bibfnamefont {P.~T.~H.}\ \bibnamefont {{Pang}}}, \bibinfo
  {author} {\bibfnamefont {J.}~\bibnamefont {{Janquart}}}, \bibinfo {author}
  {\bibfnamefont {G.}~\bibnamefont {{Koekoek}}}, \ and\ \bibinfo {author}
  {\bibfnamefont {C.}~\bibnamefont {{Van Den Broeck}}},\ }\href {\doibase
  10.48550/arXiv.2205.01182} {\bibfield  {journal} {\bibinfo  {journal} {arXiv
  e-prints}\ ,\ \bibinfo {eid} {arXiv:2205.01182}} (\bibinfo {year} {2022})},\
  \Eprint {http://arxiv.org/abs/2205.01182} {arXiv:2205.01182 [gr-qc]}
  \BibitemShut {NoStop}%
\bibitem [{\citenamefont {{Ghosh}}\ \emph {et~al.}(2022)\citenamefont
  {{Ghosh}}, \citenamefont {{Biswas}},\ and\ \citenamefont
  {{Bose}}}]{2022PhRvD.106l3529G}%
  \BibitemOpen
  \bibfield  {author} {\bibinfo {author} {\bibfnamefont {T.}~\bibnamefont
  {{Ghosh}}}, \bibinfo {author} {\bibfnamefont {B.}~\bibnamefont {{Biswas}}}, \
  and\ \bibinfo {author} {\bibfnamefont {S.}~\bibnamefont {{Bose}}},\ }\href
  {\doibase 10.1103/PhysRevD.106.123529} {\bibfield  {journal} {\bibinfo
  {journal} {\prd}\ }\textbf {\bibinfo {volume} {106}},\ \bibinfo {eid}
  {123529} (\bibinfo {year} {2022})},\ \Eprint
  {http://arxiv.org/abs/2203.11756} {arXiv:2203.11756 [astro-ph.CO]}
  \BibitemShut {NoStop}%
\bibitem [{\citenamefont {{Walker}}\ \emph {et~al.}(2024)\citenamefont
  {{Walker}}, \citenamefont {{Smith}}, \citenamefont {{Thrane}},\ and\
  \citenamefont {{Reardon}}}]{2024PhRvD.110d3013W}%
  \BibitemOpen
  \bibfield  {author} {\bibinfo {author} {\bibfnamefont {K.}~\bibnamefont
  {{Walker}}}, \bibinfo {author} {\bibfnamefont {R.}~\bibnamefont {{Smith}}},
  \bibinfo {author} {\bibfnamefont {E.}~\bibnamefont {{Thrane}}}, \ and\
  \bibinfo {author} {\bibfnamefont {D.~J.}\ \bibnamefont {{Reardon}}},\ }\href
  {\doibase 10.1103/PhysRevD.110.043013} {\bibfield  {journal} {\bibinfo
  {journal} {\prd}\ }\textbf {\bibinfo {volume} {110}},\ \bibinfo {eid}
  {043013} (\bibinfo {year} {2024})},\ \Eprint
  {http://arxiv.org/abs/2401.02604} {arXiv:2401.02604 [astro-ph.HE]}
  \BibitemShut {NoStop}%
\bibitem [{\citenamefont {{Miller}}\ \emph {et~al.}(2019)\citenamefont
  {{Miller}}, \citenamefont {{Lamb}}, \citenamefont {{Dittmann}}, \citenamefont
  {{Bogdanov}}, \citenamefont {{Arzoumanian}}, \citenamefont {{Gendreau}},
  \citenamefont {{Guillot}}, \citenamefont {{Harding}}, \citenamefont {{Ho}},
  \citenamefont {{Lattimer}}, \citenamefont {{Ludlam}}, \citenamefont
  {{Mahmoodifar}}, \citenamefont {{Morsink}}, \citenamefont {{Ray}},
  \citenamefont {{Strohmayer}}, \citenamefont {{Wood}}, \citenamefont
  {{Enoto}}, \citenamefont {{Foster}}, \citenamefont {{Okajima}}, \citenamefont
  {{Prigozhin}},\ and\ \citenamefont {{Soong}}}]{2019ApJ...887L..24M}%
  \BibitemOpen
  \bibfield  {author} {\bibinfo {author} {\bibfnamefont {M.~C.}\ \bibnamefont
  {{Miller}}}, \bibinfo {author} {\bibfnamefont {F.~K.}\ \bibnamefont
  {{Lamb}}}, \bibinfo {author} {\bibfnamefont {A.~J.}\ \bibnamefont
  {{Dittmann}}}, \bibinfo {author} {\bibfnamefont {S.}~\bibnamefont
  {{Bogdanov}}}, \bibinfo {author} {\bibfnamefont {Z.}~\bibnamefont
  {{Arzoumanian}}}, \bibinfo {author} {\bibfnamefont {K.~C.}\ \bibnamefont
  {{Gendreau}}}, \bibinfo {author} {\bibfnamefont {S.}~\bibnamefont
  {{Guillot}}}, \bibinfo {author} {\bibfnamefont {A.~K.}\ \bibnamefont
  {{Harding}}}, \bibinfo {author} {\bibfnamefont {W.~C.~G.}\ \bibnamefont
  {{Ho}}}, \bibinfo {author} {\bibfnamefont {J.~M.}\ \bibnamefont
  {{Lattimer}}}, \bibinfo {author} {\bibfnamefont {R.~M.}\ \bibnamefont
  {{Ludlam}}}, \bibinfo {author} {\bibfnamefont {S.}~\bibnamefont
  {{Mahmoodifar}}}, \bibinfo {author} {\bibfnamefont {S.~M.}\ \bibnamefont
  {{Morsink}}}, \bibinfo {author} {\bibfnamefont {P.~S.}\ \bibnamefont
  {{Ray}}}, \bibinfo {author} {\bibfnamefont {T.~E.}\ \bibnamefont
  {{Strohmayer}}}, \bibinfo {author} {\bibfnamefont {K.~S.}\ \bibnamefont
  {{Wood}}}, \bibinfo {author} {\bibfnamefont {T.}~\bibnamefont {{Enoto}}},
  \bibinfo {author} {\bibfnamefont {R.}~\bibnamefont {{Foster}}}, \bibinfo
  {author} {\bibfnamefont {T.}~\bibnamefont {{Okajima}}}, \bibinfo {author}
  {\bibfnamefont {G.}~\bibnamefont {{Prigozhin}}}, \ and\ \bibinfo {author}
  {\bibfnamefont {Y.}~\bibnamefont {{Soong}}},\ }\href {\doibase
  10.3847/2041-8213/ab50c5} {\bibfield  {journal} {\bibinfo  {journal}
  {Astrophys. J. Lett.}\ }\textbf {\bibinfo {volume} {887}},\ \bibinfo {eid}
  {L24} (\bibinfo {year} {2019})},\ \Eprint {http://arxiv.org/abs/1912.05705}
  {arXiv:1912.05705 [astro-ph.HE]} \BibitemShut {NoStop}%
\bibitem [{\citenamefont {{Riley}}\ \emph {et~al.}(2019)\citenamefont
  {{Riley}}, \citenamefont {{Watts}}, \citenamefont {{Bogdanov}}, \citenamefont
  {{Ray}}, \citenamefont {{Ludlam}}, \citenamefont {{Guillot}}, \citenamefont
  {{Arzoumanian}}, \citenamefont {{Baker}}, \citenamefont {{Bilous}},
  \citenamefont {{Chakrabarty}}, \citenamefont {{Gendreau}}, \citenamefont
  {{Harding}}, \citenamefont {{Ho}}, \citenamefont {{Lattimer}}, \citenamefont
  {{Morsink}},\ and\ \citenamefont {{Strohmayer}}}]{2019ApJ...887L..21R}%
  \BibitemOpen
  \bibfield  {author} {\bibinfo {author} {\bibfnamefont {T.~E.}\ \bibnamefont
  {{Riley}}}, \bibinfo {author} {\bibfnamefont {A.~L.}\ \bibnamefont
  {{Watts}}}, \bibinfo {author} {\bibfnamefont {S.}~\bibnamefont {{Bogdanov}}},
  \bibinfo {author} {\bibfnamefont {P.~S.}\ \bibnamefont {{Ray}}}, \bibinfo
  {author} {\bibfnamefont {R.~M.}\ \bibnamefont {{Ludlam}}}, \bibinfo {author}
  {\bibfnamefont {S.}~\bibnamefont {{Guillot}}}, \bibinfo {author}
  {\bibfnamefont {Z.}~\bibnamefont {{Arzoumanian}}}, \bibinfo {author}
  {\bibfnamefont {C.~L.}\ \bibnamefont {{Baker}}}, \bibinfo {author}
  {\bibfnamefont {A.~V.}\ \bibnamefont {{Bilous}}}, \bibinfo {author}
  {\bibfnamefont {D.}~\bibnamefont {{Chakrabarty}}}, \bibinfo {author}
  {\bibfnamefont {K.~C.}\ \bibnamefont {{Gendreau}}}, \bibinfo {author}
  {\bibfnamefont {A.~K.}\ \bibnamefont {{Harding}}}, \bibinfo {author}
  {\bibfnamefont {W.~C.~G.}\ \bibnamefont {{Ho}}}, \bibinfo {author}
  {\bibfnamefont {J.~M.}\ \bibnamefont {{Lattimer}}}, \bibinfo {author}
  {\bibfnamefont {S.~M.}\ \bibnamefont {{Morsink}}}, \ and\ \bibinfo {author}
  {\bibfnamefont {T.~E.}\ \bibnamefont {{Strohmayer}}},\ }\href {\doibase
  10.3847/2041-8213/ab481c} {\bibfield  {journal} {\bibinfo  {journal}
  {Astrophys. J. Lett.}\ }\textbf {\bibinfo {volume} {887}},\ \bibinfo {eid}
  {L21} (\bibinfo {year} {2019})},\ \Eprint {http://arxiv.org/abs/1912.05702}
  {arXiv:1912.05702 [astro-ph.HE]} \BibitemShut {NoStop}%
\bibitem [{\citenamefont {{Miller}}\ \emph {et~al.}(2021)\citenamefont
  {{Miller}}, \citenamefont {{Lamb}}, \citenamefont {{Dittmann}}, \citenamefont
  {{Bogdanov}}, \citenamefont {{Arzoumanian}}, \citenamefont {{Gendreau}},
  \citenamefont {{Guillot}}, \citenamefont {{Ho}}, \citenamefont {{Lattimer}},
  \citenamefont {{Loewenstein}}, \citenamefont {{Morsink}}, \citenamefont
  {{Ray}}, \citenamefont {{Wolff}}, \citenamefont {{Baker}}, \citenamefont
  {{Cazeau}}, \citenamefont {{Manthripragada}}, \citenamefont {{Markwardt}},
  \citenamefont {{Okajima}}, \citenamefont {{Pollard}}, \citenamefont
  {{Cognard}}, \citenamefont {{Cromartie}}, \citenamefont {{Fonseca}},
  \citenamefont {{Guillemot}}, \citenamefont {{Kerr}}, \citenamefont
  {{Parthasarathy}}, \citenamefont {{Pennucci}}, \citenamefont {{Ransom}},\
  and\ \citenamefont {{Stairs}}}]{2021ApJ...918L..28M}%
  \BibitemOpen
  \bibfield  {author} {\bibinfo {author} {\bibfnamefont {M.~C.}\ \bibnamefont
  {{Miller}}}, \bibinfo {author} {\bibfnamefont {F.~K.}\ \bibnamefont
  {{Lamb}}}, \bibinfo {author} {\bibfnamefont {A.~J.}\ \bibnamefont
  {{Dittmann}}}, \bibinfo {author} {\bibfnamefont {S.}~\bibnamefont
  {{Bogdanov}}}, \bibinfo {author} {\bibfnamefont {Z.}~\bibnamefont
  {{Arzoumanian}}}, \bibinfo {author} {\bibfnamefont {K.~C.}\ \bibnamefont
  {{Gendreau}}}, \bibinfo {author} {\bibfnamefont {S.}~\bibnamefont
  {{Guillot}}}, \bibinfo {author} {\bibfnamefont {W.~C.~G.}\ \bibnamefont
  {{Ho}}}, \bibinfo {author} {\bibfnamefont {J.~M.}\ \bibnamefont
  {{Lattimer}}}, \bibinfo {author} {\bibfnamefont {M.}~\bibnamefont
  {{Loewenstein}}}, \bibinfo {author} {\bibfnamefont {S.~M.}\ \bibnamefont
  {{Morsink}}}, \bibinfo {author} {\bibfnamefont {P.~S.}\ \bibnamefont
  {{Ray}}}, \bibinfo {author} {\bibfnamefont {M.~T.}\ \bibnamefont {{Wolff}}},
  \bibinfo {author} {\bibfnamefont {C.~L.}\ \bibnamefont {{Baker}}}, \bibinfo
  {author} {\bibfnamefont {T.}~\bibnamefont {{Cazeau}}}, \bibinfo {author}
  {\bibfnamefont {S.}~\bibnamefont {{Manthripragada}}}, \bibinfo {author}
  {\bibfnamefont {C.~B.}\ \bibnamefont {{Markwardt}}}, \bibinfo {author}
  {\bibfnamefont {T.}~\bibnamefont {{Okajima}}}, \bibinfo {author}
  {\bibfnamefont {S.}~\bibnamefont {{Pollard}}}, \bibinfo {author}
  {\bibfnamefont {I.}~\bibnamefont {{Cognard}}}, \bibinfo {author}
  {\bibfnamefont {H.~T.}\ \bibnamefont {{Cromartie}}}, \bibinfo {author}
  {\bibfnamefont {E.}~\bibnamefont {{Fonseca}}}, \bibinfo {author}
  {\bibfnamefont {L.}~\bibnamefont {{Guillemot}}}, \bibinfo {author}
  {\bibfnamefont {M.}~\bibnamefont {{Kerr}}}, \bibinfo {author} {\bibfnamefont
  {A.}~\bibnamefont {{Parthasarathy}}}, \bibinfo {author} {\bibfnamefont
  {T.~T.}\ \bibnamefont {{Pennucci}}}, \bibinfo {author} {\bibfnamefont
  {S.}~\bibnamefont {{Ransom}}}, \ and\ \bibinfo {author} {\bibfnamefont
  {I.}~\bibnamefont {{Stairs}}},\ }\href {\doibase 10.3847/2041-8213/ac089b}
  {\bibfield  {journal} {\bibinfo  {journal} {Astrophys. J. Lett.}\ }\textbf
  {\bibinfo {volume} {918}},\ \bibinfo {eid} {L28} (\bibinfo {year} {2021})},\
  \Eprint {http://arxiv.org/abs/2105.06979} {arXiv:2105.06979 [astro-ph.HE]}
  \BibitemShut {NoStop}%
\bibitem [{\citenamefont {{Riley}}\ \emph {et~al.}(2021)\citenamefont
  {{Riley}}, \citenamefont {{Watts}}, \citenamefont {{Ray}}, \citenamefont
  {{Bogdanov}}, \citenamefont {{Guillot}}, \citenamefont {{Morsink}},
  \citenamefont {{Bilous}}, \citenamefont {{Arzoumanian}}, \citenamefont
  {{Choudhury}}, \citenamefont {{Deneva}}, \citenamefont {{Gendreau}},
  \citenamefont {{Harding}}, \citenamefont {{Ho}}, \citenamefont {{Lattimer}},
  \citenamefont {{Loewenstein}}, \citenamefont {{Ludlam}}, \citenamefont
  {{Markwardt}}, \citenamefont {{Okajima}}, \citenamefont
  {{Prescod-Weinstein}}, \citenamefont {{Remillard}}, \citenamefont {{Wolff}},
  \citenamefont {{Fonseca}}, \citenamefont {{Cromartie}}, \citenamefont
  {{Kerr}}, \citenamefont {{Pennucci}}, \citenamefont {{Parthasarathy}},
  \citenamefont {{Ransom}}, \citenamefont {{Stairs}}, \citenamefont
  {{Guillemot}},\ and\ \citenamefont {{Cognard}}}]{2021ApJ...918L..27R}%
  \BibitemOpen
  \bibfield  {author} {\bibinfo {author} {\bibfnamefont {T.~E.}\ \bibnamefont
  {{Riley}}}, \bibinfo {author} {\bibfnamefont {A.~L.}\ \bibnamefont
  {{Watts}}}, \bibinfo {author} {\bibfnamefont {P.~S.}\ \bibnamefont {{Ray}}},
  \bibinfo {author} {\bibfnamefont {S.}~\bibnamefont {{Bogdanov}}}, \bibinfo
  {author} {\bibfnamefont {S.}~\bibnamefont {{Guillot}}}, \bibinfo {author}
  {\bibfnamefont {S.~M.}\ \bibnamefont {{Morsink}}}, \bibinfo {author}
  {\bibfnamefont {A.~V.}\ \bibnamefont {{Bilous}}}, \bibinfo {author}
  {\bibfnamefont {Z.}~\bibnamefont {{Arzoumanian}}}, \bibinfo {author}
  {\bibfnamefont {D.}~\bibnamefont {{Choudhury}}}, \bibinfo {author}
  {\bibfnamefont {J.~S.}\ \bibnamefont {{Deneva}}}, \bibinfo {author}
  {\bibfnamefont {K.~C.}\ \bibnamefont {{Gendreau}}}, \bibinfo {author}
  {\bibfnamefont {A.~K.}\ \bibnamefont {{Harding}}}, \bibinfo {author}
  {\bibfnamefont {W.~C.~G.}\ \bibnamefont {{Ho}}}, \bibinfo {author}
  {\bibfnamefont {J.~M.}\ \bibnamefont {{Lattimer}}}, \bibinfo {author}
  {\bibfnamefont {M.}~\bibnamefont {{Loewenstein}}}, \bibinfo {author}
  {\bibfnamefont {R.~M.}\ \bibnamefont {{Ludlam}}}, \bibinfo {author}
  {\bibfnamefont {C.~B.}\ \bibnamefont {{Markwardt}}}, \bibinfo {author}
  {\bibfnamefont {T.}~\bibnamefont {{Okajima}}}, \bibinfo {author}
  {\bibfnamefont {C.}~\bibnamefont {{Prescod-Weinstein}}}, \bibinfo {author}
  {\bibfnamefont {R.~A.}\ \bibnamefont {{Remillard}}}, \bibinfo {author}
  {\bibfnamefont {M.~T.}\ \bibnamefont {{Wolff}}}, \bibinfo {author}
  {\bibfnamefont {E.}~\bibnamefont {{Fonseca}}}, \bibinfo {author}
  {\bibfnamefont {H.~T.}\ \bibnamefont {{Cromartie}}}, \bibinfo {author}
  {\bibfnamefont {M.}~\bibnamefont {{Kerr}}}, \bibinfo {author} {\bibfnamefont
  {T.~T.}\ \bibnamefont {{Pennucci}}}, \bibinfo {author} {\bibfnamefont
  {A.}~\bibnamefont {{Parthasarathy}}}, \bibinfo {author} {\bibfnamefont
  {S.}~\bibnamefont {{Ransom}}}, \bibinfo {author} {\bibfnamefont
  {I.}~\bibnamefont {{Stairs}}}, \bibinfo {author} {\bibfnamefont
  {L.}~\bibnamefont {{Guillemot}}}, \ and\ \bibinfo {author} {\bibfnamefont
  {I.}~\bibnamefont {{Cognard}}},\ }\href {\doibase 10.3847/2041-8213/ac0a81}
  {\bibfield  {journal} {\bibinfo  {journal} {Astrophys. J. Lett.}\ }\textbf
  {\bibinfo {volume} {918}},\ \bibinfo {eid} {L27} (\bibinfo {year} {2021})},\
  \Eprint {http://arxiv.org/abs/2105.06980} {arXiv:2105.06980 [astro-ph.HE]}
  \BibitemShut {NoStop}%
\bibitem [{\citenamefont {{Hu}}\ \emph {et~al.}(2020)\citenamefont {{Hu}},
  \citenamefont {{Kramer}}, \citenamefont {{Wex}}, \citenamefont {{Champion}},\
  and\ \citenamefont {{Kehl}}}]{2020MNRAS.497.3118H}%
  \BibitemOpen
  \bibfield  {author} {\bibinfo {author} {\bibfnamefont {H.}~\bibnamefont
  {{Hu}}}, \bibinfo {author} {\bibfnamefont {M.}~\bibnamefont {{Kramer}}},
  \bibinfo {author} {\bibfnamefont {N.}~\bibnamefont {{Wex}}}, \bibinfo
  {author} {\bibfnamefont {D.~J.}\ \bibnamefont {{Champion}}}, \ and\ \bibinfo
  {author} {\bibfnamefont {M.~S.}\ \bibnamefont {{Kehl}}},\ }\href {\doibase
  10.1093/mnras/staa2107} {\bibfield  {journal} {\bibinfo  {journal} {\mnras}\
  }\textbf {\bibinfo {volume} {497}},\ \bibinfo {pages} {3118} (\bibinfo {year}
  {2020})},\ \Eprint {http://arxiv.org/abs/2007.07725} {arXiv:2007.07725
  [astro-ph.SR]} \BibitemShut {NoStop}%
\bibitem [{\citenamefont {{Fonseca}}\ \emph {et~al.}(2021)\citenamefont
  {{Fonseca}}, \citenamefont {{Cromartie}}, \citenamefont {{Pennucci}},
  \citenamefont {{Ray}}, \citenamefont {{Kirichenko}}, \citenamefont
  {{Ransom}}, \citenamefont {{Demorest}}, \citenamefont {{Stairs}},
  \citenamefont {{Arzoumanian}}, \citenamefont {{Guillemot}}, \citenamefont
  {{Parthasarathy}}, \citenamefont {{Kerr}}, \citenamefont {{Cognard}},
  \citenamefont {{Baker}}, \citenamefont {{Blumer}}, \citenamefont {{Brook}},
  \citenamefont {{DeCesar}}, \citenamefont {{Dolch}}, \citenamefont {{Dong}},
  \citenamefont {{Ferrara}}, \citenamefont {{Fiore}}, \citenamefont
  {{Garver-Daniels}}, \citenamefont {{Good}}, \citenamefont {{Jennings}},
  \citenamefont {{Jones}}, \citenamefont {{Kaspi}}, \citenamefont {{Lam}},
  \citenamefont {{Lorimer}}, \citenamefont {{Luo}}, \citenamefont {{McEwen}},
  \citenamefont {{McKee}}, \citenamefont {{McLaughlin}}, \citenamefont
  {{McMann}}, \citenamefont {{Meyers}}, \citenamefont {{Naidu}}, \citenamefont
  {{Ng}}, \citenamefont {{Nice}}, \citenamefont {{Pol}}, \citenamefont
  {{Radovan}}, \citenamefont {{Shapiro-Albert}}, \citenamefont {{Tan}},
  \citenamefont {{Tendulkar}}, \citenamefont {{Swiggum}}, \citenamefont
  {{Wahl}},\ and\ \citenamefont {{Zhu}}}]{2021ApJ...915L..12F}%
  \BibitemOpen
  \bibfield  {author} {\bibinfo {author} {\bibfnamefont {E.}~\bibnamefont
  {{Fonseca}}}, \bibinfo {author} {\bibfnamefont {H.~T.}\ \bibnamefont
  {{Cromartie}}}, \bibinfo {author} {\bibfnamefont {T.~T.}\ \bibnamefont
  {{Pennucci}}}, \bibinfo {author} {\bibfnamefont {P.~S.}\ \bibnamefont
  {{Ray}}}, \bibinfo {author} {\bibfnamefont {A.~Y.}\ \bibnamefont
  {{Kirichenko}}}, \bibinfo {author} {\bibfnamefont {S.~M.}\ \bibnamefont
  {{Ransom}}}, \bibinfo {author} {\bibfnamefont {P.~B.}\ \bibnamefont
  {{Demorest}}}, \bibinfo {author} {\bibfnamefont {I.~H.}\ \bibnamefont
  {{Stairs}}}, \bibinfo {author} {\bibfnamefont {Z.}~\bibnamefont
  {{Arzoumanian}}}, \bibinfo {author} {\bibfnamefont {L.}~\bibnamefont
  {{Guillemot}}}, \bibinfo {author} {\bibfnamefont {A.}~\bibnamefont
  {{Parthasarathy}}}, \bibinfo {author} {\bibfnamefont {M.}~\bibnamefont
  {{Kerr}}}, \bibinfo {author} {\bibfnamefont {I.}~\bibnamefont {{Cognard}}},
  \bibinfo {author} {\bibfnamefont {P.~T.}\ \bibnamefont {{Baker}}}, \bibinfo
  {author} {\bibfnamefont {H.}~\bibnamefont {{Blumer}}}, \bibinfo {author}
  {\bibfnamefont {P.~R.}\ \bibnamefont {{Brook}}}, \bibinfo {author}
  {\bibfnamefont {M.}~\bibnamefont {{DeCesar}}}, \bibinfo {author}
  {\bibfnamefont {T.}~\bibnamefont {{Dolch}}}, \bibinfo {author} {\bibfnamefont
  {F.~A.}\ \bibnamefont {{Dong}}}, \bibinfo {author} {\bibfnamefont {E.~C.}\
  \bibnamefont {{Ferrara}}}, \bibinfo {author} {\bibfnamefont {W.}~\bibnamefont
  {{Fiore}}}, \bibinfo {author} {\bibfnamefont {N.}~\bibnamefont
  {{Garver-Daniels}}}, \bibinfo {author} {\bibfnamefont {D.~C.}\ \bibnamefont
  {{Good}}}, \bibinfo {author} {\bibfnamefont {R.}~\bibnamefont {{Jennings}}},
  \bibinfo {author} {\bibfnamefont {M.~L.}\ \bibnamefont {{Jones}}}, \bibinfo
  {author} {\bibfnamefont {V.~M.}\ \bibnamefont {{Kaspi}}}, \bibinfo {author}
  {\bibfnamefont {M.~T.}\ \bibnamefont {{Lam}}}, \bibinfo {author}
  {\bibfnamefont {D.~R.}\ \bibnamefont {{Lorimer}}}, \bibinfo {author}
  {\bibfnamefont {J.}~\bibnamefont {{Luo}}}, \bibinfo {author} {\bibfnamefont
  {A.}~\bibnamefont {{McEwen}}}, \bibinfo {author} {\bibfnamefont {J.~W.}\
  \bibnamefont {{McKee}}}, \bibinfo {author} {\bibfnamefont {M.~A.}\
  \bibnamefont {{McLaughlin}}}, \bibinfo {author} {\bibfnamefont
  {N.}~\bibnamefont {{McMann}}}, \bibinfo {author} {\bibfnamefont {B.~W.}\
  \bibnamefont {{Meyers}}}, \bibinfo {author} {\bibfnamefont {A.}~\bibnamefont
  {{Naidu}}}, \bibinfo {author} {\bibfnamefont {C.}~\bibnamefont {{Ng}}},
  \bibinfo {author} {\bibfnamefont {D.~J.}\ \bibnamefont {{Nice}}}, \bibinfo
  {author} {\bibfnamefont {N.}~\bibnamefont {{Pol}}}, \bibinfo {author}
  {\bibfnamefont {H.~A.}\ \bibnamefont {{Radovan}}}, \bibinfo {author}
  {\bibfnamefont {B.}~\bibnamefont {{Shapiro-Albert}}}, \bibinfo {author}
  {\bibfnamefont {C.~M.}\ \bibnamefont {{Tan}}}, \bibinfo {author}
  {\bibfnamefont {S.~P.}\ \bibnamefont {{Tendulkar}}}, \bibinfo {author}
  {\bibfnamefont {J.~K.}\ \bibnamefont {{Swiggum}}}, \bibinfo {author}
  {\bibfnamefont {H.~M.}\ \bibnamefont {{Wahl}}}, \ and\ \bibinfo {author}
  {\bibfnamefont {W.~W.}\ \bibnamefont {{Zhu}}},\ }\href {\doibase
  10.3847/2041-8213/ac03b8} {\bibfield  {journal} {\bibinfo  {journal} {\apjl}\
  }\textbf {\bibinfo {volume} {915}},\ \bibinfo {eid} {L12} (\bibinfo {year}
  {2021})},\ \Eprint {http://arxiv.org/abs/2104.00880} {arXiv:2104.00880
  [astro-ph.HE]} \BibitemShut {NoStop}%
\bibitem [{\citenamefont {{Kramer}}\ \emph {et~al.}(2021)\citenamefont
  {{Kramer}}, \citenamefont {{Stairs}}, \citenamefont {{Manchester}},
  \citenamefont {{Wex}}, \citenamefont {{Deller}}, \citenamefont {{Coles}},
  \citenamefont {{Ali}}, \citenamefont {{Burgay}}, \citenamefont {{Camilo}},
  \citenamefont {{Cognard}}, \citenamefont {{Damour}}, \citenamefont
  {{Desvignes}}, \citenamefont {{Ferdman}}, \citenamefont {{Freire}},
  \citenamefont {{Grondin}}, \citenamefont {{Guillemot}}, \citenamefont
  {{Hobbs}}, \citenamefont {{Janssen}}, \citenamefont {{Karuppusamy}},
  \citenamefont {{Lorimer}}, \citenamefont {{Lyne}}, \citenamefont {{McKee}},
  \citenamefont {{McLaughlin}}, \citenamefont {{M{\"u}nch}}, \citenamefont
  {{Perera}}, \citenamefont {{Pol}}, \citenamefont {{Possenti}}, \citenamefont
  {{Sarkissian}}, \citenamefont {{Stappers}},\ and\ \citenamefont
  {{Theureau}}}]{2021PhRvX..11d1050K}%
  \BibitemOpen
  \bibfield  {author} {\bibinfo {author} {\bibfnamefont {M.}~\bibnamefont
  {{Kramer}}}, \bibinfo {author} {\bibfnamefont {I.~H.}\ \bibnamefont
  {{Stairs}}}, \bibinfo {author} {\bibfnamefont {R.~N.}\ \bibnamefont
  {{Manchester}}}, \bibinfo {author} {\bibfnamefont {N.}~\bibnamefont {{Wex}}},
  \bibinfo {author} {\bibfnamefont {A.~T.}\ \bibnamefont {{Deller}}}, \bibinfo
  {author} {\bibfnamefont {W.~A.}\ \bibnamefont {{Coles}}}, \bibinfo {author}
  {\bibfnamefont {M.}~\bibnamefont {{Ali}}}, \bibinfo {author} {\bibfnamefont
  {M.}~\bibnamefont {{Burgay}}}, \bibinfo {author} {\bibfnamefont
  {F.}~\bibnamefont {{Camilo}}}, \bibinfo {author} {\bibfnamefont
  {I.}~\bibnamefont {{Cognard}}}, \bibinfo {author} {\bibfnamefont
  {T.}~\bibnamefont {{Damour}}}, \bibinfo {author} {\bibfnamefont
  {G.}~\bibnamefont {{Desvignes}}}, \bibinfo {author} {\bibfnamefont {R.~D.}\
  \bibnamefont {{Ferdman}}}, \bibinfo {author} {\bibfnamefont {P.~C.~C.}\
  \bibnamefont {{Freire}}}, \bibinfo {author} {\bibfnamefont {S.}~\bibnamefont
  {{Grondin}}}, \bibinfo {author} {\bibfnamefont {L.}~\bibnamefont
  {{Guillemot}}}, \bibinfo {author} {\bibfnamefont {G.~B.}\ \bibnamefont
  {{Hobbs}}}, \bibinfo {author} {\bibfnamefont {G.}~\bibnamefont {{Janssen}}},
  \bibinfo {author} {\bibfnamefont {R.}~\bibnamefont {{Karuppusamy}}}, \bibinfo
  {author} {\bibfnamefont {D.~R.}\ \bibnamefont {{Lorimer}}}, \bibinfo {author}
  {\bibfnamefont {A.~G.}\ \bibnamefont {{Lyne}}}, \bibinfo {author}
  {\bibfnamefont {J.~W.}\ \bibnamefont {{McKee}}}, \bibinfo {author}
  {\bibfnamefont {M.}~\bibnamefont {{McLaughlin}}}, \bibinfo {author}
  {\bibfnamefont {L.~E.}\ \bibnamefont {{M{\"u}nch}}}, \bibinfo {author}
  {\bibfnamefont {B.~B.~P.}\ \bibnamefont {{Perera}}}, \bibinfo {author}
  {\bibfnamefont {N.}~\bibnamefont {{Pol}}}, \bibinfo {author} {\bibfnamefont
  {A.}~\bibnamefont {{Possenti}}}, \bibinfo {author} {\bibfnamefont
  {J.}~\bibnamefont {{Sarkissian}}}, \bibinfo {author} {\bibfnamefont {B.~W.}\
  \bibnamefont {{Stappers}}}, \ and\ \bibinfo {author} {\bibfnamefont
  {G.}~\bibnamefont {{Theureau}}},\ }\href {\doibase
  10.1103/PhysRevX.11.041050} {\bibfield  {journal} {\bibinfo  {journal}
  {Physical Review X}\ }\textbf {\bibinfo {volume} {11}},\ \bibinfo {eid}
  {041050} (\bibinfo {year} {2021})},\ \Eprint
  {http://arxiv.org/abs/2112.06795} {arXiv:2112.06795 [astro-ph.HE]}
  \BibitemShut {NoStop}%
\bibitem [{\citenamefont {{Adhikari}}\ \emph {et~al.}(2021)\citenamefont
  {{Adhikari}}, \citenamefont {{Albataineh}}, \citenamefont {{Androic}},
  \citenamefont {{Aniol}}, \citenamefont {{Armstrong}}, \citenamefont
  {{Averett}}, \citenamefont {{Ayerbe Gayoso}}, \citenamefont {{Barcus}},
  \citenamefont {{Bellini}}, \citenamefont {{Beminiwattha}}, \citenamefont
  {{Benesch}}, \citenamefont {{Bhatt}}, \citenamefont {{Bhatta Pathak}},
  \citenamefont {{Bhetuwal}}, \citenamefont {{Blaikie}},\ and\ \citenamefont
  {\textit{et al}. ({PREX Collaboration})}}]{2021PhRvL.126q2502A}%
  \BibitemOpen
  \bibfield  {author} {\bibinfo {author} {\bibfnamefont {D.}~\bibnamefont
  {{Adhikari}}}, \bibinfo {author} {\bibfnamefont {H.}~\bibnamefont
  {{Albataineh}}}, \bibinfo {author} {\bibfnamefont {D.}~\bibnamefont
  {{Androic}}}, \bibinfo {author} {\bibfnamefont {K.}~\bibnamefont {{Aniol}}},
  \bibinfo {author} {\bibfnamefont {D.~S.}\ \bibnamefont {{Armstrong}}},
  \bibinfo {author} {\bibfnamefont {T.}~\bibnamefont {{Averett}}}, \bibinfo
  {author} {\bibfnamefont {C.}~\bibnamefont {{Ayerbe Gayoso}}}, \bibinfo
  {author} {\bibfnamefont {S.}~\bibnamefont {{Barcus}}}, \bibinfo {author}
  {\bibfnamefont {V.}~\bibnamefont {{Bellini}}}, \bibinfo {author}
  {\bibfnamefont {R.~S.}\ \bibnamefont {{Beminiwattha}}}, \bibinfo {author}
  {\bibfnamefont {J.~F.}\ \bibnamefont {{Benesch}}}, \bibinfo {author}
  {\bibfnamefont {H.}~\bibnamefont {{Bhatt}}}, \bibinfo {author} {\bibfnamefont
  {D.}~\bibnamefont {{Bhatta Pathak}}}, \bibinfo {author} {\bibfnamefont
  {D.}~\bibnamefont {{Bhetuwal}}}, \bibinfo {author} {\bibfnamefont
  {B.}~\bibnamefont {{Blaikie}}}, \ and\ \bibinfo {author} {\bibnamefont
  {\textit{et al}. ({PREX Collaboration})}},\ }\href {\doibase
  10.1103/PhysRevLett.126.172502} {\bibfield  {journal} {\bibinfo  {journal}
  {\prl}\ }\textbf {\bibinfo {volume} {126}},\ \bibinfo {eid} {172502}
  (\bibinfo {year} {2021})},\ \Eprint {http://arxiv.org/abs/2102.10767}
  {arXiv:2102.10767 [nucl-ex]} \BibitemShut {NoStop}%
\bibitem [{\citenamefont {{Adhikari}}\ \emph {et~al.}(2022)\citenamefont
  {{Adhikari}}, \citenamefont {{Albataineh}}, \citenamefont {{Androic}},
  \citenamefont {{Aniol}}, \citenamefont {{Armstrong}}, \citenamefont
  {{Averett}}, \citenamefont {{Ayerbe Gayoso}}, \citenamefont {{Barcus}},
  \citenamefont {{Bellini}}, \citenamefont {{Beminiwattha}}, \citenamefont
  {{Benesch}}, \citenamefont {{Bhatt}}, \citenamefont {{Bhatta Pathak}},
  \citenamefont {{Bhetuwal}}, \citenamefont {{Blaikie}},\ and\ \citenamefont
  {\textit{et al}. ({CREX Collaboration})}}]{2022PhRvL.129d2501A}%
  \BibitemOpen
  \bibfield  {author} {\bibinfo {author} {\bibfnamefont {D.}~\bibnamefont
  {{Adhikari}}}, \bibinfo {author} {\bibfnamefont {H.}~\bibnamefont
  {{Albataineh}}}, \bibinfo {author} {\bibfnamefont {D.}~\bibnamefont
  {{Androic}}}, \bibinfo {author} {\bibfnamefont {K.~A.}\ \bibnamefont
  {{Aniol}}}, \bibinfo {author} {\bibfnamefont {D.~S.}\ \bibnamefont
  {{Armstrong}}}, \bibinfo {author} {\bibfnamefont {T.}~\bibnamefont
  {{Averett}}}, \bibinfo {author} {\bibfnamefont {C.}~\bibnamefont {{Ayerbe
  Gayoso}}}, \bibinfo {author} {\bibfnamefont {S.~K.}\ \bibnamefont
  {{Barcus}}}, \bibinfo {author} {\bibfnamefont {V.}~\bibnamefont {{Bellini}}},
  \bibinfo {author} {\bibfnamefont {R.~S.}\ \bibnamefont {{Beminiwattha}}},
  \bibinfo {author} {\bibfnamefont {J.~F.}\ \bibnamefont {{Benesch}}}, \bibinfo
  {author} {\bibfnamefont {H.}~\bibnamefont {{Bhatt}}}, \bibinfo {author}
  {\bibfnamefont {D.}~\bibnamefont {{Bhatta Pathak}}}, \bibinfo {author}
  {\bibfnamefont {D.}~\bibnamefont {{Bhetuwal}}}, \bibinfo {author}
  {\bibfnamefont {B.}~\bibnamefont {{Blaikie}}}, \ and\ \bibinfo {author}
  {\bibnamefont {\textit{et al}. ({CREX Collaboration})}},\ }\href {\doibase
  10.1103/PhysRevLett.129.042501} {\bibfield  {journal} {\bibinfo  {journal}
  {\prl}\ }\textbf {\bibinfo {volume} {129}},\ \bibinfo {eid} {042501}
  (\bibinfo {year} {2022})},\ \Eprint {http://arxiv.org/abs/2205.11593}
  {arXiv:2205.11593 [nucl-ex]} \BibitemShut {NoStop}%
\bibitem [{\citenamefont {{Walecka}}(1974)}]{1974AnPhy..83..491W}%
  \BibitemOpen
  \bibfield  {author} {\bibinfo {author} {\bibfnamefont {J.~D.}\ \bibnamefont
  {{Walecka}}},\ }\href {\doibase 10.1016/0003-4916(74)90208-5} {\bibfield
  {journal} {\bibinfo  {journal} {Annals of Physics}\ }\textbf {\bibinfo
  {volume} {83}},\ \bibinfo {pages} {491} (\bibinfo {year} {1974})}\BibitemShut
  {NoStop}%
\bibitem [{\citenamefont {{Shen}}\ \emph {et~al.}(1998)\citenamefont {{Shen}},
  \citenamefont {{Toki}}, \citenamefont {{Oyamatsu}},\ and\ \citenamefont
  {{Sumiyoshi}}}]{1998NuPhA.637..435S}%
  \BibitemOpen
  \bibfield  {author} {\bibinfo {author} {\bibfnamefont {H.}~\bibnamefont
  {{Shen}}}, \bibinfo {author} {\bibfnamefont {H.}~\bibnamefont {{Toki}}},
  \bibinfo {author} {\bibfnamefont {K.}~\bibnamefont {{Oyamatsu}}}, \ and\
  \bibinfo {author} {\bibfnamefont {K.}~\bibnamefont {{Sumiyoshi}}},\ }\href
  {\doibase 10.1016/S0375-9474(98)00236-X} {\bibfield  {journal} {\bibinfo
  {journal} {\nphysa}\ }\textbf {\bibinfo {volume} {637}},\ \bibinfo {pages}
  {435} (\bibinfo {year} {1998})},\ \Eprint
  {http://arxiv.org/abs/nucl-th/9805035} {arXiv:nucl-th/9805035 [nucl-th]}
  \BibitemShut {NoStop}%
\bibitem [{\citenamefont {{Li}}\ \emph {et~al.}(2008)\citenamefont {{Li}},
  \citenamefont {{Chen}},\ and\ \citenamefont {{Ko}}}]{2008PhR...464..113L}%
  \BibitemOpen
  \bibfield  {author} {\bibinfo {author} {\bibfnamefont {B.-A.}\ \bibnamefont
  {{Li}}}, \bibinfo {author} {\bibfnamefont {L.-W.}\ \bibnamefont {{Chen}}}, \
  and\ \bibinfo {author} {\bibfnamefont {C.~M.}\ \bibnamefont {{Ko}}},\ }\href
  {\doibase 10.1016/j.physrep.2008.04.005} {\bibfield  {journal} {\bibinfo
  {journal} {\physrep}\ }\textbf {\bibinfo {volume} {464}},\ \bibinfo {pages}
  {113} (\bibinfo {year} {2008})},\ \Eprint {http://arxiv.org/abs/0804.3580}
  {arXiv:0804.3580 [nucl-th]} \BibitemShut {NoStop}%
\bibitem [{\citenamefont {{Alford}}\ \emph {et~al.}(2022)\citenamefont
  {{Alford}}, \citenamefont {{Brodie}}, \citenamefont {{Haber}},\ and\
  \citenamefont {{Tews}}}]{2022PhRvC.106e5804A}%
  \BibitemOpen
  \bibfield  {author} {\bibinfo {author} {\bibfnamefont {M.~G.}\ \bibnamefont
  {{Alford}}}, \bibinfo {author} {\bibfnamefont {L.}~\bibnamefont {{Brodie}}},
  \bibinfo {author} {\bibfnamefont {A.}~\bibnamefont {{Haber}}}, \ and\
  \bibinfo {author} {\bibfnamefont {I.}~\bibnamefont {{Tews}}},\ }\href
  {\doibase 10.1103/PhysRevC.106.055804} {\bibfield  {journal} {\bibinfo
  {journal} {\prc}\ }\textbf {\bibinfo {volume} {106}},\ \bibinfo {eid}
  {055804} (\bibinfo {year} {2022})},\ \Eprint
  {http://arxiv.org/abs/2205.10283} {arXiv:2205.10283 [nucl-th]} \BibitemShut
  {NoStop}%
\bibitem [{\citenamefont {{Fischer}}\ \emph {et~al.}(2011)\citenamefont
  {{Fischer}}, \citenamefont {{Sagert}}, \citenamefont {{Pagliara}},
  \citenamefont {{Hempel}}, \citenamefont {{Schaffner-Bielich}}, \citenamefont
  {{Rauscher}}, \citenamefont {{Thielemann}}, \citenamefont {{K{\"a}ppeli}},
  \citenamefont {{Mart{\'\i}nez-Pinedo}},\ and\ \citenamefont
  {{Liebend{\"o}rfer}}}]{2011ApJS..194...39F}%
  \BibitemOpen
  \bibfield  {author} {\bibinfo {author} {\bibfnamefont {T.}~\bibnamefont
  {{Fischer}}}, \bibinfo {author} {\bibfnamefont {I.}~\bibnamefont {{Sagert}}},
  \bibinfo {author} {\bibfnamefont {G.}~\bibnamefont {{Pagliara}}}, \bibinfo
  {author} {\bibfnamefont {M.}~\bibnamefont {{Hempel}}}, \bibinfo {author}
  {\bibfnamefont {J.}~\bibnamefont {{Schaffner-Bielich}}}, \bibinfo {author}
  {\bibfnamefont {T.}~\bibnamefont {{Rauscher}}}, \bibinfo {author}
  {\bibfnamefont {F.~K.}\ \bibnamefont {{Thielemann}}}, \bibinfo {author}
  {\bibfnamefont {R.}~\bibnamefont {{K{\"a}ppeli}}}, \bibinfo {author}
  {\bibfnamefont {G.}~\bibnamefont {{Mart{\'\i}nez-Pinedo}}}, \ and\ \bibinfo
  {author} {\bibfnamefont {M.}~\bibnamefont {{Liebend{\"o}rfer}}},\ }\href
  {\doibase 10.1088/0067-0049/194/2/39} {\bibfield  {journal} {\bibinfo
  {journal} {\apjs}\ }\textbf {\bibinfo {volume} {194}},\ \bibinfo {eid} {39}
  (\bibinfo {year} {2011})},\ \Eprint {http://arxiv.org/abs/1011.3409}
  {arXiv:1011.3409 [astro-ph.HE]} \BibitemShut {NoStop}%
\bibitem [{\citenamefont {{Steiner}}\ \emph {et~al.}(2013)\citenamefont
  {{Steiner}}, \citenamefont {{Hempel}},\ and\ \citenamefont
  {{Fischer}}}]{2013ApJ...774...17S}%
  \BibitemOpen
  \bibfield  {author} {\bibinfo {author} {\bibfnamefont {A.~W.}\ \bibnamefont
  {{Steiner}}}, \bibinfo {author} {\bibfnamefont {M.}~\bibnamefont {{Hempel}}},
  \ and\ \bibinfo {author} {\bibfnamefont {T.}~\bibnamefont {{Fischer}}},\
  }\href {\doibase 10.1088/0004-637X/774/1/17} {\bibfield  {journal} {\bibinfo
  {journal} {\apj}\ }\textbf {\bibinfo {volume} {774}},\ \bibinfo {eid} {17}
  (\bibinfo {year} {2013})},\ \Eprint {http://arxiv.org/abs/1207.2184}
  {arXiv:1207.2184 [astro-ph.SR]} \BibitemShut {NoStop}%
\bibitem [{\citenamefont {{Zhu}}\ and\ \citenamefont
  {{Rezzolla}}(2021)}]{2021PhRvD.104h3004Z}%
  \BibitemOpen
  \bibfield  {author} {\bibinfo {author} {\bibfnamefont {Z.}~\bibnamefont
  {{Zhu}}}\ and\ \bibinfo {author} {\bibfnamefont {L.}~\bibnamefont
  {{Rezzolla}}},\ }\href {\doibase 10.1103/PhysRevD.104.083004} {\bibfield
  {journal} {\bibinfo  {journal} {\prd}\ }\textbf {\bibinfo {volume} {104}},\
  \bibinfo {eid} {083004} (\bibinfo {year} {2021})},\ \Eprint
  {http://arxiv.org/abs/2102.07721} {arXiv:2102.07721 [astro-ph.HE]}
  \BibitemShut {NoStop}%
\bibitem [{\citenamefont {{Raithel}}\ \emph {et~al.}(2021)\citenamefont
  {{Raithel}}, \citenamefont {{Paschalidis}},\ and\ \citenamefont
  {{{\"O}zel}}}]{2021PhRvD.104f3016R}%
  \BibitemOpen
  \bibfield  {author} {\bibinfo {author} {\bibfnamefont {C.~A.}\ \bibnamefont
  {{Raithel}}}, \bibinfo {author} {\bibfnamefont {V.}~\bibnamefont
  {{Paschalidis}}}, \ and\ \bibinfo {author} {\bibfnamefont {F.}~\bibnamefont
  {{{\"O}zel}}},\ }\href {\doibase 10.1103/PhysRevD.104.063016} {\bibfield
  {journal} {\bibinfo  {journal} {\prd}\ }\textbf {\bibinfo {volume} {104}},\
  \bibinfo {eid} {063016} (\bibinfo {year} {2021})},\ \Eprint
  {http://arxiv.org/abs/2104.07226} {arXiv:2104.07226 [astro-ph.HE]}
  \BibitemShut {NoStop}%
\bibitem [{\citenamefont {{Camilletti}}\ \emph {et~al.}(2022)\citenamefont
  {{Camilletti}}, \citenamefont {{Chiesa}}, \citenamefont {{Ricigliano}},
  \citenamefont {{Perego}}, \citenamefont {{Lippold}}, \citenamefont
  {{Padamata}}, \citenamefont {{Bernuzzi}}, \citenamefont {{Radice}},
  \citenamefont {{Logoteta}},\ and\ \citenamefont
  {{Guercilena}}}]{2022MNRAS.516.4760C}%
  \BibitemOpen
  \bibfield  {author} {\bibinfo {author} {\bibfnamefont {A.}~\bibnamefont
  {{Camilletti}}}, \bibinfo {author} {\bibfnamefont {L.}~\bibnamefont
  {{Chiesa}}}, \bibinfo {author} {\bibfnamefont {G.}~\bibnamefont
  {{Ricigliano}}}, \bibinfo {author} {\bibfnamefont {A.}~\bibnamefont
  {{Perego}}}, \bibinfo {author} {\bibfnamefont {L.~C.}\ \bibnamefont
  {{Lippold}}}, \bibinfo {author} {\bibfnamefont {S.}~\bibnamefont
  {{Padamata}}}, \bibinfo {author} {\bibfnamefont {S.}~\bibnamefont
  {{Bernuzzi}}}, \bibinfo {author} {\bibfnamefont {D.}~\bibnamefont
  {{Radice}}}, \bibinfo {author} {\bibfnamefont {D.}~\bibnamefont
  {{Logoteta}}}, \ and\ \bibinfo {author} {\bibfnamefont {F.~M.}\ \bibnamefont
  {{Guercilena}}},\ }\href {\doibase 10.1093/mnras/stac2333} {\bibfield
  {journal} {\bibinfo  {journal} {\mnras}\ }\textbf {\bibinfo {volume} {516}},\
  \bibinfo {pages} {4760} (\bibinfo {year} {2022})},\ \Eprint
  {http://arxiv.org/abs/2204.05336} {arXiv:2204.05336 [astro-ph.HE]}
  \BibitemShut {NoStop}%
\bibitem [{\citenamefont {{Zha}}\ \emph {et~al.}(2022)\citenamefont {{Zha}},
  \citenamefont {{O'Connor}}, \citenamefont {{Couch}}, \citenamefont
  {{Leung}},\ and\ \citenamefont {{Nomoto}}}]{2022MNRAS.513.1317Z}%
  \BibitemOpen
  \bibfield  {author} {\bibinfo {author} {\bibfnamefont {S.}~\bibnamefont
  {{Zha}}}, \bibinfo {author} {\bibfnamefont {E.~P.}\ \bibnamefont
  {{O'Connor}}}, \bibinfo {author} {\bibfnamefont {S.~M.}\ \bibnamefont
  {{Couch}}}, \bibinfo {author} {\bibfnamefont {S.-C.}\ \bibnamefont
  {{Leung}}}, \ and\ \bibinfo {author} {\bibfnamefont {K.}~\bibnamefont
  {{Nomoto}}},\ }\href {\doibase 10.1093/mnras/stac1035} {\bibfield  {journal}
  {\bibinfo  {journal} {\mnras}\ }\textbf {\bibinfo {volume} {513}},\ \bibinfo
  {pages} {1317} (\bibinfo {year} {2022})},\ \Eprint
  {http://arxiv.org/abs/2112.15257} {arXiv:2112.15257 [astro-ph.HE]}
  \BibitemShut {NoStop}%
\bibitem [{\citenamefont {{Zhu}}\ \emph {et~al.}(2025)\citenamefont {{Zhu}},
  \citenamefont {{Zha}},\ and\ \citenamefont {{Han}}}]{2025arXiv250617569Z}%
  \BibitemOpen
  \bibfield  {author} {\bibinfo {author} {\bibfnamefont {Z.}~\bibnamefont
  {{Zhu}}}, \bibinfo {author} {\bibfnamefont {S.}~\bibnamefont {{Zha}}}, \ and\
  \bibinfo {author} {\bibfnamefont {S.}~\bibnamefont {{Han}}},\ }\href
  {\doibase 10.48550/arXiv.2506.17569} {\bibfield  {journal} {\bibinfo
  {journal} {arXiv e-prints}\ ,\ \bibinfo {eid} {arXiv:2506.17569}} (\bibinfo
  {year} {2025})},\ \Eprint {http://arxiv.org/abs/2506.17569} {arXiv:2506.17569
  [astro-ph.HE]} \BibitemShut {NoStop}%
\bibitem [{\citenamefont {{Zhu}}\ and\ \citenamefont
  {{Li}}(2018)}]{2018PhRvC..97c5805Z}%
  \BibitemOpen
  \bibfield  {author} {\bibinfo {author} {\bibfnamefont {Z.-Y.}\ \bibnamefont
  {{Zhu}}}\ and\ \bibinfo {author} {\bibfnamefont {A.}~\bibnamefont {{Li}}},\
  }\href {\doibase 10.1103/PhysRevC.97.035805} {\bibfield  {journal} {\bibinfo
  {journal} {\prc}\ }\textbf {\bibinfo {volume} {97}},\ \bibinfo {eid} {035805}
  (\bibinfo {year} {2018})},\ \Eprint {http://arxiv.org/abs/1802.07441}
  {arXiv:1802.07441 [nucl-th]} \BibitemShut {NoStop}%
\bibitem [{\citenamefont {{Zhu}}\ \emph {et~al.}(2018)\citenamefont {{Zhu}},
  \citenamefont {{Zhou}},\ and\ \citenamefont {{Li}}}]{2018ApJ...862...98Z}%
  \BibitemOpen
  \bibfield  {author} {\bibinfo {author} {\bibfnamefont {Z.-Y.}\ \bibnamefont
  {{Zhu}}}, \bibinfo {author} {\bibfnamefont {E.-P.}\ \bibnamefont {{Zhou}}}, \
  and\ \bibinfo {author} {\bibfnamefont {A.}~\bibnamefont {{Li}}},\ }\href
  {\doibase 10.3847/1538-4357/aacc28} {\bibfield  {journal} {\bibinfo
  {journal} {\apj}\ }\textbf {\bibinfo {volume} {862}},\ \bibinfo {eid} {98}
  (\bibinfo {year} {2018})},\ \Eprint {http://arxiv.org/abs/1802.05510}
  {arXiv:1802.05510 [nucl-th]} \BibitemShut {NoStop}%
\bibitem [{\citenamefont {{Zhu}}\ \emph
  {et~al.}(2023{\natexlab{a}})\citenamefont {{Zhu}}, \citenamefont {{Li}},
  \citenamefont {{Hu}},\ and\ \citenamefont {{Shen}}}]{2023PhRvC.108b5809Z}%
  \BibitemOpen
  \bibfield  {author} {\bibinfo {author} {\bibfnamefont {Z.}~\bibnamefont
  {{Zhu}}}, \bibinfo {author} {\bibfnamefont {A.}~\bibnamefont {{Li}}},
  \bibinfo {author} {\bibfnamefont {J.}~\bibnamefont {{Hu}}}, \ and\ \bibinfo
  {author} {\bibfnamefont {H.}~\bibnamefont {{Shen}}},\ }\href {\doibase
  10.1103/PhysRevC.108.025809} {\bibfield  {journal} {\bibinfo  {journal}
  {\prc}\ }\textbf {\bibinfo {volume} {108}},\ \bibinfo {eid} {025809}
  (\bibinfo {year} {2023}{\natexlab{a}})},\ \Eprint
  {http://arxiv.org/abs/2305.16058} {arXiv:2305.16058 [nucl-th]} \BibitemShut
  {NoStop}%
\bibitem [{\citenamefont {{Traversi}}\ \emph {et~al.}(2020)\citenamefont
  {{Traversi}}, \citenamefont {{Char}},\ and\ \citenamefont
  {{Pagliara}}}]{2020ApJ...897..165T}%
  \BibitemOpen
  \bibfield  {author} {\bibinfo {author} {\bibfnamefont {S.}~\bibnamefont
  {{Traversi}}}, \bibinfo {author} {\bibfnamefont {P.}~\bibnamefont {{Char}}},
  \ and\ \bibinfo {author} {\bibfnamefont {G.}~\bibnamefont {{Pagliara}}},\
  }\href {\doibase 10.3847/1538-4357/ab99c1} {\bibfield  {journal} {\bibinfo
  {journal} {\apj}\ }\textbf {\bibinfo {volume} {897}},\ \bibinfo {eid} {165}
  (\bibinfo {year} {2020})},\ \Eprint {http://arxiv.org/abs/2002.08951}
  {arXiv:2002.08951 [astro-ph.HE]} \BibitemShut {NoStop}%
\bibitem [{\citenamefont {{Zhu}}\ \emph
  {et~al.}(2023{\natexlab{b}})\citenamefont {{Zhu}}, \citenamefont {{Li}},\
  and\ \citenamefont {{Liu}}}]{2023ApJ...943..163Z}%
  \BibitemOpen
  \bibfield  {author} {\bibinfo {author} {\bibfnamefont {Z.}~\bibnamefont
  {{Zhu}}}, \bibinfo {author} {\bibfnamefont {A.}~\bibnamefont {{Li}}}, \ and\
  \bibinfo {author} {\bibfnamefont {T.}~\bibnamefont {{Liu}}},\ }\href
  {\doibase 10.3847/1538-4357/acac1f} {\bibfield  {journal} {\bibinfo
  {journal} {\apj}\ }\textbf {\bibinfo {volume} {943}},\ \bibinfo {eid} {163}
  (\bibinfo {year} {2023}{\natexlab{b}})},\ \Eprint
  {http://arxiv.org/abs/2211.02007} {arXiv:2211.02007 [astro-ph.HE]}
  \BibitemShut {NoStop}%
\bibitem [{\citenamefont {{Legred}}\ \emph {et~al.}(2025)\citenamefont
  {{Legred}}, \citenamefont {{Brodie}}, \citenamefont {{Haber}}, \citenamefont
  {{Essick}},\ and\ \citenamefont {{Chatziioannou}}}]{2025arXiv250507677L}%
  \BibitemOpen
  \bibfield  {author} {\bibinfo {author} {\bibfnamefont {I.}~\bibnamefont
  {{Legred}}}, \bibinfo {author} {\bibfnamefont {L.}~\bibnamefont {{Brodie}}},
  \bibinfo {author} {\bibfnamefont {A.}~\bibnamefont {{Haber}}}, \bibinfo
  {author} {\bibfnamefont {R.}~\bibnamefont {{Essick}}}, \ and\ \bibinfo
  {author} {\bibfnamefont {K.}~\bibnamefont {{Chatziioannou}}},\ }\href
  {\doibase 10.48550/arXiv.2505.07677} {\bibfield  {journal} {\bibinfo
  {journal} {arXiv e-prints}\ ,\ \bibinfo {eid} {arXiv:2505.07677}} (\bibinfo
  {year} {2025})},\ \Eprint {http://arxiv.org/abs/2505.07677} {arXiv:2505.07677
  [nucl-th]} \BibitemShut {NoStop}%
\bibitem [{\citenamefont {{Yang}}\ \emph {et~al.}(2025)\citenamefont {{Yang}},
  \citenamefont {{Hippert}}, \citenamefont {{Speranza}},\ and\ \citenamefont
  {{Noronha}}}]{2025arXiv250407805Y}%
  \BibitemOpen
  \bibfield  {author} {\bibinfo {author} {\bibfnamefont {Y.}~\bibnamefont
  {{Yang}}}, \bibinfo {author} {\bibfnamefont {M.}~\bibnamefont {{Hippert}}},
  \bibinfo {author} {\bibfnamefont {E.}~\bibnamefont {{Speranza}}}, \ and\
  \bibinfo {author} {\bibfnamefont {J.}~\bibnamefont {{Noronha}}},\ }\href
  {\doibase 10.48550/arXiv.2504.07805} {\bibfield  {journal} {\bibinfo
  {journal} {arXiv e-prints}\ ,\ \bibinfo {eid} {arXiv:2504.07805}} (\bibinfo
  {year} {2025})},\ \Eprint {http://arxiv.org/abs/2504.07805} {arXiv:2504.07805
  [nucl-th]} \BibitemShut {NoStop}%
\bibitem [{\citenamefont {{Chabanov}}\ and\ \citenamefont
  {{Rezzolla}}(2025{\natexlab{a}})}]{2025PhRvD.111d4074C}%
  \BibitemOpen
  \bibfield  {author} {\bibinfo {author} {\bibfnamefont {M.}~\bibnamefont
  {{Chabanov}}}\ and\ \bibinfo {author} {\bibfnamefont {L.}~\bibnamefont
  {{Rezzolla}}},\ }\href {\doibase 10.1103/PhysRevD.111.044074} {\bibfield
  {journal} {\bibinfo  {journal} {\prd}\ }\textbf {\bibinfo {volume} {111}},\
  \bibinfo {eid} {044074} (\bibinfo {year} {2025}{\natexlab{a}})}\BibitemShut
  {NoStop}%
\bibitem [{\citenamefont {{Chabanov}}\ and\ \citenamefont
  {{Rezzolla}}(2025{\natexlab{b}})}]{2025PhRvL.134g1402C}%
  \BibitemOpen
  \bibfield  {author} {\bibinfo {author} {\bibfnamefont {M.}~\bibnamefont
  {{Chabanov}}}\ and\ \bibinfo {author} {\bibfnamefont {L.}~\bibnamefont
  {{Rezzolla}}},\ }\href {\doibase 10.1103/PhysRevLett.134.071402} {\bibfield
  {journal} {\bibinfo  {journal} {\prl}\ }\textbf {\bibinfo {volume} {134}},\
  \bibinfo {eid} {071402} (\bibinfo {year} {2025}{\natexlab{b}})},\ \Eprint
  {http://arxiv.org/abs/2307.10464} {arXiv:2307.10464 [gr-qc]} \BibitemShut
  {NoStop}%
\bibitem [{\citenamefont {{Raithel}}\ \emph {et~al.}(2019)\citenamefont
  {{Raithel}}, \citenamefont {{{\"O}zel}},\ and\ \citenamefont
  {{Psaltis}}}]{2019ApJ...875...12R}%
  \BibitemOpen
  \bibfield  {author} {\bibinfo {author} {\bibfnamefont {C.~A.}\ \bibnamefont
  {{Raithel}}}, \bibinfo {author} {\bibfnamefont {F.}~\bibnamefont
  {{{\"O}zel}}}, \ and\ \bibinfo {author} {\bibfnamefont {D.}~\bibnamefont
  {{Psaltis}}},\ }\href {\doibase 10.3847/1538-4357/ab08ea} {\bibfield
  {journal} {\bibinfo  {journal} {\apj}\ }\textbf {\bibinfo {volume} {875}},\
  \bibinfo {eid} {12} (\bibinfo {year} {2019})},\ \Eprint
  {http://arxiv.org/abs/1902.10735} {arXiv:1902.10735 [astro-ph.HE]}
  \BibitemShut {NoStop}%
\bibitem [{\citenamefont {{Zhang}}\ \emph {et~al.}(2018)\citenamefont
  {{Zhang}}, \citenamefont {{Li}},\ and\ \citenamefont
  {{Xu}}}]{2018ApJ...859...90Z}%
  \BibitemOpen
  \bibfield  {author} {\bibinfo {author} {\bibfnamefont {N.-B.}\ \bibnamefont
  {{Zhang}}}, \bibinfo {author} {\bibfnamefont {B.-A.}\ \bibnamefont {{Li}}}, \
  and\ \bibinfo {author} {\bibfnamefont {J.}~\bibnamefont {{Xu}}},\ }\href
  {\doibase 10.3847/1538-4357/aac027} {\bibfield  {journal} {\bibinfo
  {journal} {\apj}\ }\textbf {\bibinfo {volume} {859}},\ \bibinfo {eid} {90}
  (\bibinfo {year} {2018})},\ \Eprint {http://arxiv.org/abs/1801.06855}
  {arXiv:1801.06855 [nucl-th]} \BibitemShut {NoStop}%
\bibitem [{\citenamefont {{Essick}}\ \emph
  {et~al.}(2021{\natexlab{a}})\citenamefont {{Essick}}, \citenamefont {{Tews}},
  \citenamefont {{Landry}},\ and\ \citenamefont
  {{Schwenk}}}]{2021PhRvL.127s2701E}%
  \BibitemOpen
  \bibfield  {author} {\bibinfo {author} {\bibfnamefont {R.}~\bibnamefont
  {{Essick}}}, \bibinfo {author} {\bibfnamefont {I.}~\bibnamefont {{Tews}}},
  \bibinfo {author} {\bibfnamefont {P.}~\bibnamefont {{Landry}}}, \ and\
  \bibinfo {author} {\bibfnamefont {A.}~\bibnamefont {{Schwenk}}},\ }\href
  {\doibase 10.1103/PhysRevLett.127.192701} {\bibfield  {journal} {\bibinfo
  {journal} {\prl}\ }\textbf {\bibinfo {volume} {127}},\ \bibinfo {eid}
  {192701} (\bibinfo {year} {2021}{\natexlab{a}})},\ \Eprint
  {http://arxiv.org/abs/2102.10074} {arXiv:2102.10074 [nucl-th]} \BibitemShut
  {NoStop}%
\bibitem [{\citenamefont {{Veitch}}\ \emph {et~al.}(2015)\citenamefont
  {{Veitch}}, \citenamefont {{Raymond}}, \citenamefont {{Farr}}, \citenamefont
  {{Farr}}, \citenamefont {{Graff}}, \citenamefont {{Vitale}}, \citenamefont
  {{Aylott}}, \citenamefont {{Blackburn}}, \citenamefont {{Christensen}},
  \citenamefont {{Coughlin}}, \citenamefont {{Del Pozzo}}, \citenamefont
  {{Feroz}}, \citenamefont {{Gair}}, \citenamefont {{Haster}}, \citenamefont
  {{Kalogera}}, \citenamefont {{Littenberg}}, \citenamefont {{Mandel}},
  \citenamefont {{O'Shaughnessy}}, \citenamefont {{Pitkin}}, \citenamefont
  {{Rodriguez}}, \citenamefont {{R{\"o}ver}}, \citenamefont {{Sidery}},
  \citenamefont {{Smith}}, \citenamefont {{Van Der Sluys}}, \citenamefont
  {{Vecchio}}, \citenamefont {{Vousden}},\ and\ \citenamefont
  {{Wade}}}]{2015PhRvD..91d2003V}%
  \BibitemOpen
  \bibfield  {author} {\bibinfo {author} {\bibfnamefont {J.}~\bibnamefont
  {{Veitch}}}, \bibinfo {author} {\bibfnamefont {V.}~\bibnamefont {{Raymond}}},
  \bibinfo {author} {\bibfnamefont {B.}~\bibnamefont {{Farr}}}, \bibinfo
  {author} {\bibfnamefont {W.}~\bibnamefont {{Farr}}}, \bibinfo {author}
  {\bibfnamefont {P.}~\bibnamefont {{Graff}}}, \bibinfo {author} {\bibfnamefont
  {S.}~\bibnamefont {{Vitale}}}, \bibinfo {author} {\bibfnamefont
  {B.}~\bibnamefont {{Aylott}}}, \bibinfo {author} {\bibfnamefont
  {K.}~\bibnamefont {{Blackburn}}}, \bibinfo {author} {\bibfnamefont
  {N.}~\bibnamefont {{Christensen}}}, \bibinfo {author} {\bibfnamefont
  {M.}~\bibnamefont {{Coughlin}}}, \bibinfo {author} {\bibfnamefont
  {W.}~\bibnamefont {{Del Pozzo}}}, \bibinfo {author} {\bibfnamefont
  {F.}~\bibnamefont {{Feroz}}}, \bibinfo {author} {\bibfnamefont
  {J.}~\bibnamefont {{Gair}}}, \bibinfo {author} {\bibfnamefont {C.~J.}\
  \bibnamefont {{Haster}}}, \bibinfo {author} {\bibfnamefont {V.}~\bibnamefont
  {{Kalogera}}}, \bibinfo {author} {\bibfnamefont {T.}~\bibnamefont
  {{Littenberg}}}, \bibinfo {author} {\bibfnamefont {I.}~\bibnamefont
  {{Mandel}}}, \bibinfo {author} {\bibfnamefont {R.}~\bibnamefont
  {{O'Shaughnessy}}}, \bibinfo {author} {\bibfnamefont {M.}~\bibnamefont
  {{Pitkin}}}, \bibinfo {author} {\bibfnamefont {C.}~\bibnamefont
  {{Rodriguez}}}, \bibinfo {author} {\bibfnamefont {C.}~\bibnamefont
  {{R{\"o}ver}}}, \bibinfo {author} {\bibfnamefont {T.}~\bibnamefont
  {{Sidery}}}, \bibinfo {author} {\bibfnamefont {R.}~\bibnamefont {{Smith}}},
  \bibinfo {author} {\bibfnamefont {M.}~\bibnamefont {{Van Der Sluys}}},
  \bibinfo {author} {\bibfnamefont {A.}~\bibnamefont {{Vecchio}}}, \bibinfo
  {author} {\bibfnamefont {W.}~\bibnamefont {{Vousden}}}, \ and\ \bibinfo
  {author} {\bibfnamefont {L.}~\bibnamefont {{Wade}}},\ }\href {\doibase
  10.1103/PhysRevD.91.042003} {\bibfield  {journal} {\bibinfo  {journal}
  {\prd}\ }\textbf {\bibinfo {volume} {91}},\ \bibinfo {eid} {042003} (\bibinfo
  {year} {2015})},\ \Eprint {http://arxiv.org/abs/1409.7215} {arXiv:1409.7215
  [gr-qc]} \BibitemShut {NoStop}%
\bibitem [{\citenamefont {Romero-Shaw}\ \emph {et~al.}(2020)\citenamefont
  {Romero-Shaw} \emph {et~al.}}]{2020MNRAS.499.3295R}%
  \BibitemOpen
  \bibfield  {author} {\bibinfo {author} {\bibfnamefont {I.~M.}\ \bibnamefont
  {Romero-Shaw}} \emph {et~al.},\ }\href {\doibase 10.1093/mnras/staa2850}
  {\bibfield  {journal} {\bibinfo  {journal} {\mnras}\ }\textbf {\bibinfo
  {volume} {499}},\ \bibinfo {pages} {3295} (\bibinfo {year} {2020})},\ \Eprint
  {http://arxiv.org/abs/2006.00714} {arXiv:2006.00714 [astro-ph.IM]}
  \BibitemShut {NoStop}%
\bibitem [{\citenamefont {{Wofford}}\ \emph {et~al.}(2023)\citenamefont
  {{Wofford}}, \citenamefont {{Yelikar}}, \citenamefont {{Gallagher}},
  \citenamefont {{Champion}}, \citenamefont {{Wysocki}}, \citenamefont
  {{Delfavero}}, \citenamefont {{Lange}}, \citenamefont {{Rose}}, \citenamefont
  {{Valsan}}, \citenamefont {{Morisaki}}, \citenamefont {{Read}}, \citenamefont
  {{Henshaw}},\ and\ \citenamefont {{O'Shaughnessy}}}]{2023PhRvD.107b4040W}%
  \BibitemOpen
  \bibfield  {author} {\bibinfo {author} {\bibfnamefont {J.}~\bibnamefont
  {{Wofford}}}, \bibinfo {author} {\bibfnamefont {A.~B.}\ \bibnamefont
  {{Yelikar}}}, \bibinfo {author} {\bibfnamefont {H.}~\bibnamefont
  {{Gallagher}}}, \bibinfo {author} {\bibfnamefont {E.}~\bibnamefont
  {{Champion}}}, \bibinfo {author} {\bibfnamefont {D.}~\bibnamefont
  {{Wysocki}}}, \bibinfo {author} {\bibfnamefont {V.}~\bibnamefont
  {{Delfavero}}}, \bibinfo {author} {\bibfnamefont {J.}~\bibnamefont
  {{Lange}}}, \bibinfo {author} {\bibfnamefont {C.}~\bibnamefont {{Rose}}},
  \bibinfo {author} {\bibfnamefont {V.}~\bibnamefont {{Valsan}}}, \bibinfo
  {author} {\bibfnamefont {S.}~\bibnamefont {{Morisaki}}}, \bibinfo {author}
  {\bibfnamefont {J.}~\bibnamefont {{Read}}}, \bibinfo {author} {\bibfnamefont
  {C.}~\bibnamefont {{Henshaw}}}, \ and\ \bibinfo {author} {\bibfnamefont
  {R.}~\bibnamefont {{O'Shaughnessy}}},\ }\href {\doibase
  10.1103/PhysRevD.107.024040} {\bibfield  {journal} {\bibinfo  {journal}
  {\prd}\ }\textbf {\bibinfo {volume} {107}},\ \bibinfo {eid} {024040}
  (\bibinfo {year} {2023})},\ \Eprint {http://arxiv.org/abs/2210.07912}
  {arXiv:2210.07912 [gr-qc]} \BibitemShut {NoStop}%
\bibitem [{\citenamefont {{Lange}}\ \emph {et~al.}(2018)\citenamefont
  {{Lange}}, \citenamefont {{O'Shaughnessy}},\ and\ \citenamefont
  {{Rizzo}}}]{2018arXiv180510457L}%
  \BibitemOpen
  \bibfield  {author} {\bibinfo {author} {\bibfnamefont {J.}~\bibnamefont
  {{Lange}}}, \bibinfo {author} {\bibfnamefont {R.}~\bibnamefont
  {{O'Shaughnessy}}}, \ and\ \bibinfo {author} {\bibfnamefont {M.}~\bibnamefont
  {{Rizzo}}},\ }\href {\doibase 10.48550/arXiv.1805.10457} {\bibfield
  {journal} {\bibinfo  {journal} {arXiv e-prints}\ ,\ \bibinfo {eid}
  {arXiv:1805.10457}} (\bibinfo {year} {2018})},\ \Eprint
  {http://arxiv.org/abs/1805.10457} {arXiv:1805.10457 [gr-qc]} \BibitemShut
  {NoStop}%
\bibitem [{\citenamefont {{Cutler}}\ and\ \citenamefont
  {{Flanagan}}(1994)}]{1994PhRvD..49.2658C}%
  \BibitemOpen
  \bibfield  {author} {\bibinfo {author} {\bibfnamefont {C.}~\bibnamefont
  {{Cutler}}}\ and\ \bibinfo {author} {\bibfnamefont {{\'E}.~E.}\ \bibnamefont
  {{Flanagan}}},\ }\href {\doibase 10.1103/PhysRevD.49.2658} {\bibfield
  {journal} {\bibinfo  {journal} {\prd}\ }\textbf {\bibinfo {volume} {49}},\
  \bibinfo {pages} {2658} (\bibinfo {year} {1994})},\ \Eprint
  {http://arxiv.org/abs/gr-qc/9402014} {arXiv:gr-qc/9402014 [gr-qc]}
  \BibitemShut {NoStop}%
\bibitem [{\citenamefont {{Poisson}}\ and\ \citenamefont
  {{Will}}(1995)}]{1995PhRvD..52..848P}%
  \BibitemOpen
  \bibfield  {author} {\bibinfo {author} {\bibfnamefont {E.}~\bibnamefont
  {{Poisson}}}\ and\ \bibinfo {author} {\bibfnamefont {C.~M.}\ \bibnamefont
  {{Will}}},\ }\href {\doibase 10.1103/PhysRevD.52.848} {\bibfield  {journal}
  {\bibinfo  {journal} {\prd}\ }\textbf {\bibinfo {volume} {52}},\ \bibinfo
  {pages} {848} (\bibinfo {year} {1995})},\ \Eprint
  {http://arxiv.org/abs/gr-qc/9502040} {arXiv:gr-qc/9502040 [gr-qc]}
  \BibitemShut {NoStop}%
\bibitem [{\citenamefont {{Cho}}\ \emph {et~al.}(2013)\citenamefont {{Cho}},
  \citenamefont {{Ochsner}}, \citenamefont {{O'Shaughnessy}}, \citenamefont
  {{Kim}},\ and\ \citenamefont {{Lee}}}]{2013PhRvD..87b4004C}%
  \BibitemOpen
  \bibfield  {author} {\bibinfo {author} {\bibfnamefont {H.-S.}\ \bibnamefont
  {{Cho}}}, \bibinfo {author} {\bibfnamefont {E.}~\bibnamefont {{Ochsner}}},
  \bibinfo {author} {\bibfnamefont {R.}~\bibnamefont {{O'Shaughnessy}}},
  \bibinfo {author} {\bibfnamefont {C.}~\bibnamefont {{Kim}}}, \ and\ \bibinfo
  {author} {\bibfnamefont {C.-H.}\ \bibnamefont {{Lee}}},\ }\href {\doibase
  10.1103/PhysRevD.87.024004} {\bibfield  {journal} {\bibinfo  {journal}
  {\prd}\ }\textbf {\bibinfo {volume} {87}},\ \bibinfo {eid} {024004} (\bibinfo
  {year} {2013})},\ \Eprint {http://arxiv.org/abs/1209.4494} {arXiv:1209.4494
  [gr-qc]} \BibitemShut {NoStop}%
\bibitem [{\citenamefont {{O'Shaughnessy}}\ \emph {et~al.}(2014)\citenamefont
  {{O'Shaughnessy}}, \citenamefont {{Farr}}, \citenamefont {{Ochsner}},
  \citenamefont {{Cho}}, \citenamefont {{Kim}},\ and\ \citenamefont
  {{Lee}}}]{2014PhRvD..89f4048O}%
  \BibitemOpen
  \bibfield  {author} {\bibinfo {author} {\bibfnamefont {R.}~\bibnamefont
  {{O'Shaughnessy}}}, \bibinfo {author} {\bibfnamefont {B.}~\bibnamefont
  {{Farr}}}, \bibinfo {author} {\bibfnamefont {E.}~\bibnamefont {{Ochsner}}},
  \bibinfo {author} {\bibfnamefont {H.-S.}\ \bibnamefont {{Cho}}}, \bibinfo
  {author} {\bibfnamefont {C.}~\bibnamefont {{Kim}}}, \ and\ \bibinfo {author}
  {\bibfnamefont {C.-H.}\ \bibnamefont {{Lee}}},\ }\href {\doibase
  10.1103/PhysRevD.89.064048} {\bibfield  {journal} {\bibinfo  {journal}
  {\prd}\ }\textbf {\bibinfo {volume} {89}},\ \bibinfo {eid} {064048} (\bibinfo
  {year} {2014})},\ \Eprint {http://arxiv.org/abs/1308.4704} {arXiv:1308.4704
  [gr-qc]} \BibitemShut {NoStop}%
\bibitem [{\citenamefont {{Wang}}\ \emph {et~al.}(2020)\citenamefont {{Wang}},
  \citenamefont {{Zhu}}, \citenamefont {{Li}},\ and\ \citenamefont
  {{Zhao}}}]{2020ApJS..250....6W}%
  \BibitemOpen
  \bibfield  {author} {\bibinfo {author} {\bibfnamefont {B.}~\bibnamefont
  {{Wang}}}, \bibinfo {author} {\bibfnamefont {Z.}~\bibnamefont {{Zhu}}},
  \bibinfo {author} {\bibfnamefont {A.}~\bibnamefont {{Li}}}, \ and\ \bibinfo
  {author} {\bibfnamefont {W.}~\bibnamefont {{Zhao}}},\ }\href {\doibase
  10.3847/1538-4365/aba2f3} {\bibfield  {journal} {\bibinfo  {journal} {\apjs}\
  }\textbf {\bibinfo {volume} {250}},\ \bibinfo {eid} {6} (\bibinfo {year}
  {2020})},\ \Eprint {http://arxiv.org/abs/2005.12875} {arXiv:2005.12875
  [gr-qc]} \BibitemShut {NoStop}%
\bibitem [{\citenamefont {{Iacovelli}}\ \emph
  {et~al.}(2022{\natexlab{a}})\citenamefont {{Iacovelli}}, \citenamefont
  {{Mancarella}}, \citenamefont {{Foffa}},\ and\ \citenamefont
  {{Maggiore}}}]{2022ApJS..263....2I}%
  \BibitemOpen
  \bibfield  {author} {\bibinfo {author} {\bibfnamefont {F.}~\bibnamefont
  {{Iacovelli}}}, \bibinfo {author} {\bibfnamefont {M.}~\bibnamefont
  {{Mancarella}}}, \bibinfo {author} {\bibfnamefont {S.}~\bibnamefont
  {{Foffa}}}, \ and\ \bibinfo {author} {\bibfnamefont {M.}~\bibnamefont
  {{Maggiore}}},\ }\href {\doibase 10.3847/1538-4365/ac9129} {\bibfield
  {journal} {\bibinfo  {journal} {\apjs}\ }\textbf {\bibinfo {volume} {263}},\
  \bibinfo {eid} {2} (\bibinfo {year} {2022}{\natexlab{a}})},\ \Eprint
  {http://arxiv.org/abs/2207.06910} {arXiv:2207.06910 [astro-ph.IM]}
  \BibitemShut {NoStop}%
\bibitem [{\citenamefont {{Iacovelli}}\ \emph
  {et~al.}(2022{\natexlab{b}})\citenamefont {{Iacovelli}}, \citenamefont
  {{Mancarella}}, \citenamefont {{Foffa}},\ and\ \citenamefont
  {{Maggiore}}}]{2022ApJ...941..208I}%
  \BibitemOpen
  \bibfield  {author} {\bibinfo {author} {\bibfnamefont {F.}~\bibnamefont
  {{Iacovelli}}}, \bibinfo {author} {\bibfnamefont {M.}~\bibnamefont
  {{Mancarella}}}, \bibinfo {author} {\bibfnamefont {S.}~\bibnamefont
  {{Foffa}}}, \ and\ \bibinfo {author} {\bibfnamefont {M.}~\bibnamefont
  {{Maggiore}}},\ }\href {\doibase 10.3847/1538-4357/ac9cd4} {\bibfield
  {journal} {\bibinfo  {journal} {\apj}\ }\textbf {\bibinfo {volume} {941}},\
  \bibinfo {eid} {208} (\bibinfo {year} {2022}{\natexlab{b}})},\ \Eprint
  {http://arxiv.org/abs/2207.02771} {arXiv:2207.02771 [gr-qc]} \BibitemShut
  {NoStop}%
\bibitem [{\citenamefont {{Huxford}}\ \emph {et~al.}(2024)\citenamefont
  {{Huxford}}, \citenamefont {{Kashyap}}, \citenamefont {{Borhanian}},
  \citenamefont {{Dhani}}, \citenamefont {{Gupta}},\ and\ \citenamefont
  {{Sathyaprakash}}}]{2024PhRvD.109j3035H}%
  \BibitemOpen
  \bibfield  {author} {\bibinfo {author} {\bibfnamefont {R.}~\bibnamefont
  {{Huxford}}}, \bibinfo {author} {\bibfnamefont {R.}~\bibnamefont
  {{Kashyap}}}, \bibinfo {author} {\bibfnamefont {S.}~\bibnamefont
  {{Borhanian}}}, \bibinfo {author} {\bibfnamefont {A.}~\bibnamefont
  {{Dhani}}}, \bibinfo {author} {\bibfnamefont {I.}~\bibnamefont {{Gupta}}}, \
  and\ \bibinfo {author} {\bibfnamefont {B.~S.}\ \bibnamefont
  {{Sathyaprakash}}},\ }\href {\doibase 10.1103/PhysRevD.109.103035} {\bibfield
   {journal} {\bibinfo  {journal} {\prd}\ }\textbf {\bibinfo {volume} {109}},\
  \bibinfo {eid} {103035} (\bibinfo {year} {2024})},\ \Eprint
  {http://arxiv.org/abs/2307.05376} {arXiv:2307.05376 [gr-qc]} \BibitemShut
  {NoStop}%
\bibitem [{\citenamefont {{Abbott}}\ \emph {et~al.}(2023)\citenamefont
  {{Abbott}}, \citenamefont {{Abbott}}, \citenamefont {{Abbott}}, \citenamefont
  {{Acernese}}, \citenamefont {{Ackley}}, \citenamefont {{Adams}},
  \citenamefont {{Adams}}, \citenamefont {{Addesso}}, \citenamefont
  {{Adhikari}}, \citenamefont {{Adya}},\ and\ \citenamefont
  {et~al.}}]{2023PhRvX..13a1048A}%
  \BibitemOpen
  \bibfield  {author} {\bibinfo {author} {\bibfnamefont {B.~P.}\ \bibnamefont
  {{Abbott}}}, \bibinfo {author} {\bibfnamefont {R.}~\bibnamefont {{Abbott}}},
  \bibinfo {author} {\bibfnamefont {T.~D.}\ \bibnamefont {{Abbott}}}, \bibinfo
  {author} {\bibfnamefont {F.}~\bibnamefont {{Acernese}}}, \bibinfo {author}
  {\bibfnamefont {K.}~\bibnamefont {{Ackley}}}, \bibinfo {author}
  {\bibfnamefont {C.}~\bibnamefont {{Adams}}}, \bibinfo {author} {\bibfnamefont
  {T.}~\bibnamefont {{Adams}}}, \bibinfo {author} {\bibfnamefont
  {P.}~\bibnamefont {{Addesso}}}, \bibinfo {author} {\bibfnamefont {R.~X.}\
  \bibnamefont {{Adhikari}}}, \bibinfo {author} {\bibfnamefont {V.~B.}\
  \bibnamefont {{Adya}}}, \ and\ \bibinfo {author} {\bibnamefont {et~al.}},\
  }\href {\doibase 10.1103/PhysRevX.13.011048} {\bibfield  {journal} {\bibinfo
  {journal} {Physical Review X}\ }\textbf {\bibinfo {volume} {13}},\ \bibinfo
  {eid} {011048} (\bibinfo {year} {2023})},\ \Eprint
  {http://arxiv.org/abs/2111.03634} {arXiv:2111.03634 [astro-ph.HE]}
  \BibitemShut {NoStop}%
\bibitem [{\citenamefont {{Forbes}}\ \emph {et~al.}(2019)\citenamefont
  {{Forbes}}, \citenamefont {{Bose}}, \citenamefont {{Reddy}}, \citenamefont
  {{Zhou}}, \citenamefont {{Mukherjee}},\ and\ \citenamefont
  {{De}}}]{2019PhRvD.100h3010F}%
  \BibitemOpen
  \bibfield  {author} {\bibinfo {author} {\bibfnamefont {M.~M.}\ \bibnamefont
  {{Forbes}}}, \bibinfo {author} {\bibfnamefont {S.}~\bibnamefont {{Bose}}},
  \bibinfo {author} {\bibfnamefont {S.}~\bibnamefont {{Reddy}}}, \bibinfo
  {author} {\bibfnamefont {D.}~\bibnamefont {{Zhou}}}, \bibinfo {author}
  {\bibfnamefont {A.}~\bibnamefont {{Mukherjee}}}, \ and\ \bibinfo {author}
  {\bibfnamefont {S.}~\bibnamefont {{De}}},\ }\href {\doibase
  10.1103/PhysRevD.100.083010} {\bibfield  {journal} {\bibinfo  {journal}
  {\prd}\ }\textbf {\bibinfo {volume} {100}},\ \bibinfo {eid} {083010}
  (\bibinfo {year} {2019})}\BibitemShut {NoStop}%
\bibitem [{\citenamefont {{Iacovelli}}\ \emph {et~al.}(2023)\citenamefont
  {{Iacovelli}}, \citenamefont {{Mancarella}}, \citenamefont {{Mondal}},
  \citenamefont {{Puecher}}, \citenamefont {{Dietrich}}, \citenamefont
  {{Gulminelli}}, \citenamefont {{Maggiore}},\ and\ \citenamefont
  {{Oertel}}}]{2023PhRvD.108l2006I}%
  \BibitemOpen
  \bibfield  {author} {\bibinfo {author} {\bibfnamefont {F.}~\bibnamefont
  {{Iacovelli}}}, \bibinfo {author} {\bibfnamefont {M.}~\bibnamefont
  {{Mancarella}}}, \bibinfo {author} {\bibfnamefont {C.}~\bibnamefont
  {{Mondal}}}, \bibinfo {author} {\bibfnamefont {A.}~\bibnamefont {{Puecher}}},
  \bibinfo {author} {\bibfnamefont {T.}~\bibnamefont {{Dietrich}}}, \bibinfo
  {author} {\bibfnamefont {F.}~\bibnamefont {{Gulminelli}}}, \bibinfo {author}
  {\bibfnamefont {M.}~\bibnamefont {{Maggiore}}}, \ and\ \bibinfo {author}
  {\bibfnamefont {M.}~\bibnamefont {{Oertel}}},\ }\href {\doibase
  10.1103/PhysRevD.108.122006} {\bibfield  {journal} {\bibinfo  {journal}
  {\prd}\ }\textbf {\bibinfo {volume} {108}},\ \bibinfo {eid} {122006}
  (\bibinfo {year} {2023})},\ \Eprint {http://arxiv.org/abs/2308.12378}
  {arXiv:2308.12378 [gr-qc]} \BibitemShut {NoStop}%
\bibitem [{\citenamefont {{Alsing}}\ \emph {et~al.}(2018)\citenamefont
  {{Alsing}}, \citenamefont {{Silva}},\ and\ \citenamefont
  {{Berti}}}]{2018MNRAS.478.1377A}%
  \BibitemOpen
  \bibfield  {author} {\bibinfo {author} {\bibfnamefont {J.}~\bibnamefont
  {{Alsing}}}, \bibinfo {author} {\bibfnamefont {H.~O.}\ \bibnamefont
  {{Silva}}}, \ and\ \bibinfo {author} {\bibfnamefont {E.}~\bibnamefont
  {{Berti}}},\ }\href {\doibase 10.1093/mnras/sty1065} {\bibfield  {journal}
  {\bibinfo  {journal} {\mnras}\ }\textbf {\bibinfo {volume} {478}},\ \bibinfo
  {pages} {1377} (\bibinfo {year} {2018})},\ \Eprint
  {http://arxiv.org/abs/1709.07889} {arXiv:1709.07889 [astro-ph.HE]}
  \BibitemShut {NoStop}%
\bibitem [{\citenamefont {{Hinderer}}\ \emph {et~al.}(2010)\citenamefont
  {{Hinderer}}, \citenamefont {{Lackey}}, \citenamefont {{Lang}},\ and\
  \citenamefont {{Read}}}]{2010PhRvD..81l3016H}%
  \BibitemOpen
  \bibfield  {author} {\bibinfo {author} {\bibfnamefont {T.}~\bibnamefont
  {{Hinderer}}}, \bibinfo {author} {\bibfnamefont {B.~D.}\ \bibnamefont
  {{Lackey}}}, \bibinfo {author} {\bibfnamefont {R.~N.}\ \bibnamefont
  {{Lang}}}, \ and\ \bibinfo {author} {\bibfnamefont {J.~S.}\ \bibnamefont
  {{Read}}},\ }\href {\doibase 10.1103/PhysRevD.81.123016} {\bibfield
  {journal} {\bibinfo  {journal} {\prd}\ }\textbf {\bibinfo {volume} {81}},\
  \bibinfo {eid} {123016} (\bibinfo {year} {2010})},\ \Eprint
  {http://arxiv.org/abs/0911.3535} {arXiv:0911.3535 [astro-ph.HE]} \BibitemShut
  {NoStop}%
\bibitem [{\citenamefont {{Zhu}}\ \emph {et~al.}(2020)\citenamefont {{Zhu}},
  \citenamefont {{Li}},\ and\ \citenamefont
  {{Rezzolla}}}]{2020PhRvD.102h4058Z}%
  \BibitemOpen
  \bibfield  {author} {\bibinfo {author} {\bibfnamefont {Z.}~\bibnamefont
  {{Zhu}}}, \bibinfo {author} {\bibfnamefont {A.}~\bibnamefont {{Li}}}, \ and\
  \bibinfo {author} {\bibfnamefont {L.}~\bibnamefont {{Rezzolla}}},\ }\href
  {\doibase 10.1103/PhysRevD.102.084058} {\bibfield  {journal} {\bibinfo
  {journal} {\prd}\ }\textbf {\bibinfo {volume} {102}},\ \bibinfo {eid}
  {084058} (\bibinfo {year} {2020})},\ \Eprint
  {http://arxiv.org/abs/2005.02677} {arXiv:2005.02677 [astro-ph.HE]}
  \BibitemShut {NoStop}%
\bibitem [{\citenamefont {{Li}}\ \emph {et~al.}(2025)\citenamefont {{Li}},
  \citenamefont {{Han}}, \citenamefont {{Lin}}, \citenamefont {{Wang}},
  \citenamefont {{Zhou}},\ and\ \citenamefont {{Shi}}}]{2025PhRvD.111g4026L}%
  \BibitemOpen
  \bibfield  {author} {\bibinfo {author} {\bibfnamefont {R.}~\bibnamefont
  {{Li}}}, \bibinfo {author} {\bibfnamefont {S.}~\bibnamefont {{Han}}},
  \bibinfo {author} {\bibfnamefont {Z.}~\bibnamefont {{Lin}}}, \bibinfo
  {author} {\bibfnamefont {L.}~\bibnamefont {{Wang}}}, \bibinfo {author}
  {\bibfnamefont {K.}~\bibnamefont {{Zhou}}}, \ and\ \bibinfo {author}
  {\bibfnamefont {S.}~\bibnamefont {{Shi}}},\ }\href {\doibase
  10.1103/PhysRevD.111.074026} {\bibfield  {journal} {\bibinfo  {journal}
  {\prd}\ }\textbf {\bibinfo {volume} {111}},\ \bibinfo {eid} {074026}
  (\bibinfo {year} {2025})},\ \Eprint {http://arxiv.org/abs/2501.15810}
  {arXiv:2501.15810 [nucl-th]} \BibitemShut {NoStop}%
\bibitem [{\citenamefont {{Lindblom}}(1992)}]{1992ApJ...398..569L}%
  \BibitemOpen
  \bibfield  {author} {\bibinfo {author} {\bibfnamefont {L.}~\bibnamefont
  {{Lindblom}}},\ }\href {\doibase 10.1086/171882} {\bibfield  {journal}
  {\bibinfo  {journal} {\apj}\ }\textbf {\bibinfo {volume} {398}},\ \bibinfo
  {pages} {569} (\bibinfo {year} {1992})}\BibitemShut {NoStop}%
\bibitem [{\citenamefont {{Lindblom}}(1998)}]{1998PhRvD..58b4008L}%
  \BibitemOpen
  \bibfield  {author} {\bibinfo {author} {\bibfnamefont {L.}~\bibnamefont
  {{Lindblom}}},\ }\href {\doibase 10.1103/PhysRevD.58.024008} {\bibfield
  {journal} {\bibinfo  {journal} {\prd}\ }\textbf {\bibinfo {volume} {58}},\
  \bibinfo {eid} {024008} (\bibinfo {year} {1998})},\ \Eprint
  {http://arxiv.org/abs/gr-qc/9802072} {arXiv:gr-qc/9802072 [gr-qc]}
  \BibitemShut {NoStop}%
\bibitem [{\citenamefont {{Hinderer}}(2008)}]{2008ApJ...677.1216H}%
  \BibitemOpen
  \bibfield  {author} {\bibinfo {author} {\bibfnamefont {T.}~\bibnamefont
  {{Hinderer}}},\ }\href {\doibase 10.1086/533487} {\bibfield  {journal}
  {\bibinfo  {journal} {\apj}\ }\textbf {\bibinfo {volume} {677}},\ \bibinfo
  {pages} {1216} (\bibinfo {year} {2008})},\ \Eprint
  {http://arxiv.org/abs/0711.2420} {arXiv:0711.2420 [astro-ph]} \BibitemShut
  {NoStop}%
\bibitem [{\citenamefont {{Postnikov}}(2010)}]{2010PhDT........18P}%
  \BibitemOpen
  \bibfield  {author} {\bibinfo {author} {\bibfnamefont {S.}~\bibnamefont
  {{Postnikov}}},\ }\emph {\bibinfo {title} {{Topics in the physics and
  astrophysics of neutron stars}}},\ \href@noop {} {Ph.D. thesis},\ \bibinfo
  {school} {Ohio University} (\bibinfo {year} {2010})\BibitemShut {NoStop}%
\bibitem [{\citenamefont {Steil}(2017)}]{steil2017structure}%
  \BibitemOpen
  \bibfield  {author} {\bibinfo {author} {\bibnamefont {Steil}},\ }\emph
  {\bibinfo {title} {Structure of slowly rotating magnetized neutron stars in a
  perturbative approach}},\ \href
  {https://theorie.ikp.physik.tu-darmstadt.de/nhq/downloads/thesis/master.steil.pdf}
  {Master's thesis},\ \bibinfo  {school} {Technische Universität Darmstadt},
  \bibinfo {address} {Darmstadt, Germany} (\bibinfo {year} {2017})\BibitemShut
  {NoStop}%
\bibitem [{\citenamefont {{Baym}}\ \emph {et~al.}(1971)\citenamefont {{Baym}},
  \citenamefont {{Pethick}},\ and\ \citenamefont
  {{Sutherland}}}]{1971ApJ...170..299B}%
  \BibitemOpen
  \bibfield  {author} {\bibinfo {author} {\bibfnamefont {G.}~\bibnamefont
  {{Baym}}}, \bibinfo {author} {\bibfnamefont {C.}~\bibnamefont {{Pethick}}}, \
  and\ \bibinfo {author} {\bibfnamefont {P.}~\bibnamefont {{Sutherland}}},\
  }\href {\doibase 10.1086/151216} {\bibfield  {journal} {\bibinfo  {journal}
  {\apj}\ }\textbf {\bibinfo {volume} {170}},\ \bibinfo {pages} {299} (\bibinfo
  {year} {1971})}\BibitemShut {NoStop}%
\bibitem [{\citenamefont {{You}}\ \emph {et~al.}(2026)\citenamefont {{You}},
  \citenamefont {{Zhu}}, \citenamefont {{Han}},\ and\ \citenamefont
  {{O'Shaughnessy}}}]{zhu2025inprep}%
  \BibitemOpen
  \bibfield  {author} {\bibinfo {author} {\bibfnamefont {H.-S.}\ \bibnamefont
  {{You}}}, \bibinfo {author} {\bibfnamefont {Z.}~\bibnamefont {{Zhu}}},
  \bibinfo {author} {\bibfnamefont {S.}~\bibnamefont {{Han}}}, \ and\ \bibinfo
  {author} {\bibfnamefont {R.}~\bibnamefont {{O'Shaughnessy}}},\ }\href@noop {}
  {} (\bibinfo {year} {2026}),\ \bibinfo {note} {in preparation}\BibitemShut
  {NoStop}%
\bibitem [{\citenamefont {{Khan}}\ \emph {et~al.}(2016)\citenamefont {{Khan}},
  \citenamefont {{Husa}}, \citenamefont {{Hannam}}, \citenamefont {{Ohme}},
  \citenamefont {{P{\"u}rrer}}, \citenamefont {{Forteza}},\ and\ \citenamefont
  {{Boh{\'e}}}}]{2016PhRvD..93d4007K}%
  \BibitemOpen
  \bibfield  {author} {\bibinfo {author} {\bibfnamefont {S.}~\bibnamefont
  {{Khan}}}, \bibinfo {author} {\bibfnamefont {S.}~\bibnamefont {{Husa}}},
  \bibinfo {author} {\bibfnamefont {M.}~\bibnamefont {{Hannam}}}, \bibinfo
  {author} {\bibfnamefont {F.}~\bibnamefont {{Ohme}}}, \bibinfo {author}
  {\bibfnamefont {M.}~\bibnamefont {{P{\"u}rrer}}}, \bibinfo {author}
  {\bibfnamefont {X.~J.}\ \bibnamefont {{Forteza}}}, \ and\ \bibinfo {author}
  {\bibfnamefont {A.}~\bibnamefont {{Boh{\'e}}}},\ }\href {\doibase
  10.1103/PhysRevD.93.044007} {\bibfield  {journal} {\bibinfo  {journal}
  {\prd}\ }\textbf {\bibinfo {volume} {93}},\ \bibinfo {eid} {044007} (\bibinfo
  {year} {2016})},\ \Eprint {http://arxiv.org/abs/1508.07253} {arXiv:1508.07253
  [gr-qc]} \BibitemShut {NoStop}%
\bibitem [{\citenamefont {{Vallisneri}}(2008)}]{2008PhRvD..77d2001V}%
  \BibitemOpen
  \bibfield  {author} {\bibinfo {author} {\bibfnamefont {M.}~\bibnamefont
  {{Vallisneri}}},\ }\href {\doibase 10.1103/PhysRevD.77.042001} {\bibfield
  {journal} {\bibinfo  {journal} {\prd}\ }\textbf {\bibinfo {volume} {77}},\
  \bibinfo {eid} {042001} (\bibinfo {year} {2008})},\ \Eprint
  {http://arxiv.org/abs/gr-qc/0703086} {arXiv:gr-qc/0703086 [gr-qc]}
  \BibitemShut {NoStop}%
\bibitem [{\citenamefont {{Li}}\ and\ \citenamefont
  {{Han}}(2013)}]{2013PhLB..727..276L}%
  \BibitemOpen
  \bibfield  {author} {\bibinfo {author} {\bibfnamefont {B.-A.}\ \bibnamefont
  {{Li}}}\ and\ \bibinfo {author} {\bibfnamefont {X.}~\bibnamefont {{Han}}},\
  }\href {\doibase 10.1016/j.physletb.2013.10.006} {\bibfield  {journal}
  {\bibinfo  {journal} {Physics Letters B}\ }\textbf {\bibinfo {volume}
  {727}},\ \bibinfo {pages} {276} (\bibinfo {year} {2013})},\ \Eprint
  {http://arxiv.org/abs/1304.3368} {arXiv:1304.3368 [nucl-th]} \BibitemShut
  {NoStop}%
\bibitem [{\citenamefont {{Fortin}}\ \emph {et~al.}(2016)\citenamefont
  {{Fortin}}, \citenamefont {{Provid{\^e}ncia}}, \citenamefont {{Raduta}},
  \citenamefont {{Gulminelli}}, \citenamefont {{Zdunik}}, \citenamefont
  {{Haensel}},\ and\ \citenamefont {{Bejger}}}]{2016PhRvC..94c5804F}%
  \BibitemOpen
  \bibfield  {author} {\bibinfo {author} {\bibfnamefont {M.}~\bibnamefont
  {{Fortin}}}, \bibinfo {author} {\bibfnamefont {C.}~\bibnamefont
  {{Provid{\^e}ncia}}}, \bibinfo {author} {\bibfnamefont {A.~R.}\ \bibnamefont
  {{Raduta}}}, \bibinfo {author} {\bibfnamefont {F.}~\bibnamefont
  {{Gulminelli}}}, \bibinfo {author} {\bibfnamefont {J.~L.}\ \bibnamefont
  {{Zdunik}}}, \bibinfo {author} {\bibfnamefont {P.}~\bibnamefont {{Haensel}}},
  \ and\ \bibinfo {author} {\bibfnamefont {M.}~\bibnamefont {{Bejger}}},\
  }\href {\doibase 10.1103/PhysRevC.94.035804} {\bibfield  {journal} {\bibinfo
  {journal} {\prc}\ }\textbf {\bibinfo {volume} {94}},\ \bibinfo {eid} {035804}
  (\bibinfo {year} {2016})},\ \Eprint {http://arxiv.org/abs/1604.01944}
  {arXiv:1604.01944 [astro-ph.SR]} \BibitemShut {NoStop}%
\bibitem [{\citenamefont {{Perot}}\ \emph {et~al.}(2020)\citenamefont
  {{Perot}}, \citenamefont {{Chamel}},\ and\ \citenamefont
  {{Sourie}}}]{2020PhRvC.101a5806P}%
  \BibitemOpen
  \bibfield  {author} {\bibinfo {author} {\bibfnamefont {L.}~\bibnamefont
  {{Perot}}}, \bibinfo {author} {\bibfnamefont {N.}~\bibnamefont {{Chamel}}}, \
  and\ \bibinfo {author} {\bibfnamefont {A.}~\bibnamefont {{Sourie}}},\ }\href
  {\doibase 10.1103/PhysRevC.101.015806} {\bibfield  {journal} {\bibinfo
  {journal} {\prc}\ }\textbf {\bibinfo {volume} {101}},\ \bibinfo {eid}
  {015806} (\bibinfo {year} {2020})},\ \Eprint
  {http://arxiv.org/abs/2001.11068} {arXiv:2001.11068 [astro-ph.HE]}
  \BibitemShut {NoStop}%
\bibitem [{\citenamefont {{Suleiman}}\ \emph {et~al.}(2021)\citenamefont
  {{Suleiman}}, \citenamefont {{Fortin}}, \citenamefont {{Zdunik}},\ and\
  \citenamefont {{Haensel}}}]{2021PhRvC.104a5801S}%
  \BibitemOpen
  \bibfield  {author} {\bibinfo {author} {\bibfnamefont {L.}~\bibnamefont
  {{Suleiman}}}, \bibinfo {author} {\bibfnamefont {M.}~\bibnamefont
  {{Fortin}}}, \bibinfo {author} {\bibfnamefont {J.~L.}\ \bibnamefont
  {{Zdunik}}}, \ and\ \bibinfo {author} {\bibfnamefont {P.}~\bibnamefont
  {{Haensel}}},\ }\href {\doibase 10.1103/PhysRevC.104.015801} {\bibfield
  {journal} {\bibinfo  {journal} {\prc}\ }\textbf {\bibinfo {volume} {104}},\
  \bibinfo {eid} {015801} (\bibinfo {year} {2021})},\ \Eprint
  {http://arxiv.org/abs/2106.12845} {arXiv:2106.12845 [astro-ph.HE]}
  \BibitemShut {NoStop}%
\bibitem [{\citenamefont {{Drischler}}\ \emph {et~al.}(2016)\citenamefont
  {{Drischler}}, \citenamefont {{Hebeler}},\ and\ \citenamefont
  {{Schwenk}}}]{2016PhRvC..93e4314D}%
  \BibitemOpen
  \bibfield  {author} {\bibinfo {author} {\bibfnamefont {C.}~\bibnamefont
  {{Drischler}}}, \bibinfo {author} {\bibfnamefont {K.}~\bibnamefont
  {{Hebeler}}}, \ and\ \bibinfo {author} {\bibfnamefont {A.}~\bibnamefont
  {{Schwenk}}},\ }\href {\doibase 10.1103/PhysRevC.93.054314} {\bibfield
  {journal} {\bibinfo  {journal} {\prc}\ }\textbf {\bibinfo {volume} {93}},\
  \bibinfo {eid} {054314} (\bibinfo {year} {2016})},\ \Eprint
  {http://arxiv.org/abs/1510.06728} {arXiv:1510.06728 [nucl-th]} \BibitemShut
  {NoStop}%
\bibitem [{\citenamefont {{Davis}}\ \emph {et~al.}(2024)\citenamefont
  {{Davis}}, \citenamefont {{Dinh Thi}}, \citenamefont {{Fantina}},
  \citenamefont {{Gulminelli}}, \citenamefont {{Oertel}},\ and\ \citenamefont
  {{Suleiman}}}]{2024A&A...687A..44D}%
  \BibitemOpen
  \bibfield  {author} {\bibinfo {author} {\bibfnamefont {P.~J.}\ \bibnamefont
  {{Davis}}}, \bibinfo {author} {\bibfnamefont {H.}~\bibnamefont {{Dinh Thi}}},
  \bibinfo {author} {\bibfnamefont {A.~F.}\ \bibnamefont {{Fantina}}}, \bibinfo
  {author} {\bibfnamefont {F.}~\bibnamefont {{Gulminelli}}}, \bibinfo {author}
  {\bibfnamefont {M.}~\bibnamefont {{Oertel}}}, \ and\ \bibinfo {author}
  {\bibfnamefont {L.}~\bibnamefont {{Suleiman}}},\ }\href {\doibase
  10.1051/0004-6361/202348402} {\bibfield  {journal} {\bibinfo  {journal}
  {\aap}\ }\textbf {\bibinfo {volume} {687}},\ \bibinfo {eid} {A44} (\bibinfo
  {year} {2024})},\ \Eprint {http://arxiv.org/abs/2406.14906} {arXiv:2406.14906
  [astro-ph.HE]} \BibitemShut {NoStop}%
\bibitem [{\citenamefont {{Scurto}}\ \emph {et~al.}(2024)\citenamefont
  {{Scurto}}, \citenamefont {{Pais}},\ and\ \citenamefont
  {{Gulminelli}}}]{2024PhRvD.109j3015S}%
  \BibitemOpen
  \bibfield  {author} {\bibinfo {author} {\bibfnamefont {L.}~\bibnamefont
  {{Scurto}}}, \bibinfo {author} {\bibfnamefont {H.}~\bibnamefont {{Pais}}}, \
  and\ \bibinfo {author} {\bibfnamefont {F.}~\bibnamefont {{Gulminelli}}},\
  }\href {\doibase 10.1103/PhysRevD.109.103015} {\bibfield  {journal} {\bibinfo
   {journal} {\prd}\ }\textbf {\bibinfo {volume} {109}},\ \bibinfo {eid}
  {103015} (\bibinfo {year} {2024})},\ \Eprint
  {http://arxiv.org/abs/2402.15548} {arXiv:2402.15548 [nucl-th]} \BibitemShut
  {NoStop}%
\bibitem [{\citenamefont {{Klausner}}\ \emph {et~al.}(2025)\citenamefont
  {{Klausner}}, \citenamefont {{Antonelli}},\ and\ \citenamefont
  {{Gulminelli}}}]{2025arXiv250516929K}%
  \BibitemOpen
  \bibfield  {author} {\bibinfo {author} {\bibfnamefont {P.}~\bibnamefont
  {{Klausner}}}, \bibinfo {author} {\bibfnamefont {M.}~\bibnamefont
  {{Antonelli}}}, \ and\ \bibinfo {author} {\bibfnamefont {F.}~\bibnamefont
  {{Gulminelli}}},\ }\href {\doibase 10.48550/arXiv.2505.16929} {\bibfield
  {journal} {\bibinfo  {journal} {arXiv e-prints}\ ,\ \bibinfo {eid}
  {arXiv:2505.16929}} (\bibinfo {year} {2025})},\ \Eprint
  {http://arxiv.org/abs/2505.16929} {arXiv:2505.16929 [nucl-th]} \BibitemShut
  {NoStop}%
\bibitem [{\citenamefont {{Brown}}\ \emph {et~al.}(2025)\citenamefont
  {{Brown}}, \citenamefont {{Gade}}, \citenamefont {{Stroberg}}, \citenamefont
  {{Escher}}, \citenamefont {{Fossez}}, \citenamefont {{Giuliani}},
  \citenamefont {{Hoffman}}, \citenamefont {{Nazarewicz}}, \citenamefont
  {{Seng}}, \citenamefont {{Sorensen}}, \citenamefont {{Vassh}}, \citenamefont
  {{Bazin}}, \citenamefont {{Brown}}, \citenamefont {{Caprio}}, \citenamefont
  {{Crawford}}, \citenamefont {{Danielewicz}}, \citenamefont {{Drischler}},
  \citenamefont {{Garcia Ruiz}}, \citenamefont {{Godbey}}, \citenamefont
  {{Grzywacz}}, \citenamefont {{Hlophe}}, \citenamefont {{Holt}}, \citenamefont
  {{Iwasaki}}, \citenamefont {{Lee}}, \citenamefont {{Lenzi}}, \citenamefont
  {{Liddick}}, \citenamefont {{Lubna}}, \citenamefont {{Macchiavelli}},
  \citenamefont {{Mart{\'\i}nez-Pinedo}}, \citenamefont {{McCoy}},
  \citenamefont {{Mercenne}}, \citenamefont {{Minamisono}}, \citenamefont
  {{Monteagudo}}, \citenamefont {{Navratil}}, \citenamefont {{Ringle}},
  \citenamefont {{Sargsyan}}, \citenamefont {{Schatz}}, \citenamefont
  {{Spieker}}, \citenamefont {{Volya}}, \citenamefont {{Zegers}}, \citenamefont
  {{Zelevinsky}},\ and\ \citenamefont {{Zhang}}}]{2025JPhG...52e0501B}%
  \BibitemOpen
  \bibfield  {author} {\bibinfo {author} {\bibfnamefont {B.~A.}\ \bibnamefont
  {{Brown}}}, \bibinfo {author} {\bibfnamefont {A.}~\bibnamefont {{Gade}}},
  \bibinfo {author} {\bibfnamefont {S.~R.}\ \bibnamefont {{Stroberg}}},
  \bibinfo {author} {\bibfnamefont {J.~E.}\ \bibnamefont {{Escher}}}, \bibinfo
  {author} {\bibfnamefont {K.}~\bibnamefont {{Fossez}}}, \bibinfo {author}
  {\bibfnamefont {P.}~\bibnamefont {{Giuliani}}}, \bibinfo {author}
  {\bibfnamefont {C.~R.}\ \bibnamefont {{Hoffman}}}, \bibinfo {author}
  {\bibfnamefont {W.}~\bibnamefont {{Nazarewicz}}}, \bibinfo {author}
  {\bibfnamefont {C.-Y.}\ \bibnamefont {{Seng}}}, \bibinfo {author}
  {\bibfnamefont {A.}~\bibnamefont {{Sorensen}}}, \bibinfo {author}
  {\bibfnamefont {N.}~\bibnamefont {{Vassh}}}, \bibinfo {author} {\bibfnamefont
  {D.}~\bibnamefont {{Bazin}}}, \bibinfo {author} {\bibfnamefont {K.~W.}\
  \bibnamefont {{Brown}}}, \bibinfo {author} {\bibfnamefont {M.~A.}\
  \bibnamefont {{Caprio}}}, \bibinfo {author} {\bibfnamefont {H.}~\bibnamefont
  {{Crawford}}}, \bibinfo {author} {\bibfnamefont {P.}~\bibnamefont
  {{Danielewicz}}}, \bibinfo {author} {\bibfnamefont {C.}~\bibnamefont
  {{Drischler}}}, \bibinfo {author} {\bibfnamefont {R.~F.}\ \bibnamefont
  {{Garcia Ruiz}}}, \bibinfo {author} {\bibfnamefont {K.}~\bibnamefont
  {{Godbey}}}, \bibinfo {author} {\bibfnamefont {R.}~\bibnamefont
  {{Grzywacz}}}, \bibinfo {author} {\bibfnamefont {L.}~\bibnamefont
  {{Hlophe}}}, \bibinfo {author} {\bibfnamefont {J.~W.}\ \bibnamefont
  {{Holt}}}, \bibinfo {author} {\bibfnamefont {H.}~\bibnamefont {{Iwasaki}}},
  \bibinfo {author} {\bibfnamefont {D.}~\bibnamefont {{Lee}}}, \bibinfo
  {author} {\bibfnamefont {S.~M.}\ \bibnamefont {{Lenzi}}}, \bibinfo {author}
  {\bibfnamefont {S.}~\bibnamefont {{Liddick}}}, \bibinfo {author}
  {\bibfnamefont {R.}~\bibnamefont {{Lubna}}}, \bibinfo {author} {\bibfnamefont
  {A.~O.}\ \bibnamefont {{Macchiavelli}}}, \bibinfo {author} {\bibfnamefont
  {G.}~\bibnamefont {{Mart{\'\i}nez-Pinedo}}}, \bibinfo {author} {\bibfnamefont
  {A.}~\bibnamefont {{McCoy}}}, \bibinfo {author} {\bibfnamefont
  {A.}~\bibnamefont {{Mercenne}}}, \bibinfo {author} {\bibfnamefont
  {K.}~\bibnamefont {{Minamisono}}}, \bibinfo {author} {\bibfnamefont
  {B.}~\bibnamefont {{Monteagudo}}}, \bibinfo {author} {\bibfnamefont
  {P.}~\bibnamefont {{Navratil}}}, \bibinfo {author} {\bibfnamefont
  {R.}~\bibnamefont {{Ringle}}}, \bibinfo {author} {\bibfnamefont {G.~H.}\
  \bibnamefont {{Sargsyan}}}, \bibinfo {author} {\bibfnamefont
  {H.}~\bibnamefont {{Schatz}}}, \bibinfo {author} {\bibfnamefont {M.-C.}\
  \bibnamefont {{Spieker}}}, \bibinfo {author} {\bibfnamefont {A.}~\bibnamefont
  {{Volya}}}, \bibinfo {author} {\bibfnamefont {R.~G.~T.}\ \bibnamefont
  {{Zegers}}}, \bibinfo {author} {\bibfnamefont {V.}~\bibnamefont
  {{Zelevinsky}}}, \ and\ \bibinfo {author} {\bibfnamefont {X.}~\bibnamefont
  {{Zhang}}},\ }\href {\doibase 10.1088/1361-6471/adb449} {\bibfield  {journal}
  {\bibinfo  {journal} {Journal of Physics G Nuclear Physics}\ }\textbf
  {\bibinfo {volume} {52}},\ \bibinfo {eid} {050501} (\bibinfo {year}
  {2025})},\ \Eprint {http://arxiv.org/abs/2410.06144} {arXiv:2410.06144
  [nucl-th]} \BibitemShut {NoStop}%
\bibitem [{\citenamefont {{Hornick}}\ \emph {et~al.}(2018)\citenamefont
  {{Hornick}}, \citenamefont {{Tolos}}, \citenamefont {{Zacchi}}, \citenamefont
  {{Christian}},\ and\ \citenamefont
  {{Schaffner-Bielich}}}]{2018PhRvC..98f5804H}%
  \BibitemOpen
  \bibfield  {author} {\bibinfo {author} {\bibfnamefont {N.}~\bibnamefont
  {{Hornick}}}, \bibinfo {author} {\bibfnamefont {L.}~\bibnamefont {{Tolos}}},
  \bibinfo {author} {\bibfnamefont {A.}~\bibnamefont {{Zacchi}}}, \bibinfo
  {author} {\bibfnamefont {J.-E.}\ \bibnamefont {{Christian}}}, \ and\ \bibinfo
  {author} {\bibfnamefont {J.}~\bibnamefont {{Schaffner-Bielich}}},\ }\href
  {\doibase 10.1103/PhysRevC.98.065804} {\bibfield  {journal} {\bibinfo
  {journal} {\prc}\ }\textbf {\bibinfo {volume} {98}},\ \bibinfo {eid} {065804}
  (\bibinfo {year} {2018})},\ \Eprint {http://arxiv.org/abs/1808.06808}
  {arXiv:1808.06808 [astro-ph.HE]} \BibitemShut {NoStop}%
\bibitem [{\citenamefont {{Villar}}\ \emph {et~al.}(2017)\citenamefont
  {{Villar}}, \citenamefont {{Guillochon}}, \citenamefont {{Berger}},
  \citenamefont {{Metzger}}, \citenamefont {{Cowperthwaite}}, \citenamefont
  {{Nicholl}}, \citenamefont {{Alexander}}, \citenamefont {{Blanchard}},
  \citenamefont {{Chornock}}, \citenamefont {{Eftekhari}}, \citenamefont
  {{Fong}}, \citenamefont {{Margutti}},\ and\ \citenamefont
  {{Williams}}}]{2017ApJ...851L..21V}%
  \BibitemOpen
  \bibfield  {author} {\bibinfo {author} {\bibfnamefont {V.~A.}\ \bibnamefont
  {{Villar}}}, \bibinfo {author} {\bibfnamefont {J.}~\bibnamefont
  {{Guillochon}}}, \bibinfo {author} {\bibfnamefont {E.}~\bibnamefont
  {{Berger}}}, \bibinfo {author} {\bibfnamefont {B.~D.}\ \bibnamefont
  {{Metzger}}}, \bibinfo {author} {\bibfnamefont {P.~S.}\ \bibnamefont
  {{Cowperthwaite}}}, \bibinfo {author} {\bibfnamefont {M.}~\bibnamefont
  {{Nicholl}}}, \bibinfo {author} {\bibfnamefont {K.~D.}\ \bibnamefont
  {{Alexander}}}, \bibinfo {author} {\bibfnamefont {P.~K.}\ \bibnamefont
  {{Blanchard}}}, \bibinfo {author} {\bibfnamefont {R.}~\bibnamefont
  {{Chornock}}}, \bibinfo {author} {\bibfnamefont {T.}~\bibnamefont
  {{Eftekhari}}}, \bibinfo {author} {\bibfnamefont {W.}~\bibnamefont {{Fong}}},
  \bibinfo {author} {\bibfnamefont {R.}~\bibnamefont {{Margutti}}}, \ and\
  \bibinfo {author} {\bibfnamefont {P.~K.~G.}\ \bibnamefont {{Williams}}},\
  }\href {\doibase 10.3847/2041-8213/aa9c84} {\bibfield  {journal} {\bibinfo
  {journal} {\apjl}\ }\textbf {\bibinfo {volume} {851}},\ \bibinfo {eid} {L21}
  (\bibinfo {year} {2017})},\ \Eprint {http://arxiv.org/abs/1710.11576}
  {arXiv:1710.11576 [astro-ph.HE]} \BibitemShut {NoStop}%
\bibitem [{\citenamefont {{Mondal}}\ and\ \citenamefont
  {{Gulminelli}}(2022)}]{2022PhRvD.105h3016M}%
  \BibitemOpen
  \bibfield  {author} {\bibinfo {author} {\bibfnamefont {C.}~\bibnamefont
  {{Mondal}}}\ and\ \bibinfo {author} {\bibfnamefont {F.}~\bibnamefont
  {{Gulminelli}}},\ }\href {\doibase 10.1103/PhysRevD.105.083016} {\bibfield
  {journal} {\bibinfo  {journal} {\prd}\ }\textbf {\bibinfo {volume} {105}},\
  \bibinfo {eid} {083016} (\bibinfo {year} {2022})},\ \Eprint
  {http://arxiv.org/abs/2111.04520} {arXiv:2111.04520 [nucl-th]} \BibitemShut
  {NoStop}%
\bibitem [{\citenamefont {{Imam}}\ \emph {et~al.}(2024)\citenamefont {{Imam}},
  \citenamefont {{Mukherjee}}, \citenamefont {{Agrawal}},\ and\ \citenamefont
  {{Banerjee}}}]{2024PhRvC.109b5804I}%
  \BibitemOpen
  \bibfield  {author} {\bibinfo {author} {\bibfnamefont {S.~M.~A.}\
  \bibnamefont {{Imam}}}, \bibinfo {author} {\bibfnamefont {A.}~\bibnamefont
  {{Mukherjee}}}, \bibinfo {author} {\bibfnamefont {B.~K.}\ \bibnamefont
  {{Agrawal}}}, \ and\ \bibinfo {author} {\bibfnamefont {G.}~\bibnamefont
  {{Banerjee}}},\ }\href {\doibase 10.1103/PhysRevC.109.025804} {\bibfield
  {journal} {\bibinfo  {journal} {\prc}\ }\textbf {\bibinfo {volume} {109}},\
  \bibinfo {eid} {025804} (\bibinfo {year} {2024})},\ \Eprint
  {http://arxiv.org/abs/2305.11007} {arXiv:2305.11007 [nucl-th]} \BibitemShut
  {NoStop}%
\bibitem [{\citenamefont {{Wouters}}\ \emph {et~al.}(2025)\citenamefont
  {{Wouters}}, \citenamefont {{Pang}}, \citenamefont {{Koehn}}, \citenamefont
  {{Rose}}, \citenamefont {{Somasundaram}}, \citenamefont {{Tews}},
  \citenamefont {{Dietrich}},\ and\ \citenamefont {{Van Den
  Broeck}}}]{2025arXiv250415893W}%
  \BibitemOpen
  \bibfield  {author} {\bibinfo {author} {\bibfnamefont {T.}~\bibnamefont
  {{Wouters}}}, \bibinfo {author} {\bibfnamefont {P.~T.~H.}\ \bibnamefont
  {{Pang}}}, \bibinfo {author} {\bibfnamefont {H.}~\bibnamefont {{Koehn}}},
  \bibinfo {author} {\bibfnamefont {H.}~\bibnamefont {{Rose}}}, \bibinfo
  {author} {\bibfnamefont {R.}~\bibnamefont {{Somasundaram}}}, \bibinfo
  {author} {\bibfnamefont {I.}~\bibnamefont {{Tews}}}, \bibinfo {author}
  {\bibfnamefont {T.}~\bibnamefont {{Dietrich}}}, \ and\ \bibinfo {author}
  {\bibfnamefont {C.}~\bibnamefont {{Van Den Broeck}}},\ }\href {\doibase
  10.48550/arXiv.2504.15893} {\bibfield  {journal} {\bibinfo  {journal} {arXiv
  e-prints}\ ,\ \bibinfo {eid} {arXiv:2504.15893}} (\bibinfo {year} {2025})},\
  \Eprint {http://arxiv.org/abs/2504.15893} {arXiv:2504.15893 [astro-ph.HE]}
  \BibitemShut {NoStop}%
\bibitem [{\citenamefont {{Essick}}\ \emph
  {et~al.}(2021{\natexlab{b}})\citenamefont {{Essick}}, \citenamefont
  {{Landry}}, \citenamefont {{Schwenk}},\ and\ \citenamefont
  {{Tews}}}]{2021PhRvC.104f5804E}%
  \BibitemOpen
  \bibfield  {author} {\bibinfo {author} {\bibfnamefont {R.}~\bibnamefont
  {{Essick}}}, \bibinfo {author} {\bibfnamefont {P.}~\bibnamefont {{Landry}}},
  \bibinfo {author} {\bibfnamefont {A.}~\bibnamefont {{Schwenk}}}, \ and\
  \bibinfo {author} {\bibfnamefont {I.}~\bibnamefont {{Tews}}},\ }\href
  {\doibase 10.1103/PhysRevC.104.065804} {\bibfield  {journal} {\bibinfo
  {journal} {\prc}\ }\textbf {\bibinfo {volume} {104}},\ \bibinfo {eid}
  {065804} (\bibinfo {year} {2021}{\natexlab{b}})},\ \Eprint
  {http://arxiv.org/abs/2107.05528} {arXiv:2107.05528 [nucl-th]} \BibitemShut
  {NoStop}%
\end{thebibliography}%

\appendix

\onecolumngrid
\section{Appendix: the $\beta$-equilibrium of neutron stars}
\label{sec:appendix}

The general cold equation of state (EoS) of nuclear matter is a two-dimensional function that depends on both the baryon number density $n_{_B}$ and the proton fraction $Y_p$. It can be written as:
\begin{eqnarray}
  \label{eq:mass_dist}
  E(n_{_B}, Y_p) & = & E_0(n_{_B}) + E_{\rm sym}(n_{_B})(1-2Y_p)^2 + O(Y_P^4),
\end{eqnarray}
where $E(n_{_B}, Y_p)$ represents the energy per baryon of nuclear matter. The first term $E_0(n_{_B})$ is the energy per baryon of symmetric nuclear matter, while the second term $E_{\rm sym}(n_{_B})$ denotes the symmetry energy characterizing the energy cost associated with varying proton fraction. They can be expanded around the saturation density $n_0$ as:
\begin{eqnarray}
  \label{eq:esym_expand}
  E_0(n_{_B}) & = & E_0 + \frac{K_0}{2} \left( \frac{n_{_B}-n_0}{3n_0} \right)^2 + \frac{J_0}{6} \left( \frac{n_{_B}-n_0}{3n_0} \right)^3  + O\biggl((n_{_B}-n_0)^4\biggr), \\
  E_{\rm sym}(n_{_B}) & = & E_{\rm sym} + L_{\rm sym} \left( \frac{n_{_B}-n_0}{3n_0} \right) + \frac{K_{\rm sym}}{2} \left( \frac{n_{_B}-n_0}{3n_0} \right)^2  + O\biggl((n_{_B}-n_0)^3\biggr).
\end{eqnarray}

In contrast, the cold neutron star EoS is barotropic due to the $\beta$-equilibrium condition, which includes the contribution of electrons and fixes the proton fraction $Y_p$ at each density by minimizing the energy per baryon. This $\beta$-equilibrium condition reduces the two-dimensional EoS $E(n_{_B}, Y_p)$ to a one-dimensional function $E(n_{_B}, Y_p(n_{_B}))$, leading to the degeneracies among nuclear properties discussed in the main text. Therefore, it is of interest to present the posteriors of proton fraction and examine how it is constrained by the inspiral GW signals.

\begin{figure}
\vspace{-0.3cm}
{\centering
\includegraphics[width=0.49\textwidth]{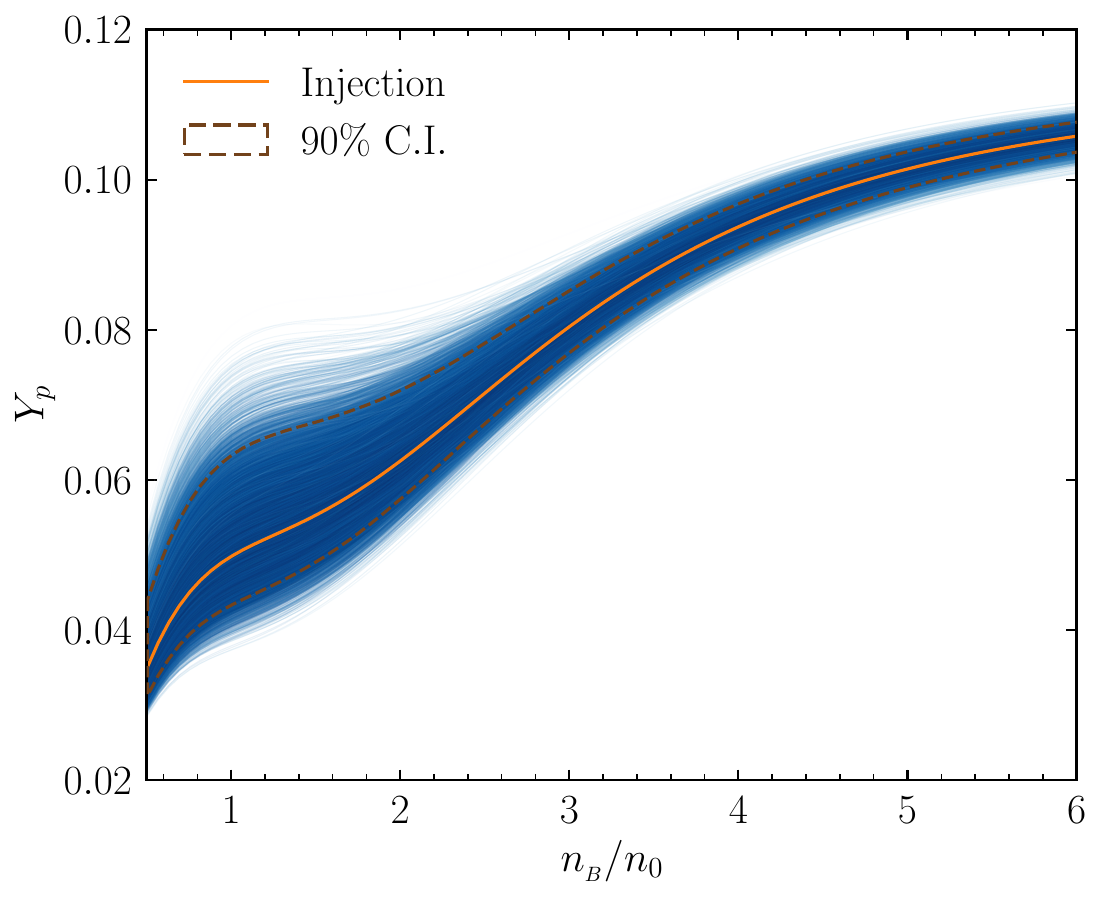}}
\caption{Posterior electron fraction as a function of number density, with notations as in Fig.~\ref{fig:posteriors}.}
\label{fig:Ye_post}
\vspace{-0.1cm}
\end{figure}

We display the posterior proton fraction as a function of number density in Fig.~\ref{fig:Ye_post}. Similar to the symmetry energy, the proton fraction is also poorly constrained. The uncertainties can reach up to $40\%$ near the saturation density, and remain above $10\%$ even at $3n_0$. The highly uncertain proton fraction reflects the degeneracies in the symmetry energy. If we look at the $\beta$-equilibrium condition of NS
\begin{eqnarray}
  \label{eq:mass_dist}
  \mu_e & = & \mu_n - \mu_p = 4 E_{\rm sym}(n_{_B})(1-2Y_p) + O(Y_P^4) \approx \left(3 \pi^2 n_{_B} Y_p\right)^{1/3},
\end{eqnarray}
where we neglect the mass of the electron in the last step. This relation shows that the proton fraction is solely determined by the symmetry energy. The tidal deformability measured by inspiral GW observations can only determine or constrain the EoS at $\beta$-equilibrium, which is a combination of symmetric nuclear matter EoS and the symmetry energy. Therefore, different combinations of symmetry energy, symmetric nuclear matter EoS, and proton fractions can yield nearly identical neutron star EoSs and tidal deformabilities, resulting in the large uncertainties in all of them.

On the other hand, the large uncertainties in proton fraction can be significant for those phenomena that depend on the composition and out-of-equilibrium effects, \eg the bulk viscosity, matter ejection, and post-merger GW signals. These signals serve as crucial probes to break the degeneracies and provide complementary information about dense matter properties.

\end{document}